\documentclass[sigconf,10pt]{acmart}

\usepackage{graphicx}
\usepackage{balance}  
\usepackage{booktabs} 
\usepackage{tabularx}
\usepackage{epstopdf}
\usepackage{multirow,multicol,enumitem,epsfig,url,makecell}
\usepackage[ruled, vlined, linesnumbered]{algorithm2e}
\usepackage{subfig}

\newcommand{\eat}[1]{}

\def\done{\hspace*{\fill} {$\square$}}
\def\header{\vspace{2.5mm} \noindent}

\def\mymathhyphen{{\hbox{-}}}

\newcommand{\E}{{\mathbb E}\xspace}
\newcommand{\ar}{{\widehat{\boldsymbol{\rho}}_s}\xspace}
\newcommand{\qs}{{\mathbf{q}_s}\xspace}
\newcommand{\rks}{{\mathbf{r}^{(k)}_s}\xspace}

\newcommand{\rs}{{\boldsymbol{\rho}}_s\xspace}

\newcommand{\as}{\mathbf{a}_s\xspace}
\newcommand{\bs}{\mathbf{b}_s\xspace}

\newcommand{\rbs}{\mathbf{rb}^{(k)}_s\xspace}

\newcommand{\nib}{$\mathsf{Nibble}$\xspace}
\newcommand{\prnib}{$\mathsf{PR\mymathhyphen Nibble}$\xspace}

\newcommand{\simlocal}{$\mathsf{SimpleLocal}$\xspace}
\newcommand{\crd}{$\mathsf{CRD}$\xspace}

\newcommand{\mctxt}{Monte\textrm{-}Carlo}
\newcommand{\mc}{$\mathsf{\mctxt}$\xspace}

\newcommand{\rswk}{$k\textrm{-}\mathsf{RandomWalk}$\xspace}
\newcommand{\hkpushtxt}{HK\mymathhyphen Push}
\newcommand{\hkpush}{$\mathsf{\hkpushtxt}$\xspace}
\newcommand{\hkpushb}{$\boldsymbol{\mathsf{\hkpushtxt}}$\xspace}
\newcommand{\hkpushplus}{$\mathsf{HK\mymathhyphen Push+}$\xspace}
\newcommand{\pukratxt}{TEA}
\newcommand{\pukra}{$\mathsf{\pukratxt}$\xspace}
\newcommand{\pukrab}{$\boldsymbol{\mathsf{\pukratxt}}$\xspace}
\newcommand{\pukraplus}{$\mathsf{\pukratxt+}$\xspace}
\newcommand{\pukraplusb}{$\boldsymbol{\mathsf{\pukratxt+}}$\xspace}
\newcommand{\chkprtxt}{ClusterHKPR}
\newcommand{\chkpr}{$\mathsf{\chkprtxt}$\xspace}
\newcommand{\chkprb}{$\boldsymbol{\mathsf{\chkprtxt}}$\xspace}
\newcommand{\hkrelaxtxt}{HK\mymathhyphen Relax}
\newcommand{\hkrelax}{$\mathsf{\hkrelaxtxt}$\xspace}
\newcommand{\hkrelaxb}{$\boldsymbol{\mathsf{\hkrelaxtxt}}$\xspace}

\newtheorem{definition}{Definition}
\newtheorem{theorem}{Theorem}
\newtheorem{lemma}{Lemma}

\let\oldnl\nl
\newcommand{\nonl}{\renewcommand{\nl}{\let\nl\oldnl}}

\newtheorem{example}{Example}

\usepackage{balance}

\def\BibTeX{{\rm B\kern-.05em{\sc i\kern-.025em b}\kern-.08emT\kern-.1667em\lower.7ex\hbox{E}\kern-.125emX}}

\copyrightyear{2019} 
\acmYear{2019} 
\setcopyright{acmcopyright}
\acmConference[SIGMOD '19]{2019 International Conference on Management of Data}{June 30-July 5, 2019}{Amsterdam, Netherlands}
\acmBooktitle{2019 International Conference on Management of Data (SIGMOD '19), June 30-July 5, 2019, Amsterdam, Netherlands}
\acmPrice{15.00}
\acmDOI{10.1145/3299869.3319886}
\acmISBN{978-1-4503-5643-5/19/06}

\begin{document}
\title{Efficient Estimation of Heat Kernel PageRank for Local Clustering}
\subtitle{[Technical Report]}

\author{Renchi Yang}
\affiliation{\institution{Nanyang Technological University}}
\email{yang0461@e.ntu.edu.sg}

\author{Xiaokui Xiao}
\affiliation{\institution{National University of Singapore}}
\email{xkxiao@nus.edu.sg}

\author{Zhewei Wei}
\affiliation{\institution{Renmin University of China}}
\email{zhewei@ruc.edu.cn}
\authornote{Work partially done at Beijing Key Laboratory of Big Data Management and Analysis Methods.}

\author{Sourav S Bhowmick}
\affiliation{\institution{Nanyang Technological University}}
\email{assourav@ntu.edu.sg}

\author{Jun Zhao}
\affiliation{\institution{Nanyang Technological University}}
\email{junzhao@ntu.edu.sg}

\author{Rong-Hua Li}
\affiliation{\institution{Beijing Institute of Technology}}
\email{lironghuascut@gmail.com}

\sloppy

\begin{abstract}
Given an undirected graph $G$ and a {\em seed} node $s$, the {\em local clustering} problem aims to identify a high-quality cluster containing $s$ in time roughly proportional to the size of the cluster, regardless of the size of $G$. This problem finds numerous applications on large-scale graphs. 
Recently, {\em heat kernel PageRank} (HKPR), which is a measure of the proximity of nodes in graphs, is applied to this problem and found to be more efficient compared with prior methods. 
However, existing solutions for computing HKPR either are prohibitively expensive or provide unsatisfactory error approximation on HKPR values, rendering them impractical especially on billion-edge graphs.

In this paper, we present \pukra and \pukraplus, two novel local graph clustering algorithms based on HKPR, to address the aforementioned limitations. Specifically, these algorithms provide non-trivial theoretical guarantees in \textit{relative error} of HKPR values and the time complexity. The basic idea is to utilize deterministic graph traversal to produce a rough estimation of exact HKPR vector, and then exploit Monte-Carlo random walks to refine the results in an optimized and non-trivial way. In particular, \pukraplus offers practical efficiency and effectiveness due to non-trivial optimizations. Extensive experiments on real-world datasets demonstrate that \pukraplus outperforms the state-of-the-art algorithm by more than four times on most benchmark datasets in terms of computational time when achieving the same clustering quality, and in particular, is an order of magnitude faster on large graphs including the widely studied {\em Twitter} and {\em Friendster} datasets.
\end{abstract} 


%

\maketitle

\thispagestyle{plain}
\pagestyle{plain}

\sloppy

\section{Introduction} \label{sec:intro}

Graph clustering is a fundamental problem that finds numerous applications, \textit{e.g.,} community detection ~\cite{fortunato2010community,leskovec2010empirical,wang2015community}, image segmentation \cite{felzenszwalb2004efficient,tolliver2006graph}, and protein grouping in biological networks \cite{liao2009isorankn,voevodski2009finding}. The problem has been studied extensively in the literature, and yet, clustering massive graphs remains a challenge in terms of computation efficiency. This motivates a series of algorithms \cite{spielman2004nearly,andersen2006local,andersen2009finding,gharan2012approximating,spielman2013local,zhu2013local,chung2014hklocal,orecchia2014flow,kloster2014heat,avron2015community,veldt2016simple,wang2017capacity} for {\em local clustering}, which takes as input an undirected graph $G$ and a {\em seed} node $s$, and identifies a cluster (\textit{i.e.,} a set of nodes) containing $s$ in time depending on the size of the cluster, regardless of the size of $G$.

Local clustering algorithms have the potential to underpin interactive exploration of massive graphs. Specifically, they can facilitate exploration of a relatively small region of a large graph that is of interest to a user. For example, consider Bob, a budding entrepreneur, who is interested in exploring the local clusters of visionary entrepreneurs in \textit{Twitter}. Particularly, he wishes to begin his exploration with the cluster associated with \textsf{Elon Musk} (\textit{i.e.,} seed). Since Bob thinks that Elon Musk is an inspirational entrepreneur, he would like to explore if there are any other notable entrepreneurs (e.g., \textsf{Kevin Rose}) in Elon's local cluster (\textit{e.g.,} followers, followees) and wishes to further explore the local neighborhoods of these entrepreneurs. In order to ensure a palatable and non-disruptive interactive experience, Bob needs an efficient local clustering framework that can return high quality clusters within few seconds. \textit{Which existing local clustering framework can Bob utilize for his exploration?}

\color{black} Spielman and Teng \cite{spielman2004nearly,spielman2013local} are the first to study the local clustering problem, and they propose a random-walk-based algorithm, \nib, that optimizes the {\it conductance} \cite{bollobas1998modern} of the cluster returned. Specifically, the conductance of cluster $S$ is defined as $\Phi(S)=\frac{|\mathrm{cut}(S)|}{\min\{\mathrm{vol}(S), 2m - \mathrm{vol}(S)\}}$, where $\mathrm{vol}(S)$ is the sum of the degrees of all nodes in $S$, $m$ is the number of edges in the graph $G$, and $|\mathrm{cut}(C)|$ is number of edges with one endpoint in $S$ and another not in $S$. Intuitively, if a cluster $C$ has a small conductance, then the nodes in $S$ are better connected to each other than to the nodes apart from $S$, in which case $S$ should be considered a good cluster. This algorithm is subsequently improved in a series of work \cite{andersen2006local,andersen2009finding,gharan2012approximating,spielman2013local,zhu2013local,chung2014hklocal,orecchia2014flow,wang2017capacity,veldt2016simple} that aims to either improve the efficiency of local clustering or reduce the conductance of the cluster returned.

\begin{small}
	\begin{table*} \small
		\renewcommand{\arraystretch}{1}
		\centering
		\caption{Theoretical guarantee of our solution against that of the state-of-the-art solutions.} \label{tbl:algos} \vspace{-3.5mm}
		\begin{tabular}{|l|l|l|}
			\hline
			\multicolumn{1}{|c|}{{\bf Algorithm}} &
			\multicolumn{1}{c|}{{\bf Accuracy Guarantee}} &
			\multicolumn{1}{c|}{{\bf Time Complexity}} \\
			\hline
			{\chkpr} \cite{chung2014hklocal}& with probability at least $1-\epsilon$, $\begin{cases}
			|\ar[v]-\rs[v]| \le \epsilon\cdot \rs[v], & \textrm{if } \rs[v]> \epsilon \\[0.2mm]
			|\ar[v]-\rs[v]| \le \epsilon, & \textrm{otherwise},
			\end{cases}$ & $O\left(\frac{t\log{(n)}}{\epsilon^3}\right)$ \\
			\hline
			{\hkrelax} \cite{kloster2014heat} & $\frac{1}{d(v)}\left|\ar[v]-\rs[v]\right|< \epsilon_a$ & $O\left(\frac{te^{t}\log{(1/\epsilon_a)}}{\epsilon_a}\right)$\\
			\hline
			{Our solutions} & with probability at least $1 - p_f$, $\begin{cases}
			\frac{1}{d(v)}\left|\ar[v]-\rs[v]\right|\le \epsilon_r\cdot \frac{\rs[v]}{d(v)}, & \textrm{if } \frac{\rs[v]}{d(v)} > \delta \\[1mm]
			\frac{1}{d(v)}\left|\ar[v]-\rs[v]\right|\le \epsilon_r\cdot\delta, & \textrm{otherwise},
			\end{cases}$ & $O\left(\frac{t\log{(n/p_f)}}{\epsilon^2_r\cdot\delta}\right)$ \\
			\hline
		\end{tabular}
		\vspace{-2mm}
	\end{table*}
\end{small}

The state-of-the-art solutions~\cite{chung2014hklocal,kloster2014heat} for local clustering are based on {\em heat kernel PageRank} (HKPR) \cite{chung2007heat}, which is a measure of the proximity of nodes in graphs. Given a seed node $s$, these solutions first compute a vector $\ar$ where each element $\ar[v]$ approximates the HKPR value of a node $v$ with respect to $s$ (i.e., $\ar[v]$ approximately measures the proximity of $s$ to $v$). Then, they utilize $\ar$ to derive a local cluster $C$ containing $s$. It is shown that the quality of $S$ depends on the accuracy of $\ar$ \cite{kloster2014heat,chung2015computing}, in the sense that the conductance of $S$ tends to decrease with the approximation error in $\ar$. Therefore, existing HKPR-based solutions \cite{chung2014hklocal,kloster2014heat} all focus on striking a good trade-off between time efficiency and the accuracy of $\ar$. In particular, the current best solution \hkrelax \cite{kloster2014heat} ensures that (i) $\frac{1}{d(v)}\left|\ar[v]-\rs[v]\right|< \epsilon_a$ for any node $v$, where $\epsilon_a$ is a given threshold, $d(v)$ is the degree of node $v$ and $\rs[v]$ is the exact HKPR value of node $v$ with respect to $s$, and (ii) $\ar$ is computed in $O\left(\frac{te^{t}\log{(1/\epsilon_a)}}{\epsilon_a}\right)$ time, where $t$ is constant (referred to as the {\it heat constant}) used in the definition of HKPR.

\vspace{-1mm}
\header
{\bf Motivation.} The time complexity of \hkrelax has a large factor $e^{t}$, where $t$ (i.e., the heat constant) could be as large as  a few dozens \cite{chung2014hklocal,kloster2014heat,shun2016parallel}.  Consequently, it can be inefficient for several applications. For instance, reconsider Bob's endeavor to explore the local clusters of Elon Musk and Kevin Rose. \hkrelax consumes around 15s and 48s, respectively, to compute their local clusters. Such performance is disruptive for any interactive graph exploration. In addition, \hkrelax provides an accuracy guarantee on each $\frac{\ar[v]}{d(v)}$ in terms of its {\it absolute error}, but as we discuss in Section~\ref{sec:so}, this guarantee is less than ideal for accurate local clustering. Specifically, HKPR-based local clustering requires ranking each node $v$ in descending order of $\frac{\ar[v]}{d(v)}$, which we refer to as $v$'s {\it normalized HKPR}. To optimize this accuracy of this ranking, we observe that it is more effective to minimize the {\it relative errors} of normalized HKPR values than their absolute errors. To explain, we note that the normalized HKPR varies significantly from nodes to nodes. For the aforementioned ranking, nodes with large normalized HKPR could tolerate more absolute errors than nodes with small normalized HKPR, and hence, imposing the same absolute error guarantees on all nodes tend to produce sub-optimal results.

\vspace{-1mm}
\header
{\bf Our contributions.} Motivated by the deficiency of existing solutions, we present an in-depth study on HKPR-based local clustering, and make the following contributions. First, we formalize the problem of approximate HKPR computation with probabilistic relative error guarantees, and pinpoint why none of the existing techniques could provide an efficient solution to this problem.

Second, based on our problem formulation, we propose two new algorithms, \pukra and \pukraplus, both of which (i) take as input a seed node $s$, two thresholds $\epsilon_r, \delta$, and a failure probability $p_f$, and (ii) return an approximate HKPR vector $\ar$ where each element $\ar[v]$ with $\frac{\rs[v]}{d(v)}>\delta$ has at most $\epsilon_r$ relative error with at least $1 - p_f$ probability (i.e., all {\it significant} HKPR values are accurately approximated with high probability). The core of \pukra is an adaptive method that combines deterministic graph traversal with random walks to estimate normalized HKPR in a cost-effective manner, while \pukraplus significantly improves over \pukra in terms of practical efficiency by incorporating a number of non-trivial optimization techniques. Both algorithms have a time complexity of $O\left(\frac{t\log{(n/p_f)}}{\epsilon^2_r\cdot\delta}\right)$, which eliminates the exponential term $e^t$ in \hkrelax's efficiency bound (see Table~\ref{tbl:algos}).

Third, we experimentally evaluate them against the state of the art, using large datasets with up to 65 million nodes and 1.8 billion edges. Our results show that \pukraplus is up to an order of magnitude faster than competing methods when achieving the same clustering quality. In particular, it can compute the local clusters of Elon Musk and Kevin Rose within 1.3s and 6.1s, respectively, thereby facilitating interactive exploration.


%

\vspace{-1mm}
\header
{\bf Paper Organization.}
The rest of the paper is organized as follows. In Section~\ref{sec:bg}, we introduce background on heat kernel-based local clustering. An overview of our solution framework is presented in Section~\ref{sec:so}. We present \pukra and \pukraplus in Sections~\ref{sec:pukra} and~\ref{sec:opt}, respectively. Related work is reviewed in Section~\ref{sec:cpt}. We evaluate the practical efficiency of our algorithms against the competitors in Section~\ref{sec:exp}. Finally, Section~\ref{sec:ccl} concludes the paper. Proofs of theorems and lemmas are given in Appendix~\ref{sec:proof}. Table~\ref{tbl:notations} lists the notations that are frequently used in our paper. 

\section{Preliminaries} \label{sec:bg}

\subsection{Basic Terminology} \label{sec:notation}

Let $G=(V,E)$ be an undirected and unweighted graph, where $V$ and $E$ denote the node and edge sets, respectively. We use $d(v)$ to denote the degree of node $v$, and $\mathbf{A}$ to denote the adjacency matrix of $G$; i.e., $\mathbf{A}[i,j]=\mathbf{A}[j,i]=1$ if and only if $( v_i,v_j )\in E$. Let $\boldsymbol{\mathrm{D}}$ be the {\em diagonal degree matrix} of $G$, where $\mathbf{D}[i,i]=d(v_i)$. Then, the {\em probability transition matrix} (a.k.a.\ {\em random walk transition matrix}) for $G$ is defined as $\boldsymbol{\mathrm{P}}=\boldsymbol{\mathrm{D}}^{-1}\boldsymbol{\mathrm{A}}$. Accordingly, $\mathbf{P}^{k}[s,v]$ denotes the probability that a $k$-hop ($k\ge 1$) random walk from node $s$ would end at $v$. 


\begin{table}[!t]
	\centering
	\caption{Frequently used notations.} \label{tbl:notations}
	\vspace{-2mm}
	\renewcommand{\arraystretch}{1.2}
	\begin{small}
		\begin{tabularx}{85mm}{|c|X|}
			\hline
			{\bf Notation} &  {\bf Description}\\
			\hline
			$G$=$(V,E)$   & An undirected graph with node set $V$ and edge set $E$\\
			\hline
			$n, m$   & The numbers of nodes and edges in $G$, respectively\\
			\hline
			$N(v)$   & The set of neighbors of node $v$ \\
			\hline
			$d(v)$   & The degree of node $v$ \\
			\hline
			$\bar{d}$   & The average degree of the graph, i.e., $\frac{2m}{n}$ \\
			\hline
			$\boldsymbol{\mathrm{A}}, \boldsymbol{\mathrm{D}}, \boldsymbol{\mathrm{P}}$   & The adjacency, diagonal degree, and transition matrices of $G$\\
			\hline
			$t$ & The heat constant of HKPR\\
			\hline
			$\eta(k), \psi(k)$ & See Equation~(\ref{eq:eta}) and Equation~(\ref{eq:psi}), respectively\\
			\hline
			$\rs[v]$ & HKPR of $v$ w.r.t. $s$, defined by Equation~(\ref{eq:hkpr})\\
			\hline
			$\epsilon_r, \delta, p_f$ & Parameters of an approximate HKPR, as in Section~\ref{sec:so}\\
			\hline
			$\rks[v]$ & The $k$-hop residue of $v$ during performing push operations from $s$\\
			\hline
			$\qs[v]$ & The reserve of $v$ during performing push operations from $s$\\
			\hline
			$K$ & The maximum number of hops during performing push operations from the seed node\\
			\hline
		\end{tabularx}
		\vspace{-2mm}
	\end{small}
\end{table}


A {\it cluster} in $G$ is a node set $S \subseteq V$. Intuitively, a good cluster should be both internally cohesive and well separated from the remainder of $G$. We say that $S$ is a high-quality cluster if it has a small {\it conductance} \cite{bollobas1998modern} $\Phi(S)$, defined as:
\vspace{-1ex}\begin{equation*}
	\Phi(S)=\frac{|\mathrm{cut}(S)|}{\min\{(\mathrm{vol}(S), \mathrm{vol}(V\setminus S))\}},
\end{equation*}
where $\mathrm{vol}(S)$ is the {\it volume} of $S$, namely, the sum of the degrees of all nodes in $S$, and
$\mathrm{cut}(S)$ is the {\it cut} 
of $S$, i.e., the set of edges with one endpoint in $S$ and another not in $S$.

\subsection{Heat Kernel-based Local Clustering} \label{sec:hklc}
Given a {\em heat constant} $t$ and two nodes $u$ and $v$, the HKPR value from $u$ to $v$ is defined as the probability that a random walk of length $k$ starting from $u$ would end at $v$, where $k$ is sampled from the following Poisson distribution:
\vspace{-1ex}\begin{equation}\label{eq:eta}
\eta(k)=\frac{e^{-t}t^k}{k!}.
\end{equation}
Let $s$ be the seed node for local clustering. We define the HKPR vector $\rs$ of $s$ as an $n$-size vector, such that the $i$-th element of $\rs$ equals the HKPR value from $s$ to the $i$-th node in $G$. In addition, we use $\rs[v]$ to denote the HKPR value from $s$ to $v$, which is defined by
\vspace{-1ex}\begin{equation}\label{eq:hkpr}
\rs[v]=\sum_{k=0}^{\infty}{\eta(k)\cdot\mathbf{P}^k[s,v]}.
\end{equation}
Existing heat-kernel-based algorithms \cite{chung2014hklocal,kloster2014heat,chung2015distributed,shun2016parallel} all adopt a two-phase approach. In particular, they first compute an {\em approximate} HKPR vector $\ar$ for $s$, and then perform a {\it sweep} as follows:
\begin{enumerate}
	\item Take the set $S^*$ of nodes with non-zero values in $\widehat{\boldsymbol{\rho}}_s$.
	\item Sort the nodes $v\in S^*$ in descending order of $\frac{\ar[v]}{d(v)}$. Let $S^*_i$ be a set containing the first $i$ nodes in the sorted sequence.
	\item Inspect $S^*_i$ in ascending order of $i$. Return the set $S^*_i$ with the smallest conductance among the ones that have been inspected.
\end{enumerate}
It is shown in \cite{zhu2013local,shun2016parallel} that the above sweep can be conducted in ${O\left(|S^*|\cdot\log{|S^*|}\right)}$ time, assuming that $\widehat{\boldsymbol{\rho}}_s$ is given in a sparse representation with $O(|S^*|)$ entries. In contrast, the computation of $\widehat{\boldsymbol{\rho}}_s$ is much more costly, and hence, has been the main subject of research in existing work \cite{chung2014hklocal,kloster2014heat,chung2015distributed,shun2016parallel}.

\section{Solution Overview} \label{sec:so}

Our solution for local clustering is based on heat kernel PageRank, and it follows the same two-phase framework in the existing work \cite{chung2007heat,chung2009local,chung2014hklocal,kloster2014heat,chung2015distributed,shun2016parallel}. That is, we also compute an approximate HKPR vector $\ar$ for $s$, and then conduct a sweep on $\ar$. However, we require that $\ar$ should be a {\em $(d, \epsilon_r,\delta)$-approximate HKPR vector}, which is a criterion not considered in any existing work \cite{chung2007heat,chung2009local,chung2014hklocal,kloster2014heat,chung2015distributed,shun2016parallel}.

\begin{definition}($(d, \epsilon_r,\delta)$-approximate HKPR) \label{def:approx-hkpd}
	Let $\rs$ be the HKPR vector for a seed node $s$, and $\ar$ be an approximated version of $\rs$. $\ar$ is $(d, \epsilon_r,\delta)$-approximate if it satisfies the following conditions:
	\begin{itemize}
		\item For every $v \in V$ with $\frac{\rs[v]}{d(v)} > \delta$,
		\begin{equation*}\label{eq:hkpd-upper}
			\left| \frac{\ar[v]}{d(v)}-\frac{\rs[v]}{d(v)}\right| \le \epsilon_r \cdot \frac{\rs[v]}{d(v)};
		\end{equation*}
		\item For every $v \in V$ with $\frac{\rs[v]}{d(v)} \le \delta$,
		\begin{flalign*}\label{eq:hkpd-lower}
			&& \left| \frac{\ar[v]}{d(v)}-\frac{\rs[v]}{d(v)} \right| \le  \epsilon_r \cdot \delta. && \square
		\end{flalign*}
	\end{itemize}
\end{definition}

In other words, we require $\frac{\ar[v]}{d(v)}$ to provide a relative error guarantee when $\frac{\rs[v]}{d(v)} > \delta$, and an absolute error guarantee when $\frac{\rs[v]}{d(v)} \le \delta$. This is to ensure that when we sort the nodes in descending order or $\frac{\ar[v]}{d(v)}$ (which is a crucial step in the sweep for local clustering), the sorted sequence would be close to the one generated based on $\frac{\rs[v]}{d(v)}$. We do not consider relative error guarantees when $\frac{\rs[v]}{d(v)} \le \delta$, because (i) ensuring a small relative error for such a node $v$ requires an extremely accurate estimation of its normalized HKPR, which would incur significant computation overheads, and (ii) providing such high accuracy for $v$ is unnecessary, since $v$'s tiny normalized HKPR value indicates that it is not relevant to the result of local clustering.

By the definition of HKPR (in Equation~(\ref{eq:hkpr})), the HKPR value of $v$ w.r.t.\ $s$ is a weighted sum of $k$-hop random walk transition probabilities from $s$ to $v$, where $k$ is a Poisson distributed length. Thus, a straightforward method to compute $(d, \epsilon_r,\delta)$-approximate HKPR for seed node $s$ is to conduct Monte-Carlo simulations using a large number of random walks. Specifically, each random walk should start from $s$, and should have a length that is sampled from the Poisson distribution in Equation~(\ref{eq:eta}). Let $n_r$ be the total number of random walks, and $\ar[v]$ be the fraction of walks that end at a node $v$. Then, we can use $\ar[v]$ as an estimation of $\rs[v]$. By the Chernoff bound (see Lemma~\ref{lem:chernoff}) and union bound, it can be shown that $\ar$ is $(d, \epsilon_r,\delta)$-approximate with probability at least
\vspace{-1ex}\begin{equation*}
	1-n\cdot\exp\left(-\frac{n_r\cdot\epsilon^2_r\cdot\delta}{2(1+\epsilon_r/3)}\right).
\end{equation*}
Therefore, if we are to ensure that the above probability is at least $1 - p_f$, then we can set $n_r = \frac{2(1+\epsilon_r/3)\log{(n/p_f)}}{\epsilon^2_r\cdot\delta}$. In that case, the time required to generate the random walks is $O\left(\frac{t\log{(n/p_f)}}{\epsilon^2_r\cdot\delta}\right)$. The main issue of this Monte-Carlo approach, however, is that it incurs considerable overheads in practice (see our experimental results in Section~\ref{sec:exp-all}). To explain, consider a node $v$ with a small $\rs[v]$ but a relatively large $\frac{\rs[v]}{d(v)} > \delta$. By the requirements of $(d, \epsilon_r,\delta)$-approximation, we need to ensure that $|\ar[v] - \rs[v]| \le \epsilon_r \cdot \rs[v]$. In turn, this requires that the number $n_r$ of random walks should be large; otherwise, the number of walks that end at $v$ would be rather small, in which case the estimation of $\rs[v]$ would be inaccurate.

We also note that none of the existing methods \cite{chung2014hklocal,kloster2014heat} can be adopted to compute $(d, \epsilon_r,\delta)$-approximate HKPR efficiently. In particular, \hkrelax \cite{kloster2014heat} only ensures that for any node $v\in V$, $\frac{1}{d(v)}\left|\ar[v]-\rs[v]\right|< \epsilon_a.$ If we are to use \hkrelax for $(d, \epsilon_r,\delta)$-approximation, then we need to set $\epsilon_a=\epsilon_r\cdot\delta$, in which case its complexity would be $O\left(\frac{te^{t}\log{\left(\frac{1}{\epsilon_r\cdot\delta}\right)}}{\epsilon_r\cdot\delta}\right)$, which is considerably worse than the time complexity of the Monte-Carlo approach, due to the exponential term $e^t$. The \chkpr \cite{chung2014hklocal} algorithm suffers from a similar issue, as we point out in Section~\ref{sec:cpt}.

To mitigate the deficiencies of the aforementioned methods, we present (in Section~\ref{sec:pukra} and Section~\ref{sec:opt}) two more efficient HKPR algorithms that satisfy the following requirements:
\begin{enumerate}
	\item Return a $(d, \epsilon_r, \delta)$-approximate HKPR vector $\ar$ with at least $1 - p_f$ probability, where $p_f$ is a user-specified parameter;
	\item Run in $O\left(\frac{t\log(n/p_f)}{\epsilon^2_r \cdot \delta}\right)$ expected time.
\end{enumerate}


\section{The \pukrab Algorithm}\label{sec:pukra}


This section presents \pukra\footnote{\underline{T}wo-Phase H\underline{e}at Kernel \underline{A}pproximation}, our first-cut solution for $(d, \epsilon_r, \delta)$-approximate HKPR. \pukra is motivated by the inefficiency of the Monte-Carlo approach explained in Section~\ref{sec:so}, i.e., it requires a large number of random walks to accurately estimate HKPR values. To address this issue, we propose to combine \mc with a secondary algorithm that could help reduce the number of random walks needed. In particular, we first utilize the secondary algorithm to efficiently compute a rough estimation $\qs[v]$ of $\rs[v]$, and then perform random walks to refine each $\qs[v]$, so as to transform $\qs$ into a $(d, \epsilon_r, \delta)$-approximate HKPR vector $\ar$. Towards this end, there are several issues that we need to address:
\begin{enumerate}
	\item How to design a secondary algorithm that could generate a rough approximation of the HKPR vector at a small computation cost?
	\item How to enable \mc to leverage the output of the secondary algorithm for improved efficiency?
	\item How to ensure that the combination of \mc and the secondary algorithm still provides strong theoretical assurance in terms of time complexity and accuracy?
\end{enumerate}
To answer the above questions, we first present our choice of the secondary algorithm, referred to as \hkpush, in Section~\ref{sec:pukra-hkpush}; after that, we elaborate the integration of \hkpush and \mc in Section~\ref{sec:pukra-algo}, and then provide a theoretical analysis of the combined algorithm in Section~\ref{sec:pukra-als}.

\subsection{\hkpushb}\label{sec:pukra-hkpush}
Algorithm~\ref{alg:hkpush} shows the pseudo-code of our secondary algorithm, \hkpush, for deriving a rough approximation $\ar$ of the HKPR vector. Its basic idea is to begin with a vector $\ar$ where $\ar[s] = 1$  and $\ar[v] = 0$ for all nodes $v$ except $s$, and then perform a traversal of $G$ starting from $s$, and keep refining $\ar$ during the course of the traversal. In addition, to facilitate its combination with random walks, it not only returns an approximate HKPR vector $\qs$, but also outputs $K+1$ auxiliary vectors
$\mathbf{r}^{(0)}_s, \ldots, \mathbf{r}^{(K)}_s\in \mathbb{R}^n$ that could be used to guide the random walks conducted by \mc. We refer to $\qs$ as the {\it reserve vector} and $\mathbf{r}^{(k)}_s$ as the {\it $k$-hop residue vector}. Accordingly, for any node $v$, $\qs[v]$ and $\mathbf{r}^{(k)}_s[v]$ are referred to as the {\it reserve} and {\it $k$-hop residue} of $v$, respectively.

\begin{algorithm}[t]
	\begin{small}
		\caption{$\mathsf{HK\mymathhyphen Push}$}\label{alg:hkpush}
		\BlankLine
		\KwIn{Graph $G$, seed node $s$, threshold $r_{max}$}
		\KwOut{An approximate HKPR vector $\qs$ and $K+1$ residue vectors $\mathbf{r}^{(0)}_s, \ldots, \mathbf{r}^{(K)}_s$}
		$\qs \gets \mathbf{0}$, $\rks\gets\mathbf{0}$ for $k=0,\ldots$\;
		$\mathbf{r}^{(0)}_s[s] \gets 1$\;
		\While{$\exists{v\in V, k}$ such that $\rks[v]>r_{max}\cdot d(v)$}{
			$\qs[v] \gets \qs[v]+\frac{\eta(k)}{\psi(k)}\cdot\rks[v]$\;
			\For{$u\in N(v)$}{
				$\mathbf{r}^{(k+1)}_s[u]\gets\mathbf{r}^{(k+1)}_s[u]+\left(1-\frac{\eta(k)}{\psi(k)}\right) \cdot \frac{\rks[v]}{d(v)}$\;
			}
			$\rks[v]\gets 0$\;
		}
		$K \gets \max\left\{k \;\middle\vert\; \exists v \in V, \mathbf{r}^{(k)}_s[v] >0\right\}$\;
		\Return $\qs$ and $\mathbf{r}^{(0)}_s, \ldots, \mathbf{r}^{(K)}_s$\;
	\end{small}
\end{algorithm}

More specifically, \hkpush takes as input $G$, $s$, and a residue threshold $r_{max}$. It begins by setting all entries in $\qs$ and $\rks$ to zero, except that $\mathbf{r}^{(0)}_s[s] = 1$ (Lines 1-2). After that, it starts an iterative process to traverse $G$ from $s$ (Lines 3-7). In particular, in each iteration, it inspects the $k$-hop residue vectors to identify a node $v$ whose $k$-hop residue $\rks[v]$ is above $r_{max}\cdot d(v)$. If such a node $v$ exists, then the algorithm updates the reserve and $k$-hop residue of $v$, as well as the $(k+1)$-hop residues of $v$'s neighbors. In particular, it first adds $\frac{\eta(k)}{\psi(k)}$ fraction of $v$'s $k$-hop residue $\rks[v]$ to its reserve $\qs[v]$,  where $\eta(k)$ is as defined in Equation~(\ref{eq:eta}) and
\vspace{-1ex}
\begin{equation}\label{eq:psi}
\psi(k)=\sum_{\ell=k}^{\infty}{\eta(\ell)},
\end{equation}

\noindent
and then evenly distribute the other $1 - \frac{\eta(k)}{\psi(k)}$ fraction to the $(k+1)$-hop residues of $v$'s neighbors (Lines 4-6). For convenience, we refer to the operation of distributing a fraction of $v$'s $k$-hop residue to one of its neighbors as a {\em push} operation. The rationale of the aforementioned push operations is that, if a random walk from $s$ arrives at $v$ at the $k$-th hop, then it has $\frac{\eta(k)}{\psi(k)}$ probability to terminate at $v$, and has $1 - \frac{\eta(k)}{\psi(k)}$ probability to traverse to a randomly selected neighbor of $v$ at the next hop. After that, the algorithm sets $\rks[v] = 0$ (Line 7), and proceeds to the next iteration. After the iterative process terminates, \hkpush identifies the largest $K$ such that $\mathbf{r}^{(K)}_s$ has at least one non-zero entry, and returns $\qs$ and $\mathbf{r}^{(0)}_s, \ldots, \mathbf{r}^{(K)}_s$. The following lemma shows a crucial property of these reserve and residue vectors:
\begin{lemma}\label{lem:fwdeq}
	Consider any iteration in Algorithm~\ref{alg:hkpush}. Let $\qs$ and $\mathbf{r}^{(0)}_s, \ldots \mathbf{r}^{(K)}_s$ be the reserve and residue vectors constructed by the end of the iteration. We have
	\begin{equation}\label{eq:hkpush-invar}
	\textstyle \rs[v]=\qs[v]+\sum_{u\in V}{\sum_{k=0}^{K}{\rks[u]\cdot\mathbf{h}^{(k)}_u[v]}},
	\end{equation}
	where
	\begin{equation}\label{eq:hsum-def}
	\textstyle \mathbf{h}^{(k)}_u[v]=\sum_{\ell=0}^{\infty}{\frac{\eta(k+\ell)}{\psi(k)}\cdot\mathbf{P}^{\ell}[u,v]},
	\end{equation}
	i.e., $\mathbf{h}^{(k)}_u[v]$ is the probability that a random walk stops at $v$, conditioned on the $k$-hop of the walk is at $u$. \done
\end{lemma}

Intuitively, Lemma~\ref{lem:fwdeq} indicates that for any node $v$, $\qs[v]$ is a lower bound of $\rs[v]$ in any iteration in Algorithm~\ref{alg:hkpush}. Since each iteration of Algorithm~\ref{alg:hkpush} only increases the reserve of a selected node and never decreases any others, it guarantees that the difference between $\qs$ and $\rs$ monotonically decreases, i.e., $\qs$ becomes a better approximation of $\rs$ as the algorithm progresses. Although \hkpush may produce results that are far from satisfying the requirements of $(d, \epsilon_r, \delta)$-approximate HKPR, it is sufficient for the integration with \mc, as we show in Section~\ref{sec:pukra-algo}.

\begin{algorithm}[t]
	\begin{small}
		\caption{\rswk}\label{alg:skipped-walk}
		\BlankLine
		\KwIn{Graph $G$, node $u$, constant $k$, }
		\KwOut{An end node $v$}
		$\ell\gets k$\;
		$v_0 \gets u$\;
		\While{\bf{True}}{
			\If{$\mathsf{rand}(0,1) \le \frac{\eta(k+\ell)}{\psi(k+\ell)}$}{
				\bf{break}\;
			}\Else{
				Pick a neighbor $v_{\ell+1} \in N(v_{\ell})$ uniformly at random\;
				$\ell \gets \ell+1$\;
			}
		}
		\Return $v_{\ell}$\;
	\end{small}
\end{algorithm}

\vspace{-1ex}\subsection{Algorithm} \label{sec:pukra-algo}

\noindent
{\bf Basic Idea.} To incorporate \hkpush into \mc, we utilize Equation~(\ref{eq:hkpush-invar}), which shows that the exact HKPR vector $\rs$ can be expressed as a function of $\qs$,  $\rks$, and $\mathbf{h}^{(k)}_u[v]$ for any $u, v \in V$, and $k \in [0, K]$. Recall that $\qs$ and $\rks$ are outputs of \hkpush, while $\mathbf{h}^{(k)}_u[v]$ is the conditional probability that a random walk terminates at node $v$ given that its $k$-th hop is at node $u$. If we can accurately estimate $\mathbf{h}^{(k)}_u[v]$ for any given $u$, $v$, and $k$, then we can combine the estimated values with $\qs$ and $\rks$ to obtain an approximate version of $\rs$. Towards this end, we conduct random walks starting from $u$, and estimate $\mathbf{h}^{(k)}_u[v]$ based on the frequency that $v$ appears at the $k$-th hop of the random walks. Algorithm~\ref{alg:skipped-walk} shows the pseudo-code of our random walk generation method, referred to as \rswk.


\begin{algorithm}[t]
	\begin{small}
		\caption{\pukra}\label{alg:pukra}
		\BlankLine
		\KwIn{Graph $G$, seed node $s$, thresholds $\epsilon_r$ and $\delta$, threshold $r_{max}$, and failure probability $p_f$}
		\KwOut{A $(d,\epsilon_r,\delta)$-approximate HKPR vector $\ar$}
		\If{$\sum_{v\in V}{p_f^{d(v)-1}}\le 1$}{
			$p^{\prime}_f\gets p_f$\;
		}\Else{
			$p^{\prime}_f\gets \frac{p_f}{\sum_{v\in V}{p_f^{d(v)-1}}}$\;
		}
		$\omega \gets\frac{2(1+\epsilon_r/3)\log{(1/p^{\prime}_f)}}{\epsilon^2_r\delta}$\;
		$\left(\ar, \mathbf{r}^{(0)}_s, \ldots, \mathbf{r}^{(K)}_s \right)\gets$ \hkpush$(s, r_{max})$\;
		$\alpha \gets\sum_{k=0}^{K}{\sum_{u\in V}{\rks[u]}}$\;
		$n_r \gets \alpha \cdot \omega$\;
		\For{$i = 1$ to $n_r$}{
			Sample an entry $(u, k)$ from $V \times \{0, 1, \ldots, K\}$ with probability $\frac{\rks[u]}{\alpha}$\;
			$v \gets$ \rswk$(G, u, k)$\;
			$\ar[v] \gets \ar[v] + \frac{\alpha}{n_r}$\;
		}
		\Return $\ar$\;
	\end{small}
\end{algorithm}

%
The following lemma proves that \rswk samples each node $v$ with probability $\mathbf{h}^{(k)}_u[v]$.
\vspace{-1ex}\begin{lemma}\label{lem:srrw-approx}
	Given $G$, $u$, and $k$, for any node $v$, Algorithm~\ref{alg:skipped-walk} returns $v$ with probability $\mathbf{h}^{(k)}_u[v]$. \done
\end{lemma}
\vspace{-2mm}

\header
{\bf Details.} Algorithm~\ref{alg:pukra} illustrates the pseudo-code of \pukra, our first-cut solution that (i) incorporates \hkpush and \rswk and (ii) computes a $(d,\epsilon_r,\delta)$-approximate HKPR vector $\ar$ with at least $1 - p_f$ probability for any given seed node $s$. Given $G$, $\epsilon_r$, $\delta$, $r_{max}$, and failure probability $p_f$ as inputs, the algorithm starts by invoking \hkpush with three parameters: $G$, $s$, and $r_{max}$ (Line 6). It returns an approximate HKPR vector $\ar$ and $K+1$ residue vectors $\mathbf{r}^{(0)}_s, \ldots, \mathbf{r}^{(K)}_s$ from \hkpush. Then, \pukra proceeds to refine $\ar$ using \rswk (Lines 7-13). In particular, \pukra first computes the sum $\alpha$ of the residues in $\mathbf{r}^{(0)}_s, \ldots, \mathbf{r}^{(K)}_s$ (Line 7), and computes
\vspace{-1ex}\begin{equation*}
	\omega=\frac{2(1+\epsilon_r/3)\log{(1/p^{\prime}_f)}}{\epsilon^2_r\cdot\delta},
\end{equation*}
where
\vspace{-1ex}\begin{equation}\label{eq:pfail}
\textstyle p^{\prime}_f=\textstyle
\begin{cases}
p_f, & \textrm{if } \sum_{v\in V}{p_f^{d(v)-1}} \le 1 \\
\frac{p_f}{\sum_{v\in V}{p_f^{d(v)-1}}}, & \textrm{otherwise}.
\end{cases}
\end{equation}
Note that $p^\prime_f$ can be pre-computed when the graph $G$ is loaded.
Given $\omega$, \pukra performs $n_r = \alpha \cdot \omega$ random walks using \rswk (Lines 9-12), such that the starting point $u$ of each walk is sampled with probability $\frac{\rks[u]}{\alpha}$ (Line 10). Note that this sampling procedure can be conducted efficiently by conducting an {\it alias structure} \cite{walker1974new} on the non-zero elements in $\mathbf{r}^{(0)}_s, \ldots, \mathbf{r}^{(K)}_s$. For each random walk that ends at a node $v$, \pukra increases $\ar[v]$ by $\frac{\alpha}{n_r}$ (Line 12).

Observe that the parameter $r_{max}$ in \pukra controls the balance between its two main components: \hkpush and \rswk. In particular, by Algorithm~\ref{alg:hkpush}, the entries in $\mathbf{r}^{(0)}_s, \ldots, \mathbf{r}^{(K)}_s$ are bounded by $r_{max}$. Therefore, when $r_{max}$ is small, $\alpha = \sum_{k=0}^{K}{\sum_{u\in V}{\rks[u]}}$ would decrease, in which case the total number $\alpha \cdot \omega$ of random walks conducted by \pukra would also be small. As a trade-off, the processing cost of \hkpush would increase, as shown in the following lemma:
\vspace{-1ex}\begin{lemma}\label{lem:time-push}
	Given residue threshold $r_{max}$, Algorithm~\ref{alg:hkpush} runs in $O\left(\frac{1}{r_{max}}\right)$ time and requires $O\left(\frac{1}{r_{max}}\right)$ space (excluding the space required by the input graph). In addition, in the residue vectors $\mathbf{r}^{(0)}_s, \ldots, \mathbf{r}^{(K)}_s$ returned by Algorithm~\ref{alg:hkpush}, there are $O\left(\frac{1}{r_{max}}\right)$ non-zero elements in total.
\end{lemma}
To strike a balance between the costs incurred by \hkpush and \rswk, we set $r_{max} = O(\frac{1}{\omega \cdot t})$. In that case, the processing cost of \hkpush is $O(\omega \cdot t)$, while \rswk incurs $O(\alpha \omega t)$ expected cost, due to the following lemma:
\begin{lemma}\label{lem:time-krandwalk}
	The expected cost of each invocation of \rswk is $O(t)$.
\end{lemma}
Hence, setting $r_{max} = O(\frac{1}{\omega \cdot t})$ ensures that the overheads of \hkpush and \rswk are roughly comparable. 

\subsection{Analysis}\label{sec:pukra-als}

\noindent
{\bf Correctness.}
Let $\qs$ denote the approximate HKPR vector obtained from \hkpush in Line 6 of \pukra, and $\ar$ be the approximate HKPR vector eventually output by \pukra. In the following, we show that $\ar$ is a $(d,\epsilon_r,\delta)$-approximate HKPR vector.

First, by Lemma~\ref{lem:fwdeq}, we have the following equation for any node $v$:
\begin{equation}\label{eq:hkpr-a}
\rs[v]=\qs[v]+ \as[v],
\end{equation}
where
\begin{align}
	& \textstyle \as[v]=\alpha \cdot \sum_{k=0}^{K}{\sum_{u\in V}{\frac{\rks[u]}{\alpha}\cdot\mathbf{h}^{(k)}_u[v]}}\label{eq:pukra-as}.
\end{align}
Consider the $i$-th invocation of \rswk in \pukra. Let $(u, k)$ be the entry sampled by \pukra (in Line 10) before the invocation, and $v$ be the node returned by \rswk. Let $Y_i$ be a Bernoulli variable that equals $1$ if $v$ is returned, and $0$ otherwise. By Lemma~\ref{lem:srrw-approx},
\begin{equation}\label{eq:pukra_asum}
\textstyle \E[Y_i]=\sum_{u\in V}{\sum_{k=0}^{K}{\frac{\rks[u]}{\alpha}\cdot\mathbf{h}^{(k)}_u[v]}}.
\end{equation}
Combining Equations (\ref{eq:pukra-as}) and~(\ref{eq:pukra_asum}), we have
\begin{equation}\label{eq:pukra-exp-xi}
\textstyle \E\left[\sum_{i=1}^{n_r}{Y_i\cdot\frac{\alpha}{n_r}}\right]=\as[v],
\end{equation}
which indicates that $\sum_{i=1}^{n_r}{Y_i\cdot \frac{\alpha}{n_r}}$ is an unbiased estimator of $\as[v]$.
By the Chernoff bound (in Lemma~\ref{lem:chernoff}), we can prove that this estimator is highly accurate, based on which we obtain Theorem~\ref{thrm:pukra}.
\vspace{-1ex}\begin{lemma}[Chernoff Bound~\cite{chung2006concentration}]\label{lem:chernoff}
	Let $X_1,X_2,\cdots, X_{n_r} \in [0, 1]$ be i.i.d.\ random variables, and $X = \sum_{i=1}^{n_r}{X_i}$.
	Then,
	\begin{flalign*}\label{eq:chernoff}
		&& \textstyle \mathbb{P}[|X-\E[X]| \ge \lambda] \le \exp\left(-\frac{\lambda^2}{2\E[X]+2\lambda/3}\right). && \square
	\end{flalign*}
\end{lemma}
\vspace{-1ex}\begin{theorem}\label{thrm:pukra}
	\pukra outputs a $(d,\epsilon_r,\delta)$-approximate HKPR vector $\ar$ with probability at least $1-p_f$.
\end{theorem}

\vspace{-1ex}
\header
{\bf Time and Space Complexities.} Given $r_{max}=O\left(\frac{1}{\omega\cdot t}\right)$, \hkpush runs in $O(\omega\cdot t)$ time and $O(\omega\cdot t)$ space. In addition,  the computation of $\alpha$ as well as the construction of alias structure on $\rks$ can be done in time and space linear to the total number of non-zero entries in $\mathbf{r}^{(0)}_s, \ldots, \mathbf{r}^{(K)}_s$, which is $O(\omega\cdot t)$ (see Lemma~\ref{lem:time-push}). Furthermore, in terms of both space and time, the total expected cost incurred by the random walks in \pukra is $O(\alpha \omega t)$, where $\alpha < 1$. Therefore, the time complexity of \pukra is
\vspace{-1ex}\begin{equation*}
	O\left(\frac{1}{r_{max}}+\alpha\cdot \omega t\right)=O\left(\frac{t\log{(1/p^{\prime}_f)}}{\epsilon^2_r\cdot\delta}\right) = O\left(\frac{t\log{(n/p_f)}}{\epsilon^2_r\cdot\delta}\right),
\end{equation*}
and its space complexity is $O\left(n + m + \frac{t\log{(n/p_f)}}{\epsilon^2_r\cdot\delta}\right)$, where the $n+m$ term is due to storing of the input graph.




\begin{algorithm}[t]
	\begin{small}
		\caption{\hkpushplus}\label{alg:hkpushplus}
		\BlankLine
		\KwIn{Graph $G$, seed node $s$, thresholds $\epsilon_r$ and $\delta$, maximum hop number $K$, maximum number of pushes $n_p$}
		\KwOut{An approximate HKPR vector $\qs$ and $K+1$ residue vectors $\mathbf{r}^{(0)}_s, \ldots, \mathbf{r}^{(K)}_s$}
		$\qs \gets \mathbf{0}$, $\rks\gets\mathbf{0}$ for $k=0,\cdots,K$\;
		$\mathbf{r}^{(0)}_s[s] \gets 1$\;
		$i \gets 0$\;
		\While{$\exists{v\in V, k < K}$ such that $\rks[v]>\frac{\epsilon_r \cdot \delta}{K}\cdot d(v)$}{
			$i\gets i+d(v)$\;
			\If{$i \ge n_p$ \bf{or} $\sum_{\ell=0}^{K}{\max_{u\in V}{\left\{\frac{\mathbf{r}^{(\ell)}_s[u]}{d(u)}\right\}}}\le \epsilon_r \cdot \delta$}{
				\bf{break}\;
			}		
			\nonl Lines 8-11 are the same as Lines 4-7 in Algorithm~\ref{alg:hkpush}\;
		}
		\nonl Line 12 is the same as Line 9 in Algorithm~\ref{alg:hkpush}\;
	\end{small}
\end{algorithm}

\section{The TEA+ Algorithm}\label{sec:opt}

Although \pukra provides a strong accuracy guarantee, we observe in our experiments that it often performs a large number of random walks, which degrades its computation efficiency. One may attempt to reduce the cost of random walks by decreasing the residue threshold $r_{max}$ in \pukra (see the discussion in the end of Section~\ref{sec:pukra-algo}), but this cost reduction would be offset by the fact that \hkpush incurs a larger overhead when $r_{max}$ is small.

In this section, we present \pukraplus, an algorithm that significantly improves over \pukra in terms of practical efficiency without degrading its theoretical guarantees. \pukraplus is similar in spirit to \pukra in that it combines random walks with a variant of \hkpush, but there is a crucial difference: after \pukraplus obtains the residue vectors $\mathbf{r}^{(0)}_s, \ldots, \mathbf{r}^{(K)}_s$, it may reduce some entries in the residue vectors before performing random walks. That is, for each node $u$ that has a non-zero entry $\mathbf{r}^{(k)}_s[u]$ for some $k$, \pukraplus may choose to perform a smaller number of random walks from $u$ than \pukra does, which effectively reduces the total cost of random walks. Establishing the correctness of this pruning approach, however, is non-trivial.
In what follows, we first discuss in Section~\ref{sec:opt-hkpushplus} the extreme case where we can derive $(d, \epsilon_r,\delta)$-approximate HKPR while {\it ignoring all elements in the residue vectors $\mathbf{r}^{(0)}_s, \ldots, \mathbf{r}^{(K)}_s$}; after that, in Section~\ref{sec:opt-prune}, we generalize our discussions to the case where we reduce the non-zero entries in the residue vectors instead of completely omitting them.


\subsection{The Case without Random Walks}\label{sec:opt-hkpushplus}

Suppose that we are to let \pukra achieve $(d, \epsilon_r,\delta)$-approximation without performing any random walks. In that case, we would need to ensure that Line 6 of \pukra obtains a $(d,\epsilon_r,\delta)$-approximate HKPR vector from \hkpush. Towards this end, we present the following theorem:



\begin{theorem}\label{thrm:hkpush}
	Let $\qs$ and $\mathbf{r}^{(0)}_s, \ldots \mathbf{r}^{(K)}_s$ be the reserve and residue vectors returned by \hkpush. If
	\begin{equation} \label{eq:hkpush-con3}
	\sum_{\ell=0}^{K}{\max_{v\in V}{\left\{\frac{\mathbf{r}^{(\ell)}_s[v]}{d(v)}\right\}}}\le \epsilon_a,
	\end{equation}
	then, for any node $v$ in $G$, we have
	\begin{equation}\label{eq:abs-approx-k}
	\left|\frac{\qs[v]}{d(v)}-\frac{\rs[v]}{d(v)}\right| \le \epsilon_a.
	\end{equation}
\end{theorem}

Theorem~\ref{thrm:hkpush} provides a sufficient condition (i.e., Inequality~(\ref{eq:hkpush-con3})) for \hkpush to ensure $\epsilon_a$ absolute error in each $\frac{\qs[v]}{d(v)}$. By Definition~\ref{def:approx-hkpd}, such $\qs$ is a $(d,\epsilon_r,\delta)$-approximate HKPR vector as long as $\epsilon_a \le \epsilon_r \cdot \delta$. That said, it is rather inefficient to let \hkpush run until Inequality~(\ref{eq:hkpush-con3}) is satisfied. Instead, we propose to let \hkpush run with a fixed budget of processing cost. If it is able to satisfy Inequality~(\ref{eq:hkpush-con3}) with $\epsilon_a = \epsilon_r \cdot \delta$ before the budget is depleted, then we return $\qs$ as the final result; otherwise, we proceed to refine $\qs$ using random walks (see Section~\ref{sec:opt-prune}).

Based on the above discussion, we present \hkpushplus (in Algorithm~\ref{alg:hkpushplus}), which is a revised version of \hkpush with three major changes. First, \hkpushplus's input parameters include three thresholds $\epsilon_r$, $\delta$, and $n_p$, and it has two new termination conditions (in Line 6): (i) Inequality~(\ref{eq:hkpush-con3}) holds with $\epsilon_a = \epsilon_r \cdot \delta$; (ii) The number of push operations that it performs reaches $n_p$. Recall that a push operation refers to the operation of converting part of a node's $k$-hop residue to one of its neighbor's $(k+1)$-hop residue. In other words, \hkpushplus runs in $O(n_p)$ time and requires $O(n_p)$ space, 
and it returns a $(d,\epsilon_r,\delta)$-approximate HKPR vector whenever Inequality~(\ref{eq:hkpush-con3}) is satisfied.

Second, \hkpushplus judiciously performs push operations only on nodes $v$ with residue $\rks[v]>\frac{\epsilon_r \cdot \delta}{K}\cdot d(v)$ (Line 4), whereas \hkpush conducts push operations only when $\rks[v]$ is larger than an input given threshold $r_{max}\cdot d(v)$. The rationale is that \hkpushplus strives to reduce the $k$-hop residue of each node below $\frac{\epsilon_r \cdot \delta}{K}\cdot d(v)$, so as to satisfy Inequality~(\ref{eq:hkpush-con3}); in contrast, \hkpush is not guided by Inequality~(\ref{eq:hkpush-con3}), and hence, uses an ad hoc threshold $r_{max}$ instead.

Third, \hkpushplus makes the maximum number $K$ of hops be specified as an input parameter, whereas \hkpush does not fix $K$ in advance. We use a fixed $K$ in \hkpushplus because (i) as mentioned, Line 4 of \hkpushplus requires testing whether there exists a node $v$ with $\rks[v]>\frac{\epsilon_r \cdot \delta}{K}\cdot d(v)$, and (ii) such a test can be efficiently implemented when $K$ is fixed. Otherwise, whenever $K$ changes, we would need to recheck all nodes' residues to see if $\rks[v]>\frac{\epsilon_r \cdot \delta}{K}\cdot d(v)$ holds, which would incur considerable overheads. Meanwhile, \hkpush can afford to let $K$ dynamically change, since it uses a fixed residue threshold $r_{max}$ given as input. In our implementation of \hkpushplus, we set
\vspace{-1ex}\begin{equation*}
	\textstyle K=c\cdot \frac{\log{(\frac{1}{\epsilon_r\cdot \delta})}}{\log\left(\bar{d}\right)}, 
\end{equation*}
where $\bar{d}$ is the average degree of the nodes in $G$, and $c$ is a constant that we decide based on our experiments in Section~\ref{sec:choosec}. We refer interested readers to Appendix \ref{sec:chooseK} for the rationale of this setting of $K$.

\vspace{-1ex}\subsection{The Case with Random Walks}\label{sec:opt-prune}

Suppose that \hkpushplus depletes its computation budget $n_p$ before it satisfies Inequality~(\ref{eq:hkpush-con3}) with $\epsilon_a = \epsilon_r \cdot \delta$. In that case, the HKPR vector $\qs$ returned by \hkpushplus does not ensure $(d,\epsilon_r,\delta)$-approximation, and we need to refine $\qs$ by conducting random walks according to the residue vectors $\mathbf{r}^{(0)}_s, \ldots, \mathbf{r}^{(K)}_s$ returned by \hkpushplus. To reduce the number of random walks required, we propose to reduce the residues in $\mathbf{r}^{(0)}_s, \ldots, \mathbf{r}^{(K)}_s$ when conducting random walks, based on the following intuition.

First, the reduction of residues would incur some errors in the approximate HKPR vector, as demonstrated in Theorem~\ref{thrm:hkpush} (where we ignore all residues in $\mathbf{r}^{(k)}_s$ and using $\qs$ directly as the final approximate HKPR vector). Second, if we only reduce the residues in $\mathbf{r}^{(k)}_s$ by a small value, then the absolute errors incurred by the reduction could be so small that they would not jeopardize $(d,\epsilon_r,\delta)$-approximation. In particular, as we show in Section~\ref{sec:opt-als}, if we reduce every residue  $\mathbf{r}^{(k)}_s[v]$ by  $\beta_k \cdot \epsilon_r \delta \cdot d(v)$ ($k = 0, 1, \ldots, K$ and $v\in V$), then the absolute error in $\frac{\ar[v]}{d(v)}$ incurred by the residue reduction is at most $\epsilon_r\delta \cdot \sum_{k=0}^K \beta_k$.
In other words, if we choose $\beta_k$ such that $\sum_{k=0}^K \beta_k = 1$, then the resulting absolute error in $\frac{\ar[v]}{d(v)}$ is at most $\epsilon_r \delta$, which is permissible under $(d,\epsilon_r,\delta)$-approximation. The following example demonstrates the benefit of this residue reduction method.



\begin{table}[!t]
	\centering
	\caption{An example for \pukraplus} \label{tbl:example}
	\vspace{-2mm}
	\renewcommand{\arraystretch}{1.6}
	\begin{small}
		\begin{tabular}{|l|c|c|c|}
			\hline
			& $k \le 2$ & $k = 3$ & $k = 4$ \\
			\hline
			{$\sum_{v\in V}{\rks[v]}$}  & $0$ & $0.1$ & $0.3$ \\
			\hline
			{$\max_{v\in V}{\frac{\rks[v]}{d(v)}}$} & $0$ & $\frac{\rks[v_1]}{d(v_1)}=0.0025$ & $\frac{\rks[v_2]}{d(v_2)}=0.0076$  \\
			\hline
			{$\max_{v\in V}{\rks[v]}$} & $0$ & $\rks[v_1]=0.0025$ & $\rks[v_2]=0.076$ \\
			\hline
		\end{tabular}
		\vspace{-2mm}
	\end{small}
\end{table}

\begin{example} \label{exple:prune}
	\em Suppose that given a graph $G$, seed node $s$, $K=4$, and $\epsilon_r\cdot\delta = 0.01$, \hkpushplus returns residue vectors $\mathbf{r}^{(0)}_s, \ldots, \mathbf{r}^{(4)}_s$ that have the characteristics in Table~\ref{tbl:example}. Note that $v_1$ has the largest $3$-hop residue and degree-normalized $3$-hop residue, while $v_2$ has the largest $4$-hop residue and degree-normalized $4$-hop residue. 
	In addition, all of the $3$-hop residues except $v_1$'s are less than $0.0025$, and all of the $4$-hop residues except $v_2$'s are less than $0.075$. We can observe that
	$$\textstyle \sum_{\ell=0}^{4}{\max_{v\in V}{\left\{\frac{\mathbf{r}^{(\ell)}_s[v]}{d(v)}\right\}}}=0.0025+0.0076=0.0101,$$
	which is slightly larger than $\epsilon_r\cdot\delta$. In this case, according to Lines 7-8 in \pukra, we need to perform $\alpha\cdot \omega$ random walks, where
	$$\textstyle \alpha = \sum_{k=0}^{K}{\sum_{u\in V}{\rks[u]}} = 0.4.$$
	
	Now suppose that we apply the residue reduction method, setting $\beta_3 = 1/4$ and $\beta_4 = 3/4$. In that case, we reduce every residue $\mathbf{r}^{(3)}_s[v]$ by
	$$\beta_3\cdot\epsilon_r\delta \cdot d(v) =0.0025 \cdot d(v),$$
	and every residue $\mathbf{r}^{(4)}_s[v]$ by
	$$\beta_4\cdot\epsilon_r\delta \cdot d(v) =0.0075 \cdot d(v).$$
	Then, all residues in $\mathbf{r}^{(3)}_s$ and $\mathbf{r}^{(4)}_s$ are reduced to $0$, except that $\mathbf{r}^{(4)}_s[v_2]$ is decreased to $0.076 - \beta_4\cdot\epsilon_r\delta \cdot d(v_2) = 0.001$. Accordingly, $\alpha=\sum_{\ell=0}^{4}{\max_{v\in V}{\mathbf{r}^{(\ell)}_s[v]}}$ is reduced from 0.4 to 0.001, which implies that the number of random walks required is reduced by 400 times.
	\done
\end{example}

\vspace{-1ex}\subsection{Details of \pukraplusb}\label{sec:opt-pukraplus}


Based on the ideas described in Sections~\ref{sec:opt-hkpushplus} and ~\ref{sec:opt-prune}, we present \pukraplus, which utilizes \hkpushplus and random walks to compute a $(d,\epsilon_r,\delta)$-approximate HKPR vector $\ar$ with at least $1 - p_f$ probability for any given seed node $s$. Algorithm~\ref{alg:pukraplus} illustrates the pseudo-code of \pukraplus. The input parameters of \pukraplus are identical to those of \pukra, except that \pukraplus takes an additional parameter $c$, which, as mentioned in Section~\ref{sec:opt-hkpushplus}, is used to decide the maximum number $K$ of hops used in \hkpushplus.

\pukraplus starts by invoking \hkpushplus with the following parameters (Lines 5-6): $G$, $ \epsilon_r$, $\delta$, $K = c\cdot \frac{\log{(\frac{1}{\epsilon_r\cdot\delta})}}{\log{(\bar{d})}}$, and $n_p= \frac{\omega\cdot t}{2}$, where
$$\omega = \frac{8(1+\epsilon_r/6)\log{(1/p^{\prime}_f)}}{\epsilon^2_r \cdot \delta},$$
and $p^{\prime}_f$ is as defined in Equation~\eqref{eq:pfail} and is pre-computed when $G$ is loaded. Then, \hkpushplus returns an approximate HKPR vector $\ar$ and $K+1$ residue vectors $\mathbf{r}^{(0)}_s, \ldots, \mathbf{r}^{(K)}_s$.

If $\sum_{k=0}^{K}{\max_{u\in V}{\left\{\frac{\rks[u]}{d(u)}\right\}}} \le \epsilon_r\cdot\delta$, then by Theorem~\ref{thrm:hkpush}, $\ar$ is a $(d,\epsilon_r,\delta)$-approximate HKPR vector. In that case, \pukraplus immediately terminates and returns $\ar$ (Line 7). Otherwise, \pukraplus proceeds to refine $\ar$ using \rswk (Lines 8-20). But before that, \pukraplus first applies the residue reduction method discussed in Section~\ref{sec:opt-prune}. Specifically, it decreases each residue value $\rks[u]$ by $\beta_k\cdot\epsilon_r\delta\cdot d(u)$ (Lines 8-11), where
\vspace{-1ex}\begin{equation*}
	\beta_k = \frac{\sum_{u\in V}{\rks[u]}}{\sum_{\ell=0}^{K}{\sum_{u\in V}{\mathbf{r}^{(\ell)}_s[u]}}}.
\end{equation*}
The rationale of this choice of $\beta_k$ is as follows. First, $\sum_{k=0}^K \beta_k = 1$, which is crucial for $(d,\epsilon_r,\delta)$-approximation, as we mention in Section~\ref{sec:opt-prune}. Second, we set $\beta_k$ to be proportional to $\sum_{u\in V}{\rks[u]}$ because, intuitively, when $\sum_{u\in V}{\rks[u]}$ is large, the residue values in $\rks$ also tend to be large, in which case we need a larger reduction of the residues in $\rks$ to effectively reduce the number of random walks needed.

After reducing the residues in $\mathbf{r}^{(0)}_s, \ldots, \mathbf{r}^{(K)}_s$, \pukraplus performs random walks according to the reduced residues, in the same way as \pukra does (Lines 12-17). This results in a refined approximate HKPR vector $\ar$. Then, \pukraplus gives $\ar[v]$ a final touch by adding $\frac{\epsilon_r\cdot\delta}{2}\cdot d(v)$ to each $\ar[v]$ (Lines 18-19). The intuition of adding this offset to each $\ar[v]$ is as follows. The residue reduction method leads to an underestimation of each HKPR value, and amount of underestimation is in $[0,\epsilon_r\cdot\delta\cdot d(v)]$. By adding an offset $\frac{\epsilon_r\cdot\delta}{2}\cdot d(v)$ to $\ar[v]$, the range of the error in $\ar[v]$ becomes $[-\frac{\epsilon_r \cdot \delta}{2} \cdot d(v), \frac{\epsilon_r \cdot \delta}{2} \cdot d(v)]$, in which case the maximum absolute error in $\ar[v]$ is reduced by half, which help tightening the accuracy bound of \pukraplus.

\begin{algorithm}[t]
	\begin{small}
		\caption{\pukraplus}\label{alg:pukraplus}
		\BlankLine
		\setcounter{AlgoLine}{4}
		\KwIn{Graph $G$, seed node $s$, constant $c$, thresholds $\epsilon_r$ and $\delta$, and failure probability $p_f$}
		\KwOut{A $(d,\epsilon_r,\delta)$-approximate HKPR vector $\ar$}
		\nonl Lines 1-4 are the same as Lines 1-4 in Algorithm~\ref{alg:pukra}\;
		$\omega \gets\frac{8(1+\epsilon_r/6)\log{(1/p^{\prime}_f)}}{\epsilon^2_r\delta}, n_p\gets \frac{\omega \cdot t}{2}, K\gets c\cdot \frac{\log{(\frac{1}{\epsilon_r\cdot\delta})}}{\log{(\bar{d})}}$\;
		$\left(\ar, \mathbf{r}^{(0)}_s, \ldots, \mathbf{r}^{(K)}_s\right)\gets$ \hkpushplus$(s, \epsilon_r, \delta, K, n_p)$\;
		\lIf{$\sum_{k=0}^{K}{\max_{u\in V}{\left\{\frac{\rks[u]}{d(u)}\right\}}} \le \epsilon_r\cdot\delta$}{
			\Return $\ar$
		}
		\For{$k = 0$ to $K$}{
			$\beta_k \gets \frac{\sum_{u\in V}{\rks[u]}}{\sum_{\ell=0}^{K}{\sum_{u\in V}{\mathbf{r}^{(\ell)}_s[u]}}}$\;
			\For{any node $u$ with $\rks[u] > 0$}{
				$\rks[u] = \max\left\{0,\ {\rks[u]} - \beta_k\cdot\epsilon_r\delta\cdot d(u) \right\}$\;
			}
		}
		\nonl Lines 12-17 are the same as Lines 7-12 in Algorithm~\ref{alg:pukra}\;
		\setcounter{AlgoLine}{17}
		\For{$v \in V$}{
			$\ar[v] \gets \ar[v]+\frac{\epsilon_r\cdot \delta}{2}\cdot d(v)$\;
		}
		\Return $\ar$\;
	\end{small}
\end{algorithm}

Note that Lines 18-19 in \pukraplus can be performed in $O(1)$ time, as we can keep each $\ar[v]$ unchanged but record the value of $\frac{\epsilon_r\cdot\delta}{2}$ along with $\ar$. Then, whenever $\ar[v]$ is accessed, we can add $\frac{\epsilon_r\cdot\delta}{2}\cdot d(v)$ to $\ar[v]$ on the fly. In addition, for the purpose of local clustering, we can simply ignore this offset of $\frac{\epsilon_r\cdot\delta}{2}\cdot d(v)$ since it does not affect the ranking of nodes based on $\frac{\ar[v]}{d(v)}$. 

\subsection{Example for \pukraplusb}\label{sec:opt-toy}

\begin{figure}[t]
	\centering
	\vspace{-2mm}
	\includegraphics[height=28mm]{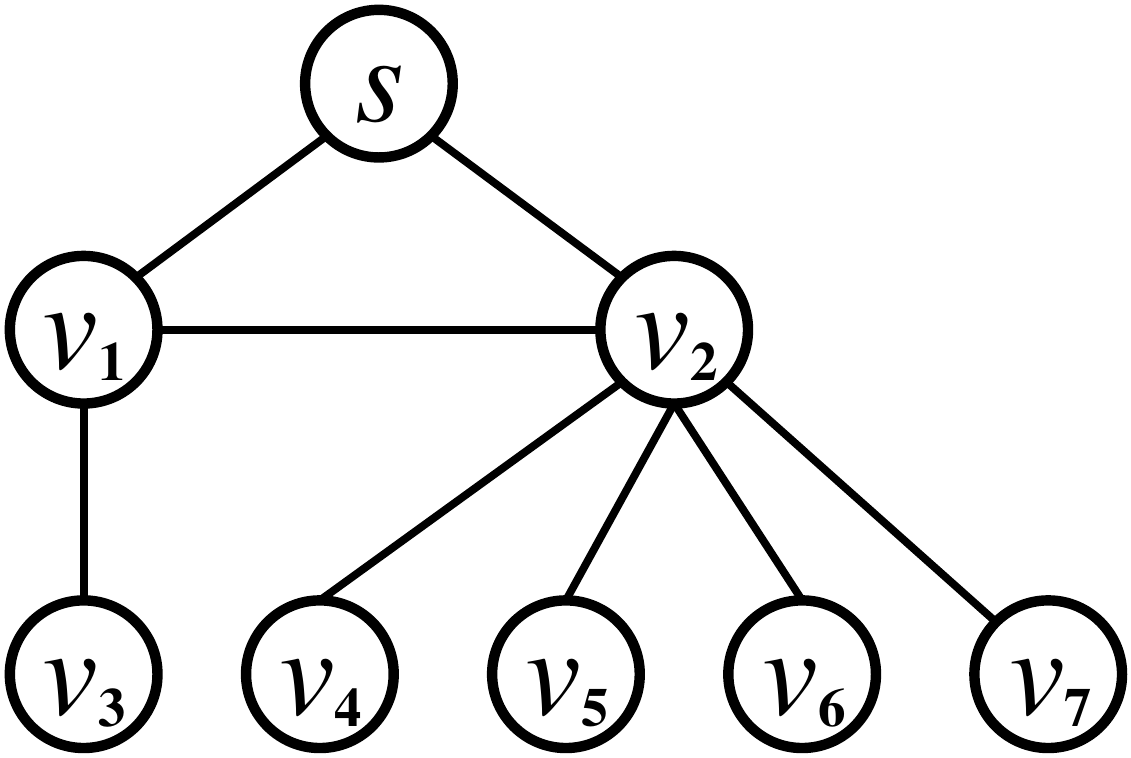}
	\caption{A graph $G'$ for illustrating \pukraplusb.}
	\label{fig:toy}
	\vspace{-2mm}
\end{figure}

\begin{table}[!t]
	\renewcommand{\arraystretch}{1}
	\centering
	\caption{The reserves and residues after the first round of push operations.}
	\label{tbl:toy-1}
	\vspace{-3mm}	
	\renewcommand{\arraystretch}{1.5}
	\begin{small}
		\begin{tabular}{|c|c|c|c|c|c|c|c|c|}
			\hline
			& $\boldsymbol{s}$ & $\boldsymbol{v_1}$ & $\boldsymbol{v_2}$ & $\boldsymbol{v_3}$ & $\boldsymbol{v_4}$ & $\boldsymbol{v_5}$ & $\boldsymbol{v_6}$ & $\boldsymbol{v_7}$ \\
			\hline
			$\mathbf{q}_s$ & $\frac{1}{e^3}$ & 0 & 0 & 0 & 0 & 0 & 0 & 0\\
			\hline
			$\mathbf{r}^{(1)}_s$ & 0 & $\frac{e^3-1}{2e^3}$ & $\frac{e^3-1}{2e^3}$ & 0 & 0 & 0 & 0 & 0 \\
			\hline
		\end{tabular}
	\end{small}
	\vspace{0mm}
\end{table}
We use the graph $G'$ in Figure~\ref{fig:toy} to illustrate \pukraplus. For ease of presentation, we set $t=3$, $p_f=10^{-2}$, $\epsilon_r=0.5$, $\delta=\frac{2\tau}{9}$, and $\tau=1-\frac{4}{e^3}\approx 0.8$, and we set $c$ to a constant such that $K=2$. Consider that we invoke Algorithm~\ref{alg:pukraplus} on $G'$ with seed node $s$ and above parameters as inputs. \pukraplus first computes $\omega \approx \frac{970}{\tau}$ and $n_p\approx \frac{1455}{\tau}$, and then invokes Algorithm~\ref{alg:hkpushplus} with seed node $s$ and parameters $\epsilon_r$, $\delta$, $K$, and $n_p$ as inputs.

Initially, Algorithm~\ref{alg:hkpushplus} sets $\mathbf{r}^{(0)}_s[s]=1$ and $k=0$. In that case, $k<K$, and $\frac{\mathbf{r}^{(0)}_s[s]}{d(s)}=0.5>\frac{\epsilon_r\delta}{K}\approx 0.0445$. Accordingly, Algorithm~\ref{alg:hkpushplus} performs push operations from $s$. It first converts $\frac{\eta(0)}{\psi(0)} =\frac{1}{e^3}$ portion of $s$'s reserve $\mathbf{r}^{(0)}_s[s]$ into its reserve $\mathbf{q}_s[s]$, and then distributes $\frac{1}{2}(1-\frac{\eta(0)}{\psi(0)})=\frac{e^3-1}{2e^3}$ portion to each of its two neighbors, i.e., $v_1$ and $v_2$. Table~\ref{tbl:toy-1} shows the reserves and residues after the first round of push operations.

\begin{table}[!t]
	\renewcommand{\arraystretch}{1}
	\centering
	\caption{The reserves and residues after the second round of push operations.}
	\vspace{-3mm}
	\label{tbl:toy-2}
	\renewcommand{\arraystretch}{1.5}
	\begin{small}
		\begin{tabular}{|c|c|c|c|c|c|c|c|c|}
			\hline
			& $\boldsymbol{s}$ & $\boldsymbol{v_1}$ & $\boldsymbol{v_2}$ & $\boldsymbol{v_3}$ & $\boldsymbol{v_4}$ & $\boldsymbol{v_5}$ & $\boldsymbol{v_6}$ & $\boldsymbol{v_7}$ \\
			\hline
			$\mathbf{q}_s$ & $\frac{1}{e^3}$ & $\frac{3}{2e^3}$ & 0 & 0 & 0 & 0 & 0 & 0\\
			\hline
			$\mathbf{r}^{(1)}_s$ & 0 & 0 & $\frac{e^3-1}{2e^3}$ & 0 & 0 & 0 & 0 & 0 \\
			\hline
			$\mathbf{r}^{(2)}_s$ & $\frac{\tau}{6}$ & 0 & $\frac{\tau}{6}$ & $\frac{\tau}{6}$ & 0 & 0 & 0 & 0\\
			\hline
		\end{tabular}
	\end{small}
	\vspace{0mm}
\end{table}

As shown in Table~\ref{tbl:toy-1} and Figure~\ref{fig:toy}, $\frac{\mathbf{r}^{(1)}_s[v_1]}{d(v_1)}\approx 0.1584$ and $\frac{\mathbf{r}^{(1)}_s[v_2]}{d(v_2)}\approx 0.0792$, both of which are greater than $\frac{\epsilon_r\delta}{K}\approx 0.0445$. In addition, $k=1<K$, and the number of push operations performed $i$ is 2, which is smaller than $n_p$. Hence, Algorithm~\ref{alg:hkpushplus} starts the second round of push operations on $\mathbf{r}^{(1)}_s[v_1]$. In that case, it converts $\frac{\eta(1)}{\psi(1)}$ portion of $\mathbf{r}^{(1)}_s[v_1]$ into $v_1$'s reserve, and distributes $\frac{1}{3}(1-\frac{\eta(1)}{\psi(1)})\cdot \mathbf{r}^{(1)}_s[v_1]=\frac{\tau}{6}$ residue  to each of $v_1$'s three neighbors $s, v_2$ and $v_3$. Table~\ref{tbl:toy-2} lists the reserves and residues after this round of push operations.

\begin{table}[!t]
	\renewcommand{\arraystretch}{1}
	\centering
	\caption{The reserves and residues after the third round of push operations.}
	\vspace{-3mm}
	\label{tbl:toy-3}
	\renewcommand{\arraystretch}{1.5}   
	\begin{small}
		\begin{tabular}{|c|c|c|c|c|c|c|c|c|}
			\hline
			& $\boldsymbol{s}$ & $\boldsymbol{v_1}$ & $\boldsymbol{v_2}$ & $\boldsymbol{v_3}$ & $\boldsymbol{v_4}$ & $\boldsymbol{v_5}$ & $\boldsymbol{v_6}$ & $\boldsymbol{v_7}$ \\
			\hline
			$\mathbf{q}_s$ & $\frac{1}{e^3}$ & $\frac{3}{2e^3}$ & $\frac{3}{2e^3}$ & 0 & 0 & 0 & 0 & 0\\
			\hline
			$\mathbf{r}^{(1)}_s$ & 0 & 0 & 0 & 0 & 0 & 0 & 0 & 0 \\
			\hline
			$\mathbf{r}^{(2)}_s$ & $\frac{\tau}{4}$ & $\frac{\tau}{12}$ & $\frac{\tau}{6}$ & $\frac{\tau}{6}$ & $\frac{\tau}{12}$ & $\frac{\tau}{12}$ & $\frac{\tau}{12}$ & $\frac{\tau}{12}$ \\
			\hline
		\end{tabular}
	\end{small}
	\vspace{0mm}
\end{table}
In the third round, Algorithm~\ref{alg:hkpushplus} converts $\frac{\eta(1)}{\psi(1)}$ portion of $\mathbf{r}^{(1)}_s[v_2]$ into the reserve of $v_2$. Meanwhile, each of $v_2$'s neighbors receives $\frac{1}{6}(1-\frac{\eta(1)}{\psi(1)})\cdot \mathbf{r}^{(1)}_s[v_2]$ residue. After this round, $k=K=2$, and hence, Algorithm~\ref{alg:hkpushplus} terminates with the reserves in Table~\ref{tbl:toy-3} as the approximate HKPR values in $\ar$.

Based on Table \ref{tbl:toy-2}~and Figure~\ref{fig:toy}, we have the following results:
\begin{equation*}
	\frac{\mathbf{r}^{(2)}_s[s]}{d(s)}=\frac{\tau}{8}, \frac{\mathbf{r}^{(2)}_s[v_1]}{d(v_1)}=\frac{\tau}{36}, \frac{\mathbf{r}^{(2)}_s[v_2]}{d(v_2)}=\frac{\tau}{36}, \frac{\mathbf{r}^{(2)}_s[v_3]}{d(v_3)}=\frac{\tau}{6},
\end{equation*}
\begin{equation*}
	\frac{\mathbf{r}^{(2)}_s[v_4]}{d(v_4)}=\frac{\mathbf{r}^{(2)}_s[v_5]}{d(v_5)}=\frac{\mathbf{r}^{(2)}_s[v_6]}{d(v_6)}=\frac{\mathbf{r}^{(2)}_s[v_6]}{d(v_6)}=\frac{\tau}{12}.
\end{equation*}
Since $\sum_{k=0}^{2}{\max_{u\in V}{\left\{\frac{\rks[u]}{d(u)}\right\}}}=\frac{\tau}{6} > \epsilon_r\cdot\delta=\frac{\tau}{9}$, Algorithm~\ref{alg:pukraplus} does not terminate, but would perform random walks to refine $\ar$. It first computes $\beta_0=\beta_1=0$ and $\beta_2=1$, and then reduces the residue of each node $v$ by $\mathbf{r}^{(2)}_s[v] = \max\left\{0,\ {\mathbf{r}^{(2)}_s[v]} - \beta_2\cdot\epsilon_r\delta\cdot d(v) \right\}$. This leads to 
\begin{align*}
	\mathbf{r}^{(2)}_s[v_3]=\mathbf{r}^{(2)}_s[v_3]-\epsilon_r\cdot\delta=\frac{\tau}{18}, \\ \mathbf{r}^{(2)}_s[s]=\mathbf{r}^{(2)}_s[s]-2\cdot \epsilon_r\delta=\frac{\tau}{36},
\end{align*}
and other residues become 0. The number of random walks required by Algorithm~\ref{alg:pukraplus} is therefore 
$$n_r=\alpha\cdot\omega=\sum_{k=0}^{2}{\sum_{u\in V}{\rks[u]}}\cdot \omega = \frac{\tau}{12}\cdot \omega \approx 81.$$ 
Before performing random walks from $s$ and $v_3$, Algorithm~\ref{alg:pukraplus} first builds an alias structure on all nodes, such that entries $(s,2)$ and $(v_3,2)$ would be sampled with probabilities $\frac{\mathbf{r}^{(2)}_s[s]}{\alpha}=\frac{1}{3}$ and $\frac{\mathbf{r}^{(2)}_s[v_3]}{\alpha}=\frac{2}{3}$, respectively. After that, it invokes Algorithm~\ref{alg:skipped-walk} for $n_r = 81$ times. In each invocation, it samples the starting entry $(v,2)$ from this alias structure, and at the $\ell$-th step, it terminates at the current node $u$ with probability $\frac{\eta(2+\ell)}{\psi(2+\ell)}$, and traverses to a randomly selected neighbor of $u$ with the other $1-\frac{\eta(2+\ell)}{\psi(2+\ell)}$ probability. Once it terminates at a node $w$, Algorithm \ref{alg:pukraplus} increases $w$'s approximate HKPR value $\ar[w]$ by $\frac{\alpha}{n_r}=\frac{\tau}{970}$. \pukraplus returns $\ar$ as an approximation of the exact HKPR vector $\rs$ after 81 random walks.

Finally, we perform a sweep over $\ar$ to find a cut with the best conductance. That is, we sort the nodes $v$ by $\frac{\ar[v]}{d(v)}$ in descending order. If the sorted order is $s,v_1,v_3,v_2,v_4-v_7$, then the sweep would produce $\{s,v_1,v_3\}$ as the best cut, as its conductance $\frac{1}{3}$ is smaller than other cuts encountered during the sweep.

\vspace{-1ex}\subsection{Analysis}\label{sec:opt-als}

\noindent
{\bf Correctness.}
\vspace{1mm}
\noindent
Let $\qs$ denote the approximate HKPR vector returned by \hkpushplus in Line 6 of \pukraplus, $\mathbf{r}^{(0)}_s, \ldots, \mathbf{r}^{(K)}_s$ be the residue vectors output by \hkpushplus at the same step, and $\ar$ be the final version of the HKPR vector output by \pukraplus. We define a residue vector $\rbs$ as:
\begin{equation}\label{eq:pukraplus-rbs}
\rbs[u] = \min\left\{\rks[u],\ \beta_k\cdot\epsilon_r\delta\cdot d(u) \right\}.
\end{equation}
Observe that $\rbs[u]$ equals the amount of residue reduction on $\rks[u]$ occurred in Lines 8-11 of \pukraplus.

Similar to the correctness analysis in Section~\ref{sec:pukra-als}, by Lemma~\ref{lem:fwdeq}, we have the following equation for any node $v$:
\vspace{-1ex}\begin{equation}\label{eq:hkpr-ab}
\textstyle \rs[v]=\qs[v]+ \as[v] + \bs[v],
\end{equation}
where
\vspace{-1ex}\begin{align}
	& \textstyle \as[v]=\alpha \cdot \sum_{k=0}^{K}{\sum_{u\in V}{\frac{\rks[u]}{\alpha}\cdot\mathbf{h}^{(k)}_u[v]}}, \textrm{ and} \label{eq:pukraplus-as}\\
	& \textstyle \bs[v]=\sum_{k=0}^{K}{\sum_{u\in V}{{\rbs[u]}\cdot{\mathbf{h}^{(k)}_u[v]}}}.\label{eq:pukraplus-bs}
\end{align}
Then, the approximation error in each $\ar[v]$ can be regarded as the sum of two approximation errors for $\as[v]$ and $\bs[v]$, respectively. The error in $\as[v]$ is caused by sampling errors in \rswk, and hence, it can be bounded using the Chernoff bound, in a way similar to the analysis in Section~\ref{sec:pukra-als}. Meanwhile, the error in $\bs[v]$ is due to the residue reduction procedure in Lines 8-11 of \pukraplus. In what follows, we present an analysis of the error in $\bs[v]$.


By Equation~(\ref{eq:pukraplus-rbs}), for any node $u\in V$ and $k\in [0,K]$, the amount of residue reduction on $\rks[u]$ satisfies $0 \le \rbs[u] \le \beta_k\cdot \epsilon_r \delta\cdot d(u)$. Combining this with Equation~(\ref{eq:pukraplus-bs}), we have
\vspace{-1ex}\begin{equation}\label{eq:b-bound}
\textstyle 0\le \bs[v] \le \sum_{k=0}^{K}\left({{\beta_k\cdot\epsilon_r \delta}\sum_{u\in V}{d(u)\cdot \mathbf{h}^{(k)}_u[v]}}\right).
\end{equation}

\begin{lemma}[\cite{pons2005computing}]\label{lem:mkrev}
	Let $u$ and $v$ be any two nodes in $G$, and $\mathbf{P}^{k}[u,v]$ (resp.\ $\mathbf{P}^{k}[v,u]$) be the probability that a length-$k$ random walk from $u$ ends at $v$ (resp.\ from $v$ ends at $u$). Then,
	\begin{flalign*}
		&& \textstyle  \frac{\mathbf{P}^{k}[u,v]}{d(v)} = \frac{\mathbf{P}^{k}[v,u]}{d(u)}. && \square
	\end{flalign*}
\end{lemma}

By Lemma~\ref{lem:mkrev} and the definition of $\mathbf{h}^{(k)}_u[v]$ in Equation~\ref{eq:hsum-def}, for any node $v\in V$ and $k\in[0,K]$, we have
\begin{align}\label{eq:hsum}
	\textstyle  \sum_{u\in V}{d(u)\cdot \mathbf{h}^{(k)}_u[v]} &= \textstyle  d(v)\cdot \sum_{u\in V}{\mathbf{h}^{(k)}_v[u]}\nonumber\\
	&\textstyle  = d(v)\cdot \sum_{\ell=0}^{\infty}{\left[\frac{\eta(k+\ell)}{\psi(k)}\cdot\sum_{u\in V}{ \mathbf{P}^{\ell}[v,u]}\right]}\nonumber\\
	&\textstyle =d(v)\cdot \sum_{\ell=0}^{\infty}{\frac{\eta(k+\ell)}{\psi(k)}}=d(v).
\end{align}
Combining Equations (\ref{eq:b-bound}) and (\ref{eq:hsum}), we have
\begin{equation}\label{eq:bsbound}
0 \le \bs[v] \le d(v)\cdot \epsilon_r \delta.
\end{equation}
Therefore, estimating $\bs[v]$ as $\frac{\epsilon_r \cdot \delta}{2} \cdot d(v)$ incurs an absolute error of at most $\frac{\epsilon_r \cdot \delta}{2} \cdot d(v)$.

Based on the above analysis, we establish the accuracy guarantee of \pukraplus as follows:
\vspace{-1ex}\begin{theorem}\label{thrm:pukraplus}
	\pukraplus outputs a $(d,\epsilon_r,\delta)$-approximate HKPR vector $\ar$ with probability at least $1-p_f$.
\end{theorem}

\noindent
{\bf Time and Space Complexities.} The time and space complexities of \pukraplus depend on its two main components: \hkpushplus and \rswk. As discussed in Section~\ref{sec:opt-hkpushplus}, both the computation and space overheads of \hkpushplus are $O(n_p)$. Since \pukraplus sets $n_p = \frac{\omega \cdot t}{2}$, its invocation of \hkpushplus incurs $O\left(\frac{t\cdot\log{(n/p_f)}}{\epsilon^2_r\cdot\delta}\right)$ time and space costs. Meanwhile, the total number of random walks conducted by \pukraplus is no more than that by \pukra, and hence, the computational and space costs of generating random walks in \pukraplus are both $O\left(\frac{t\cdot\log{(n/p_f)}}{\epsilon^2_r\cdot\delta}\right)$ in expectation. Thus, the expected time and space complexities of \pukraplus are $O\left(\frac{t\cdot\log{(n/p_f)}}{\epsilon^2_r\cdot\delta}\right)$ and $O\left(m + n + \frac{t\cdot\log{(n/p_f)}}{\epsilon^2_r\cdot\delta}\right)$, respectively, where the $m+n$ term is due to the space required by the input graph.

%
%

\section{Related Work} \label{sec:cpt}

In this section, we first review two HKPR algorithms, \chkpr and \hkrelax, that are most related to our solutions; after that, we review other work related to local clustering and HKPR computation.


\header
{\chkprb.} \chkpr~\cite{chung2014hklocal} is a random-walk-based method for computing approximate HKPR. Given a seed node $s$, it performs $\frac{16 \log n}{\epsilon^3}$ random walks from $s$, with the constraint that each walk has a length at most $K$; after that, for each node $v$, it uses the fraction of walks that end at $v$ as an estimation $\ar[v]$ of $v$'s HKPR. It is shown that with probability at least $1 - \epsilon$, \chkpr guarantees that
\begin{equation*}
	\begin{cases}
		|\ar[v]-\rs[v]|\le (1+\epsilon)\cdot\rs[v], & \textrm{if $\rs[v]>\epsilon$}  \\
		|\ar[v]-\rs[v]|\le \epsilon, & \textrm{otherwise}.
	\end{cases}
\end{equation*}
Note that for the above guarantee to be meaningful, $\epsilon \ll 1$ should hold; otherwise, the successful probability $1 - \epsilon$ of \chkpr would be too small, and there could be too many nodes having a large absolute error up to $\epsilon$. In particular, if we are to ensure that \chkpr achieves $(d, \epsilon_r, \delta)$-approximation with probability at least $1 - p_f$, then we have to set $\epsilon \le \min\left\{\epsilon_r \cdot \delta, p_f\right\}$. However, when $\epsilon \ll 1$, \chkpr incurs excessive computation cost, since its expected time complexity is inversely proportional to $\epsilon^3$. 


\header
{\hkrelaxb.}
\hkrelax \cite{kloster2014heat} is a deterministic algorithm that runs in $O\left(\frac{te^t\log{(1/\epsilon_a)}}{\epsilon_a}\right)$ time and returns an approximate HKPR vector $\ar$ such that $\left|\frac{\ar[v]}{d(v)}-\frac{\rs[v]}{d(v)} \right|\le \epsilon_a$ for any $v \in V$. \hkrelax is similar to our \hkpush algorithm in that they both (i) maintain an approximation HKPR vector $\ar$ and a number of residue vectors, and (ii) incrementally refine $\ar$ by applying push operations according to the residue of each node. However, there exist three major differences between \hkrelax and \hkpush. First, \hkrelax and \hkpush define the residue of each node in different manners, due to which \hkrelax requires more sophisticated approaches than \hkpush to update $\ar$ and the residue vectors after each push operation. Second, \hkrelax ignores all $k$-hop residues with $k > 2t\log{\frac{1}{\epsilon_a}}$, whereas \hkpush retains all residues generated for the refinement of $\ar$ via random walks. Third, \hkrelax and \hkpush have different termination conditions. Due to these differences, we are unable combine \hkrelax with random walks to achieve the same performance guarantee provided by our \pukra and \pukraplus algorithms.

%
%

\header
{\bf Other methods for local clustering.}
The first local graph clustering algorithm, \nib, is introduced in the seminal work~\cite{spielman2004nearly,spielman2013local} by Spielman and Teng. After that, Anderson~\textit{et~al}.\ \cite{andersen2006local} propose \prnib, a local clustering algorithm based on {\em personalized PageRank} \cite{haveliwala2002topic,jeh2003scaling}, and show that it improves over \nib in terms of the theoretical guarantees of both clustering quality and time complexity. In turn, Anderson~\textit{et~al}.'s method is improved in subsequent work \cite{andersen2009finding,gharan2012approximating} based on the {\it volume-biased evolving set process}~\cite{diaconis1990strong}. Subsequent work \cite{zhu2013local,orecchia2014flow,wang2017capacity,veldt2016simple} achieves further improved guarantees on the quality of local clustering, but the methods proposed are mostly of theoretical interests only, as they are difficult to implement and offer rather poor practical efficiency. As a consequence, \hkrelax remains the state-of-the-art for local clustering in terms of practical performance \cite{kloster2014heat,avron2015community,chung2015computing}.

In recent work~\cite{shun2016parallel}, Shun~\textit{et al.}\ study parallel implementations for \nib, \prnib, \chkpr, and \hkrelax, respectively, and are able to achieve significant speedup on a machine with $40$ cores. We believe that our algorithms may also exploit parallelism for higher efficiency, but a thorough treatment of this problem is beyond the scope of this paper.

\header
{\bf Methods for personalized PageRank.} We note that \pukra and \pukraplus are similar in spirit to several recent methods \cite{wang2017fora, lofgren2016personalized, wei2018topppr, wang2016hubppr} for computing personalized PageRank (PPR), since they all combine a push-operation-based algorithm with random walks. Hence, at first glance, it may seem that we can simply adopt and extend these techniques to address HKPR computation. Unfortunately, this is not the case as HKPR is inherently more sophisticated than PPR. In particular, even though both HKPR and PPR measure the proximity of a node $v$ with respect to another node $u$ by the probability that a random walk starting from $u$ would end at $v$, they differ significantly in the ways that they define random walks:

\vspace{-1mm}
\begin{itemize}
	\item PPR's random walks are {\it Markovian}: in each step of a walk, it terminates with a fixed probability $\alpha \in (0, 1)$, regardless of the previous steps.
	
	\item HKPR's random walks are {\it non-Markovian}: the termination probability of a walk at the $i$-th step is a function of $i$, i.e., the walk has to remember the number of steps that it has traversed, so as to decide whether it should terminate.
\end{itemize}
\vspace{-1mm}

The Markovianness of PPR random walks is a key property exploited in the methods in \cite{wang2017fora, lofgren2016personalized, wei2018topppr, wang2016hubppr}. Specifically, when computing the PPR $p(u, v)$ from node $u$ to node $v$, the methods in \cite{lofgren2016personalized, wei2018topppr, wang2016hubppr} require performing a {\it backward search} which starts from $v$ and traverses the incoming edges of each node in a backward manner. For each node $w$ encountered and each of its incoming neighbor $x$, the backward search needs to calculate the probability that a random walk hitting $x$ at a certain step would arrive at $w$ at the next step. For PPR random walks, this probability is a constant decided only by $\alpha$ and the number of $x$'s outgoing neighbors. Unfortunately, for HKPR random walks the probability is not a constant, but a variable depending on the number of steps that the walk has taken before reaching $x$. In other words, this variable is not unique even when $w$ and $x$ are fixed, due to which the backward search no longer can be utilized. This issue makes it unpalatable to extend the methods in \cite{lofgren2016personalized, wei2018topppr, wang2016hubppr} to compute HKPR.

Meanwhile, the FORA method in \cite{wang2017fora} does not require a backward search; instead, it combines random walks with a forward search from $u$ that is similar to the \hkpush algorithm used in \pukra. 
However, \pukra is more sophisticated than FORA as it needs to account for the non-Markovianness of HKPR, and there are three major differences between the two methods. First, \pukra requires maintaining multiple residue vectors in its invocation of \hkpush, since it needs to differentiate the residues generated at different steps of the forward search; otherwise, it would not be able to combine the results of \hkpush with random walks because of HKPR's non-Markovianness. In contrast, FORA only needs to maintain one residue vector, as the Markovianness of PPR allows it to merge the resides produced at different steps of the forward search. Second, the theoretical analysis of \pukra is more challenging than that of FORA, since it is more complicated to (i) maintain and update multiple residue vectors and (ii) combine random walks with the forward traversal in a way that takes into account the non-Markovianness of HKPR. Third, \pukra provides an accuracy guarantee in terms of each node's {\it degree-normalized} HKPR, whereas FORA's accuracy guarantee is on each node's PPR without normalization.

Last but not the least, we note that our \pukraplus algorithm, which significantly improves over \pukra in terms of practical efficiency, is based on a new optimization that is specifically designed for HKPR and our notion of $(d, \epsilon_r, \delta)$-approximation. This optimization is inapplicable for PPR computation, which further differentiates \pukraplus from FORA.




\section{Experiments} \label{sec:exp}


\subsection{Experimental Setup}\label{sec:exp-set}
We conduct all experiments on a Linux server with a Intel Xeon(R) E5-2650 v2@2.60GHz CPU and 64GB RAM. For fair comparison, all algorithms are implemented in C++ and compiled by g++ 4.8 with -O3 optimization.

We use 6 undirected real-world graphs and 2 synthetic graphs which are used in recent work~\cite{kloster2014heat,shun2016parallel,chung2015computing} as benchmark datasets (Table \ref{tbl:exp-data}). We obtain {\em DBLP}, {\em Youtube}, {\em Orkut}, {\em LiveJournal}, and {\em Friendster} from \cite{linksnap}. {\em PLC} is a synthetic graph, and it is generated by Holme and Kim algorithm for generating graphs with powerlaw degree distribution and approximate average clustering. {\em 3D-grid} is a synthetic grid graph in 3-dimensional space where every node has
six edges, each connecting it to its 2 neighbors in each dimension. {\em Twitter} is a symmetrized version of a snapshot
of the Twitter network \cite{kwak2010twitter}. For each dataset, we pick 50 seed nodes uniformly at random as our query sets.


Unless specified otherwise, following previous work~\cite{kloster2014heat,banerjee2015fast}, we set heat constant $t=5$. In addition, for all randomized algorithms, we set failure probability $p_f=10^{-6}$. We report the average query time (measured in wall-clock time) of each algorithm on each dataset with various parameter settings. Note that the $y$-axis is in log-scale and the measurement unit for running time is millisecond (ms).
%


\begin{table}[!t]
	\centering
	\begin{small}
		\caption{Statistics of graph datasets.} \label{tbl:exp-data}
		\vspace{-3mm}
		\renewcommand{\arraystretch}{1}
		\begin{tabular}{|l|r|r|c|}
			\hline
			\multicolumn{1}{|c|}{{\bf Dataset}} &
			\multicolumn{1}{c|}{{$\boldsymbol{n}$}} &
			\multicolumn{1}{c|}{{$\boldsymbol{m}$}} &
			\multicolumn{1}{c|}{{$\boldsymbol{\bar{d}}$}} \\
			\hline
			{\em DBLP}   & 317,080  & 1,049,866  & 6.62  \\
			\hline
			{\em Youtube}    & 1,134,890  & 2,987,624 &  5.27 \\
			\hline
			{\em PLC}   & 2,000,000  & 9,999,961  & 9.99  \\
			\hline
			{\em Orkut} & 3,072,441 & 117,185,083  & 76.28 \\
			\hline
			{\em LiveJournal}   & 3,997,962 & 34,681,189  & 17.35 \\
			\hline
			{\em 3D-grid}   & 9,938,375 & 29,676,450  & 5.97 \\
			\hline
			{\em Twitter}   & 41,652,231 & 1,202,513,046 & 57.74 \\
			\hline
			{\em Friendster}   & 65,608,366 & 1,806,067,135  & 55.06 \\
			\hline
		\end{tabular}
		\vspace{-3mm}
	\end{small}
	\vspace{-3ex}
\end{table}

\subsection{Tuning Parameter $c$ for \pukraplusb}\label{sec:choosec}

First, we experimentally study how to set the parameter $c$ so as to obtain the best performance for \pukraplus in practice. We run \pukraplus with parameters $\epsilon_r=0.5$, $\delta=\frac{1}{n}$, and varying $c$ from $0.5$ to $5$ on all 8 graphs.

\begin{figure}[!t]
	\centering
	\begin{small}
		\includegraphics[height=7mm]{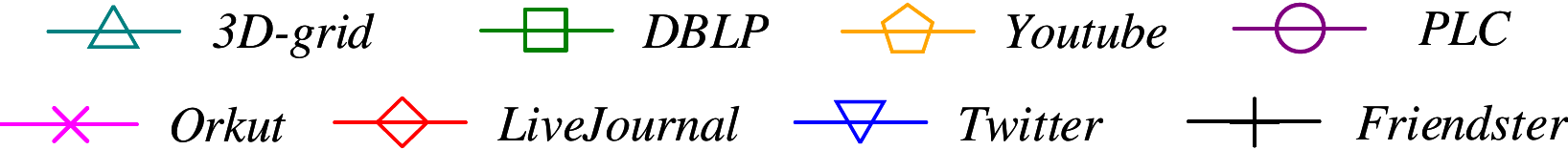}  \\
		\vspace{2mm}
		\includegraphics[height=32mm]{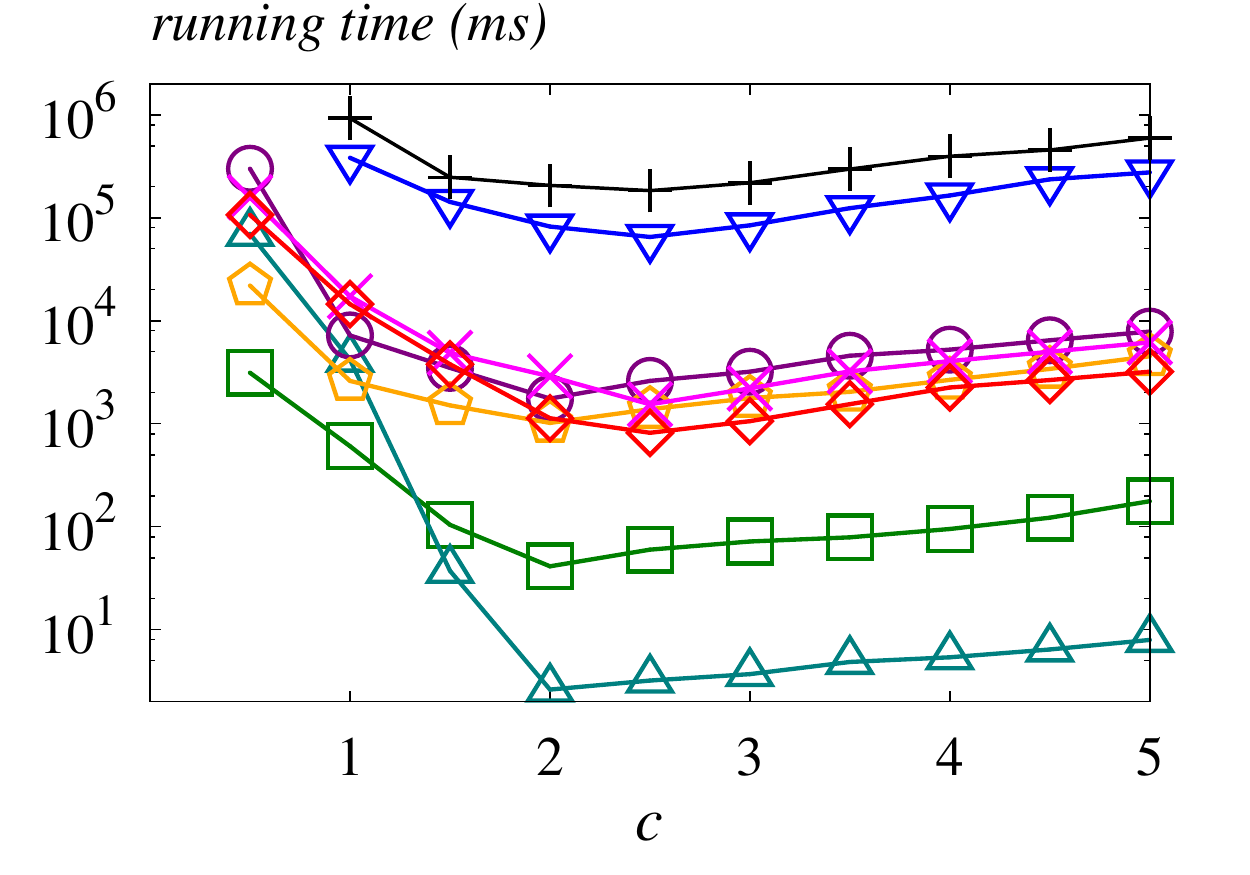}
		\vspace{-4mm}
		\caption{Running time of \pukraplusb vs $c$ (best viewed in color).} \label{fig:time-sweetpt}
		\vspace{-2mm}
	\end{small}
\end{figure}

Figure~\ref{fig:time-sweetpt} plots the running time of \pukraplus on each dataset for different $c$. We omit the results for {\em Twitter} and {\em Friendster} when $c=0.5$, because it takes several hours to finish one query. We can make the following observations. For each dataset, the running time decreases first as $c$ grows. The reason is that \pukraplus degrades to \mc when $c$ is very small, and if we keep increasing $c$, \hkpushplus will perform more push operations so as to reduce the number of random walks until $c$ balances the costs incurred for \hkpushplus and \rswk. On the other hand, when $c$ increases further, the overhead of \hkpushplus goes up gradually, disrupting the balance between \hkpushplus and \rswk. This leads to higher running time. More specifically, we can see that for graphs with small average degree including {\em DBLP}, {\em Youtube}, {\em PLC} and {\em 3D-grid}, the costs are minimized when $c$ is around $2$. On the other hand, for graphs with high average degree (\textit{e.g.,} {\em Orkut}, {\em LiveJournal}, {\em Twitter}, and {\em Friendster}), we note that $c=2.5$ achieves the best performance. Based on the above observations, a good value choice for $c$ is $2.5$, when the overheads on most of the graphs are minimized. In the sequel, we set $c=2.5$.

\subsection{Comparison between \pukrab and \pukraplusb}\label{sec:exp-pukra}
\begin{figure*}[!t]
	\centering
	\begin{small}
		\begin{tabular}{cccc}
			\multicolumn{4}{c}{
				\hspace{-4mm}\includegraphics[height=2.8mm]{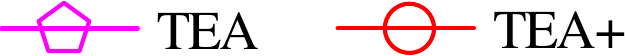}}  \\
			\hspace{-2mm} \includegraphics[height=32mm]{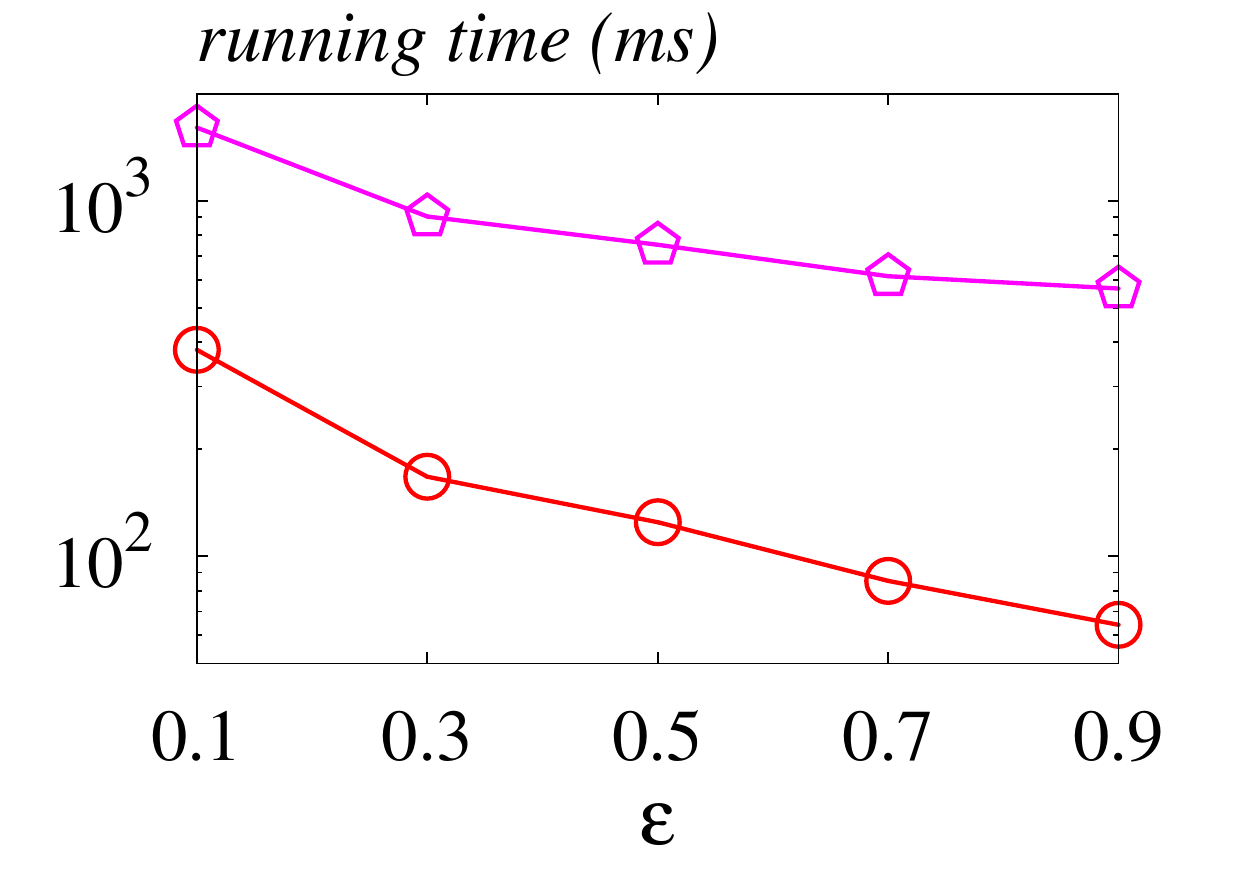} &
			\hspace{-4mm} \includegraphics[height=32mm]{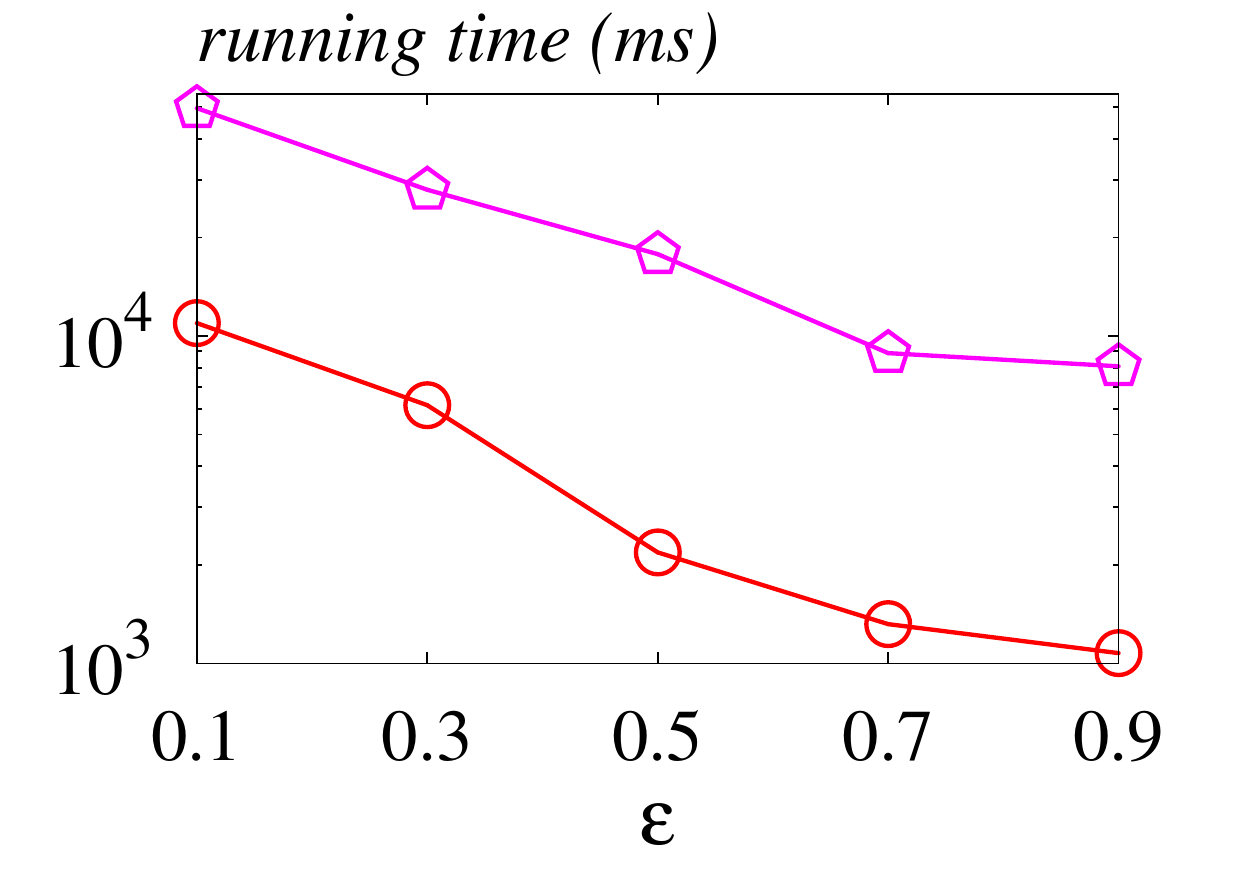} &
			\hspace{-4mm} \includegraphics[height=32mm]{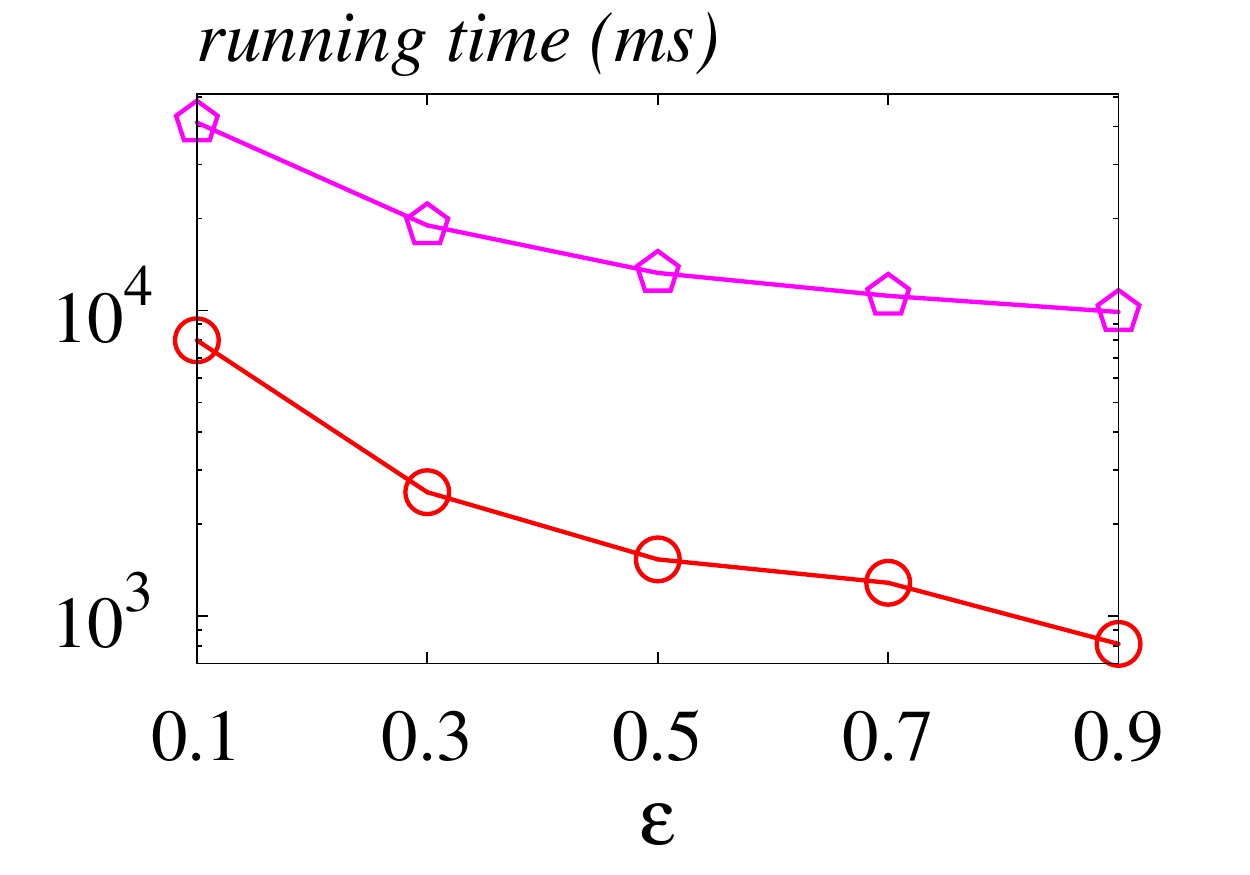} &
			\hspace{-4mm} \includegraphics[height=32mm]{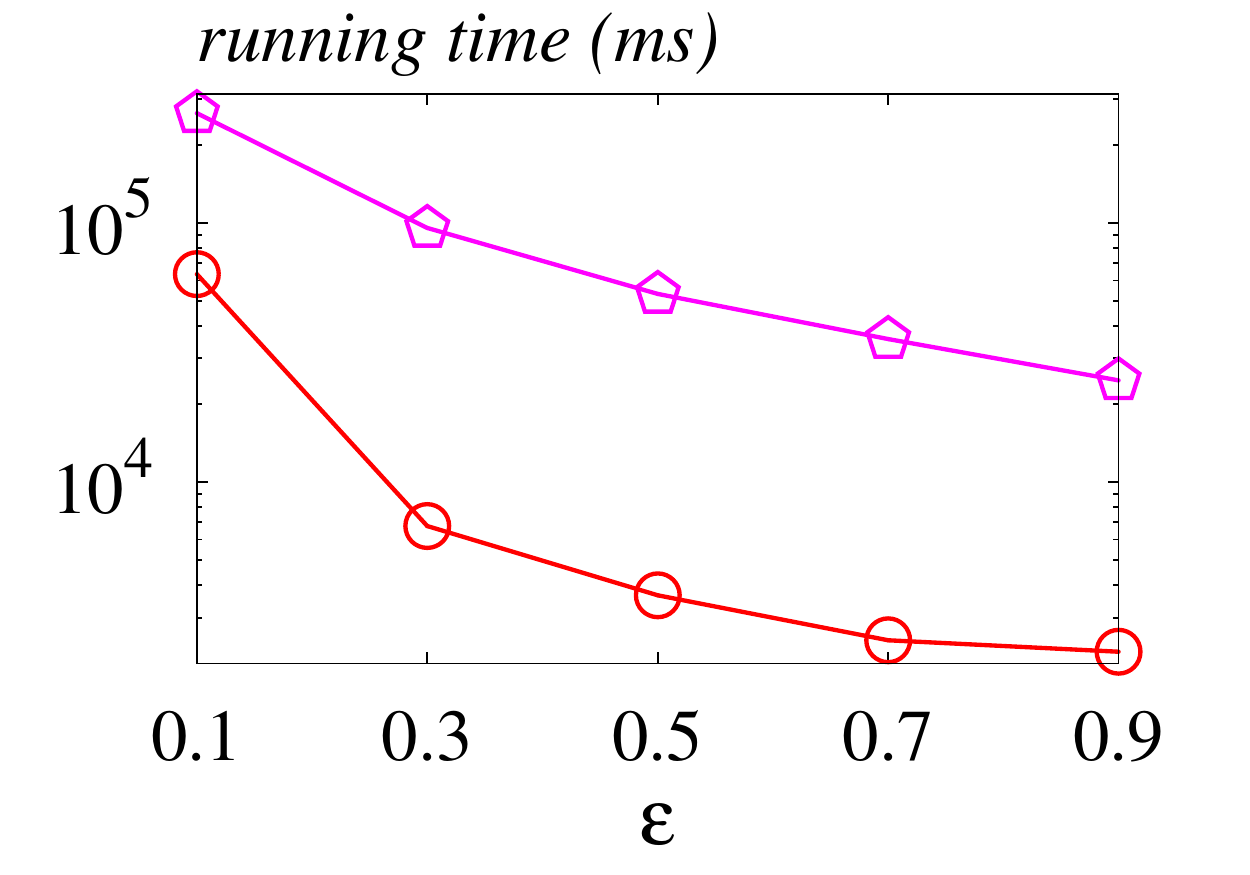}
			\\[-1mm]
			\hspace{-4mm} (a) {\em DBLP} &
			\hspace{-4mm} (b) {\em Youtube} &
			\hspace{-4mm} (c) {\em PLC} &
			\hspace{-4mm} (d) {\em Orkut}
			
			\\[1mm]
			
			\hspace{-4mm} \includegraphics[height=32mm]{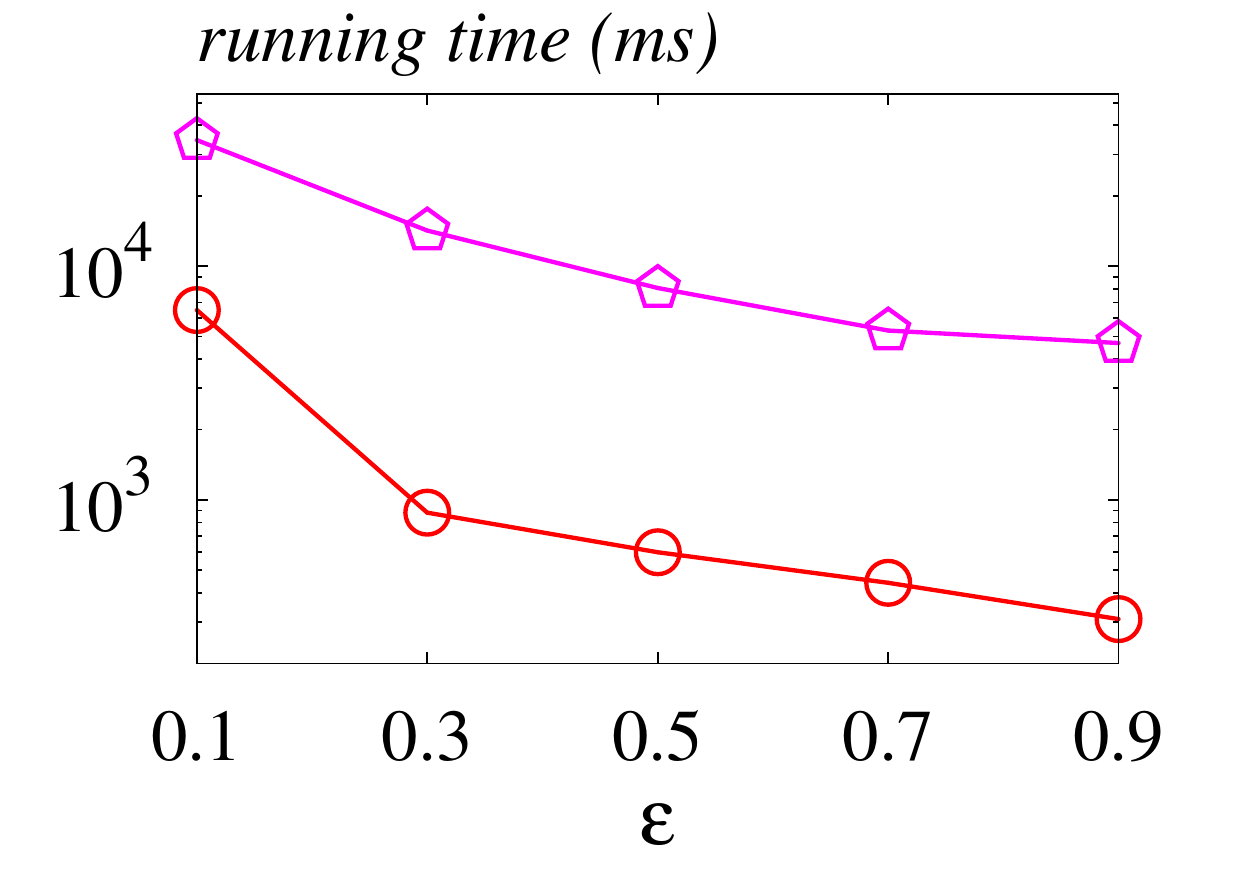} &
			\hspace{-4mm} \includegraphics[height=32mm]{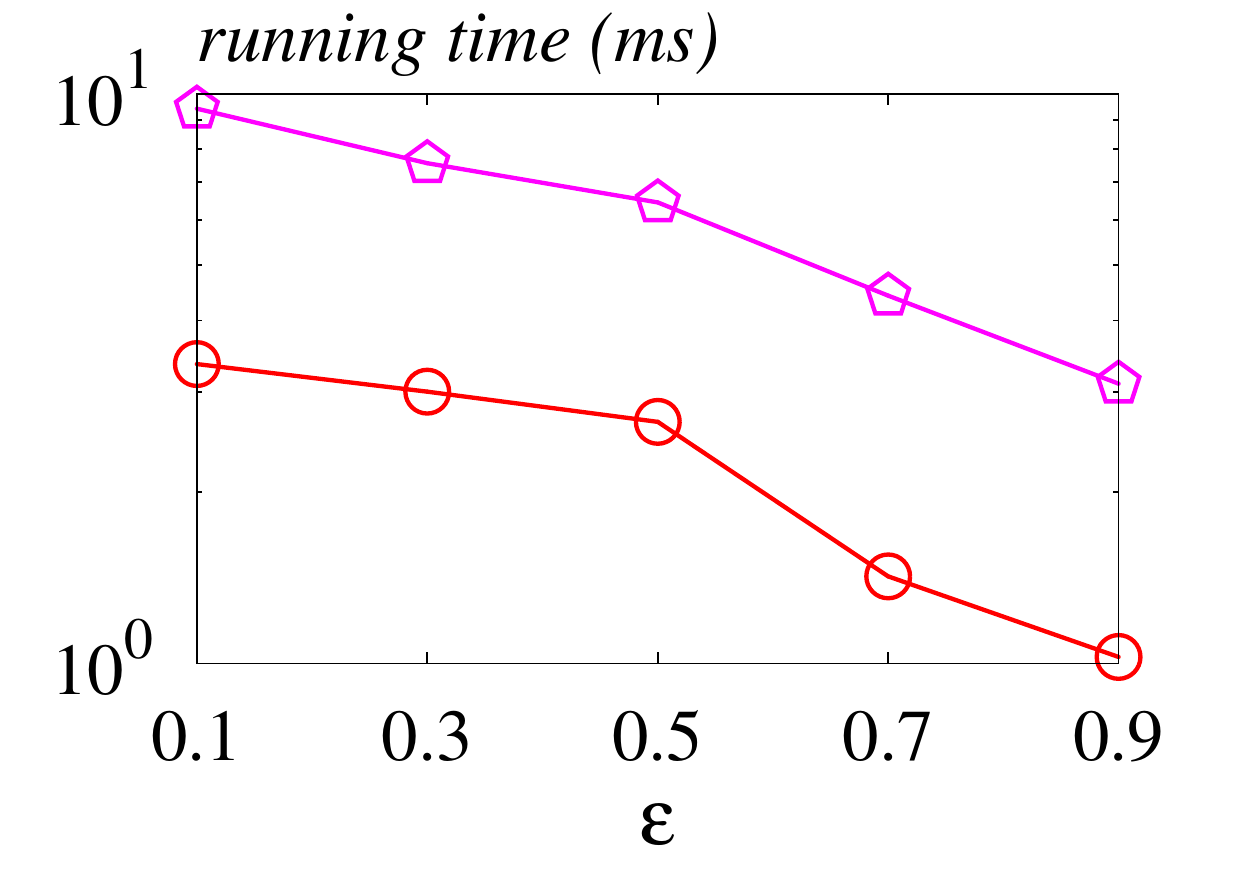} &
			\hspace{-4mm} \includegraphics[height=32mm]{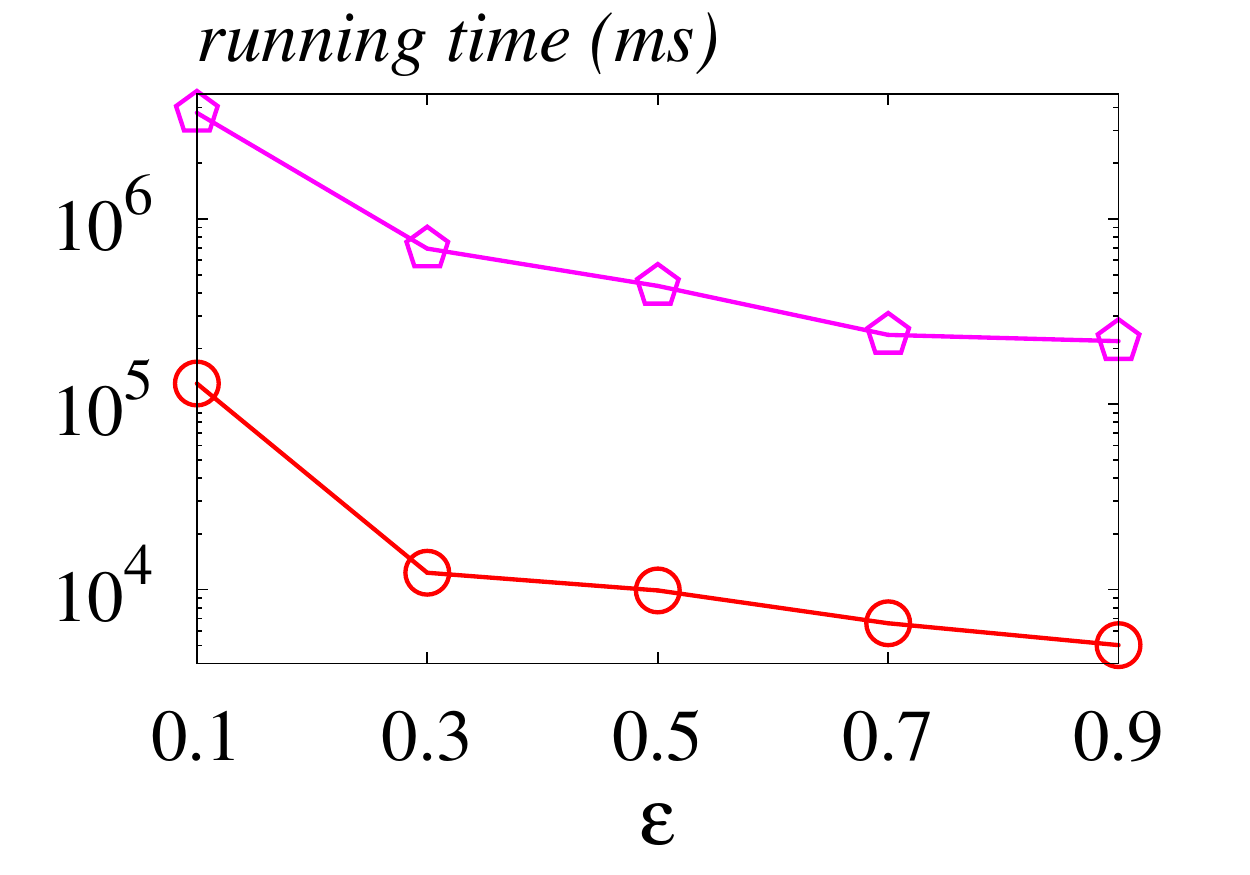} &
			\hspace{-4mm} \includegraphics[height=32mm]{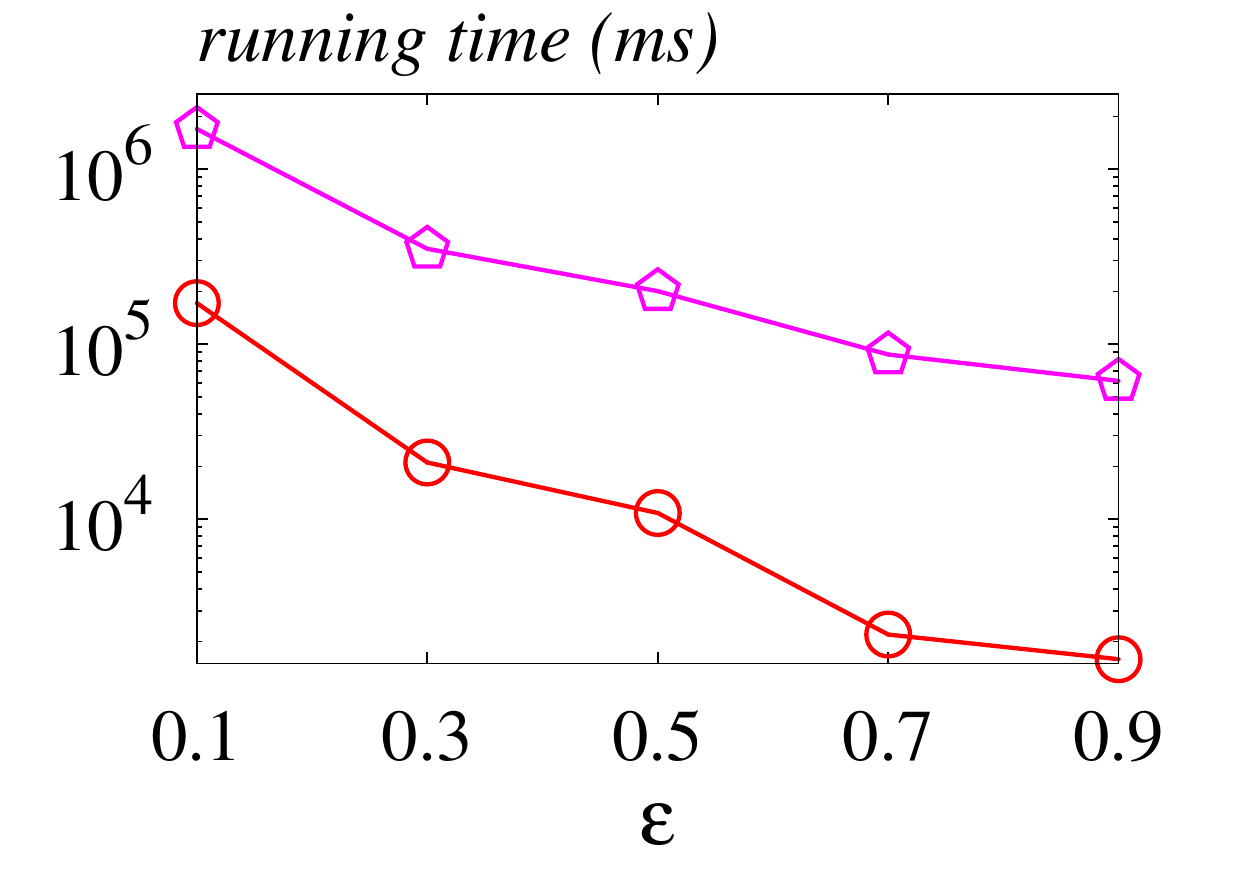}
			\\[-1mm]
			
			\hspace{-4mm} (e) {\em LiveJournal} &
			\hspace{-4mm} (f) {\em 3D-grid} &
			\hspace{-4mm} (g) {\em Twitter} &
			\hspace{-4mm} (h) {\em Friendster}
			\\[-1mm]
		\end{tabular}
		\vspace{-2mm}
		\caption{Running time \eat{of \pukrab and \pukraplusb on graphs with varying} vs $\epsilon_r$.} \label{fig:time-opt}
		\vspace{-1mm}
	\end{small}
\end{figure*}
In this set of experiments, we compare \pukraplus with \pukra based on identical theoretical accuracy guarantees. For both \pukra and \pukraplus, we set the relative error threshold $\epsilon_r=0.5$. Since the best values for $r_{max}$ vary largely for different parameter settings and various datasets, we are unable to find a universal optimal value for $r_{max}$. Instead, we tune $r_{max}$ for \pukra with different error thresholds on each dataset separately. That is, we scale $\frac{1}{\omega\cdot t}$ up or down such that the costs for \hkpush and \rswk in \pukra are roughly balanced and the total cost is minimized.

\begin{figure*}[!t]
	\centering
	\begin{small}
		\begin{tabular}{cccc}
			\multicolumn{4}{c}{
				\hspace{-4mm}\includegraphics[height=2.8mm]{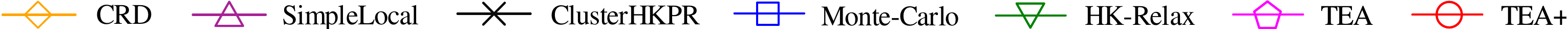}}  \\
			\hspace{-4mm} \includegraphics[height=32mm]{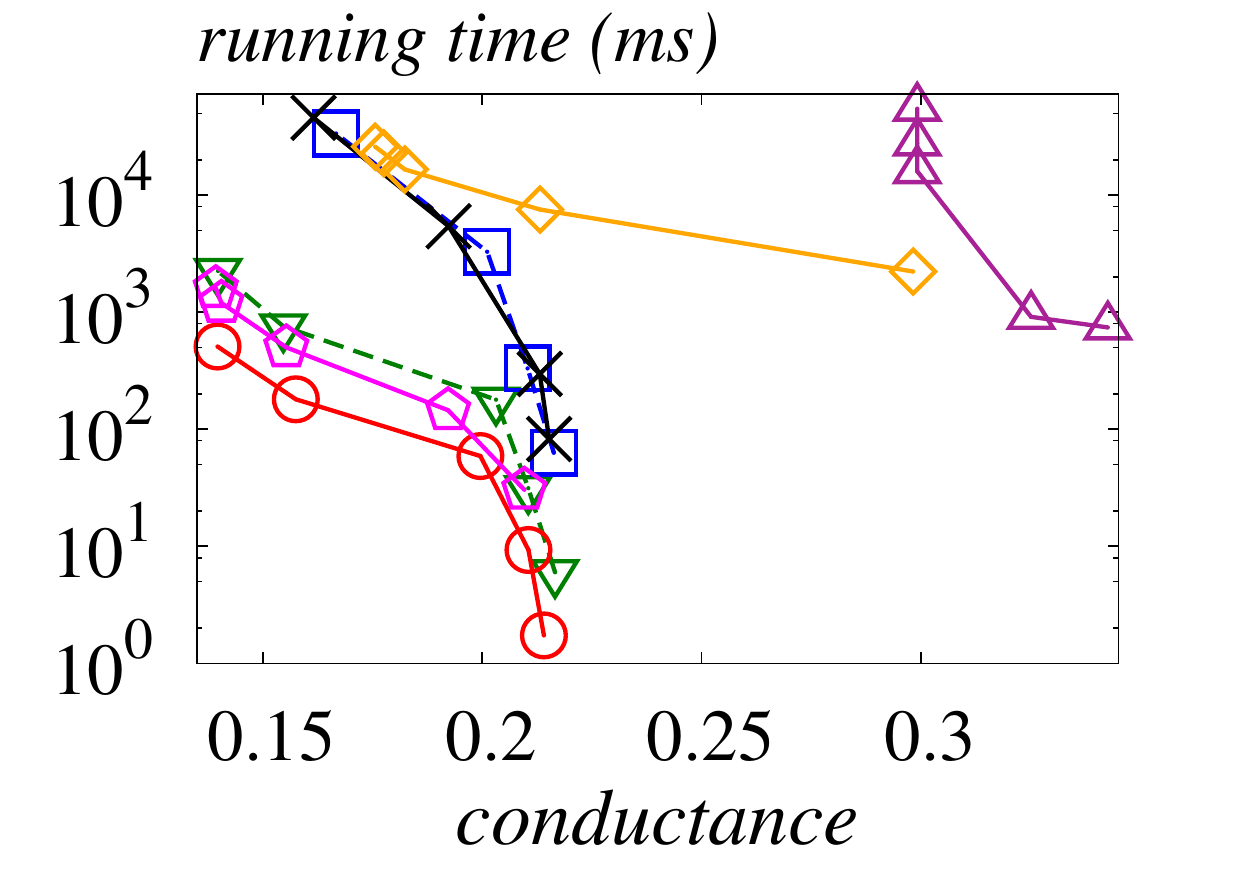} &
			\hspace{-4mm} \includegraphics[height=32mm]{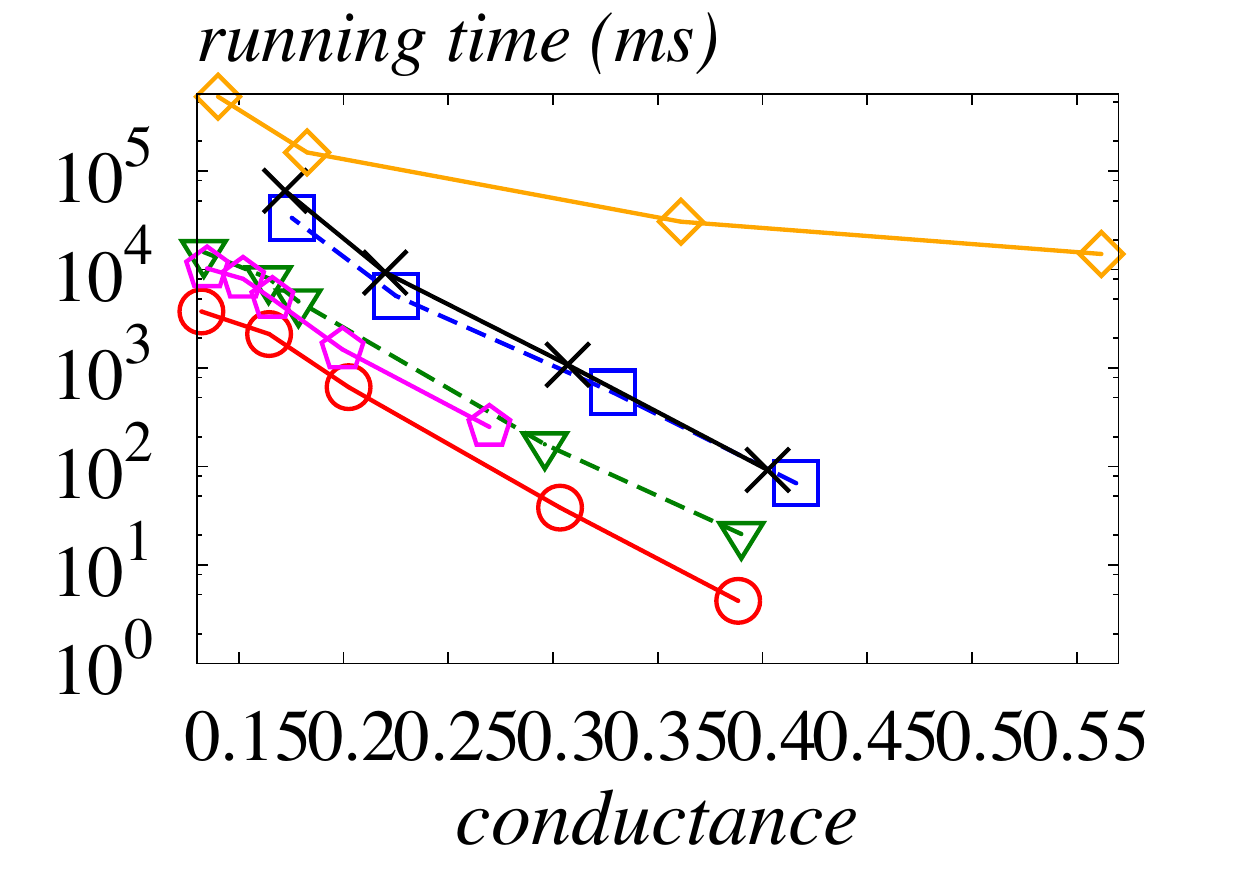} &
			\hspace{-4mm} \includegraphics[height=32mm]{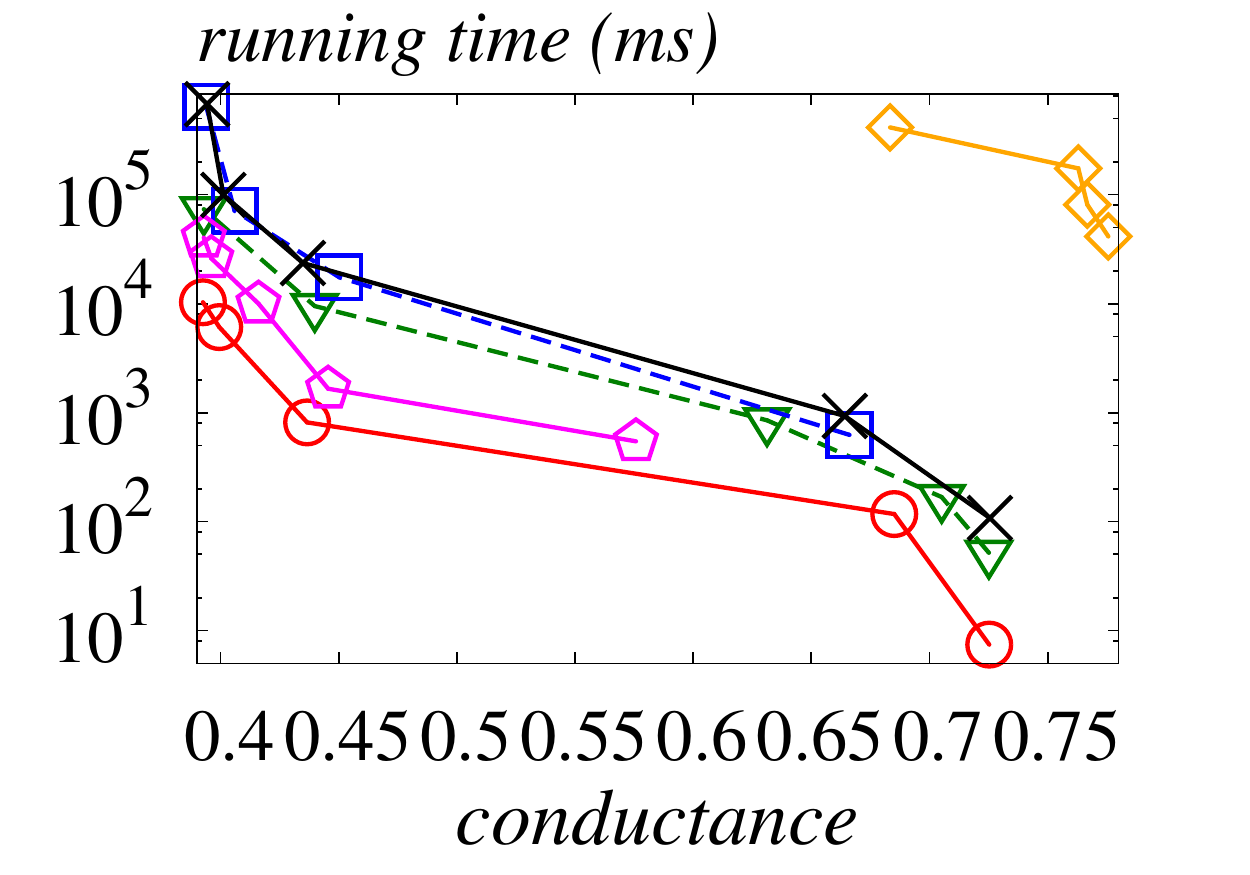} &
			\hspace{-4mm} \includegraphics[height=32mm]{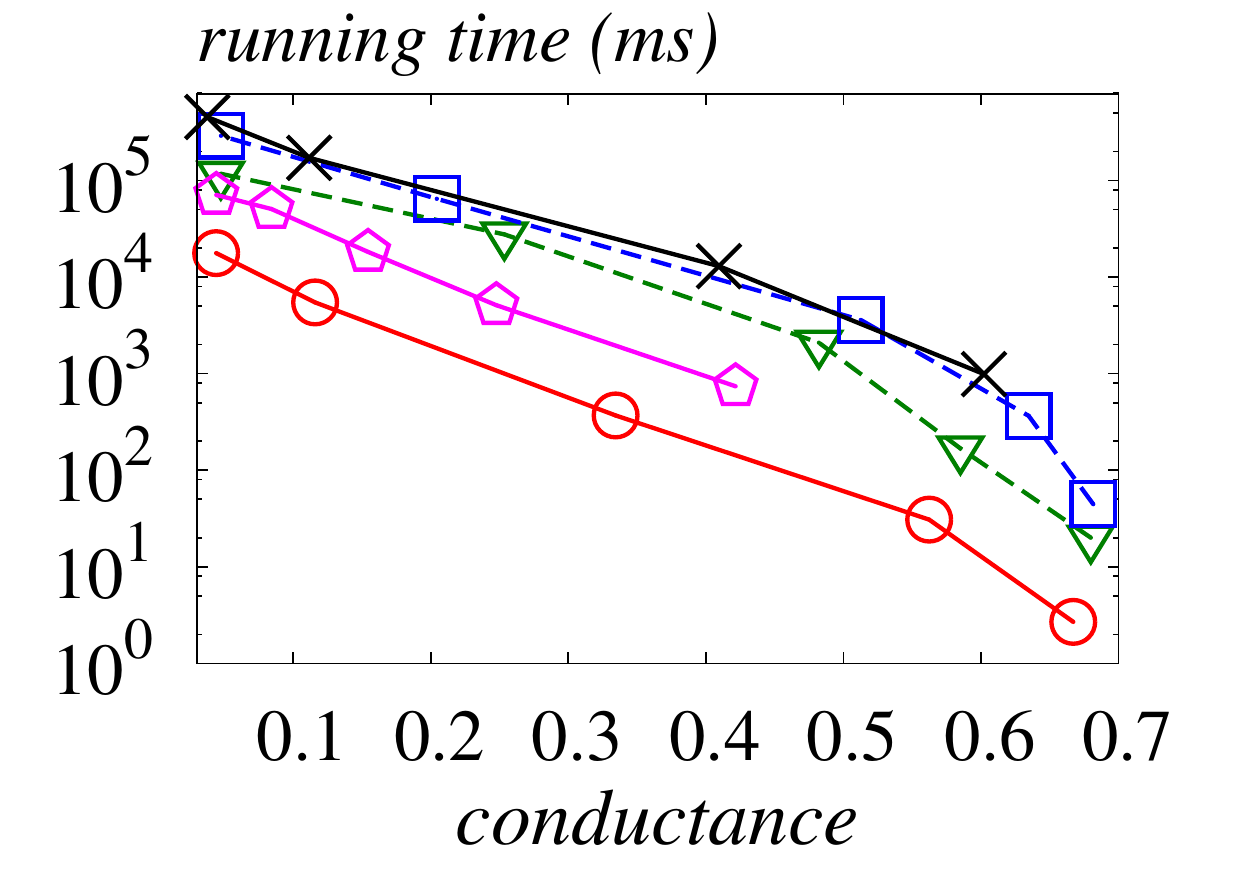}
			\\[-1mm]
			\hspace{-4mm} (a) {\em DBLP} &
			\hspace{-4mm} (b) {\em Youtube} &
			\hspace{-4mm} (c) {\em PLC} &
			\hspace{-4mm} (d) {\em Orkut}
			\\[1mm]
			\hspace{-4mm} \includegraphics[height=32mm]{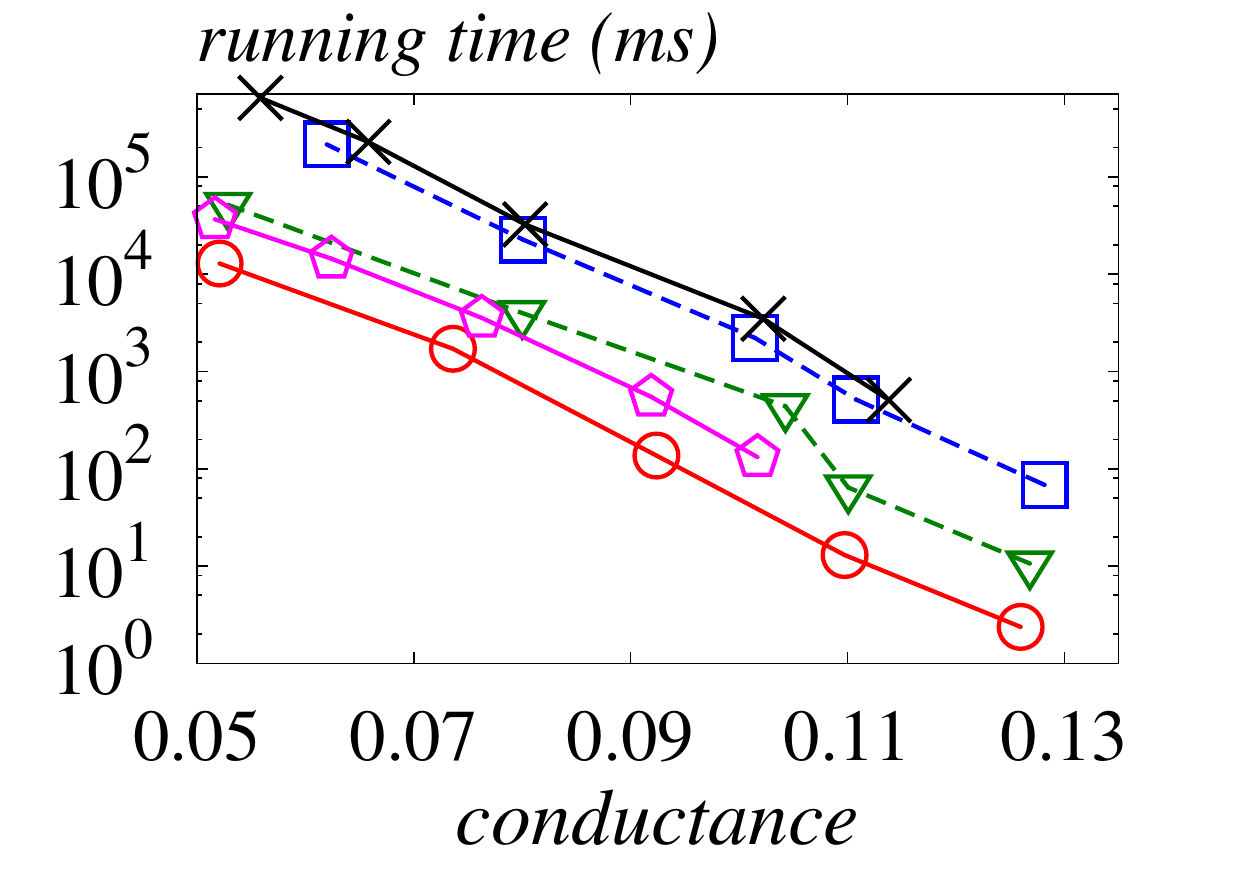} &
			\hspace{-4mm} \includegraphics[height=32mm]{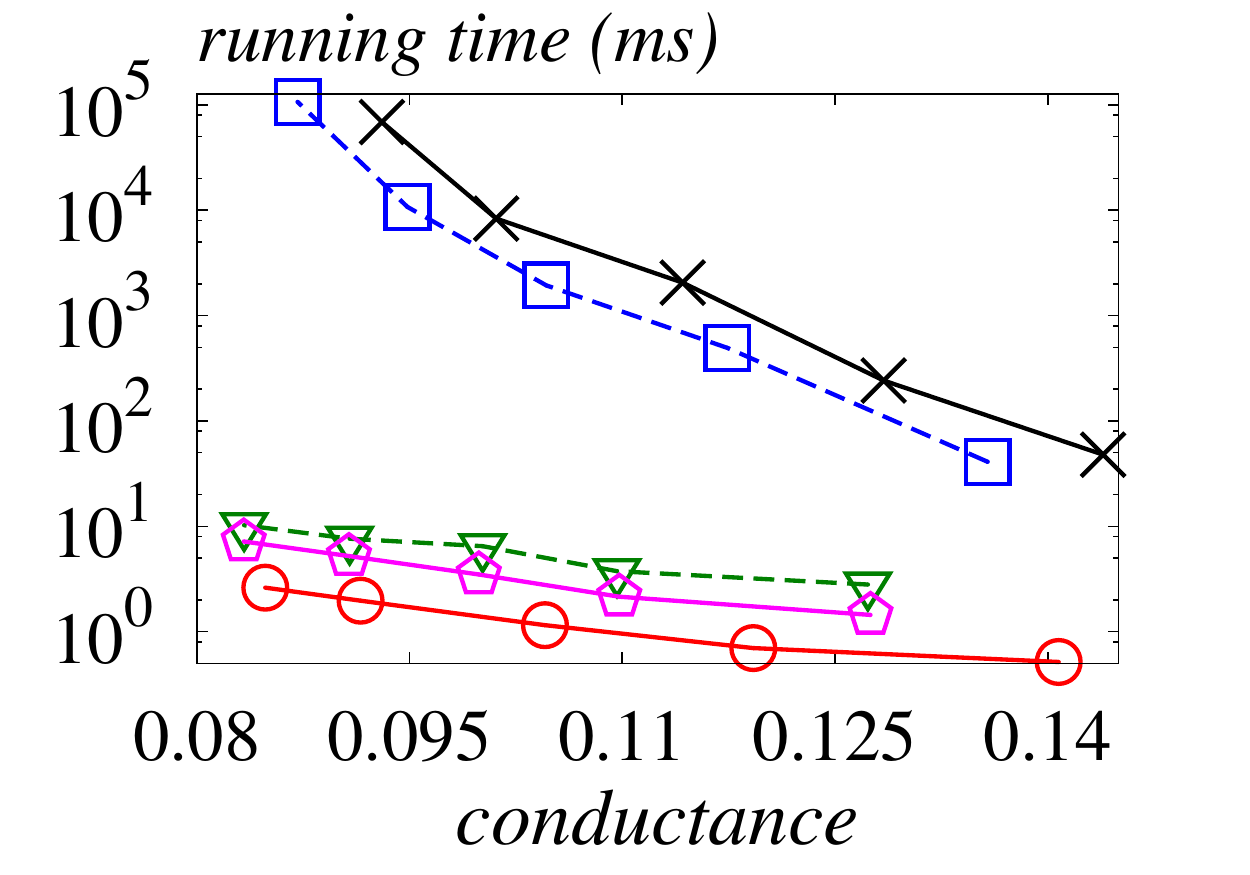} &
			\hspace{-4mm} \includegraphics[height=32mm]{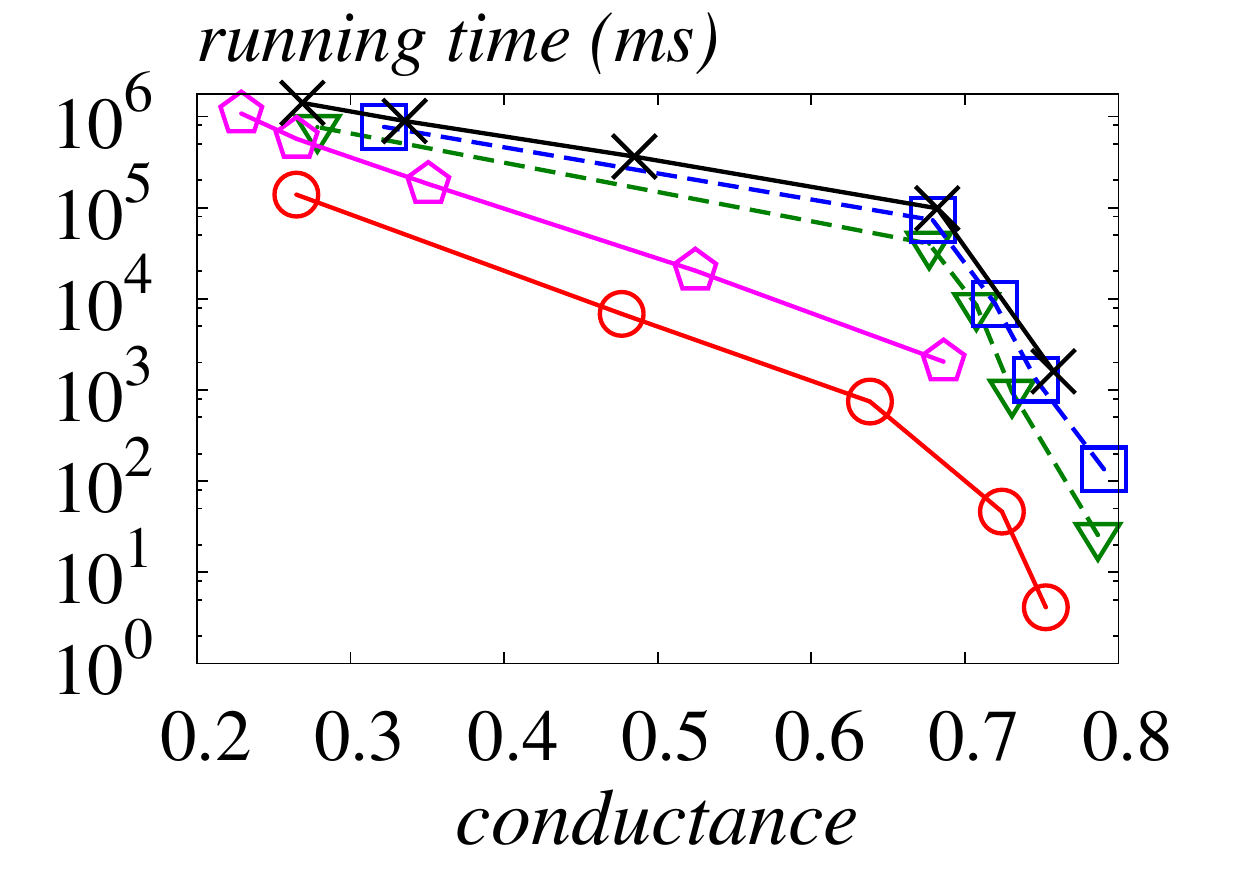} &
			\hspace{-4mm} \includegraphics[height=32mm]{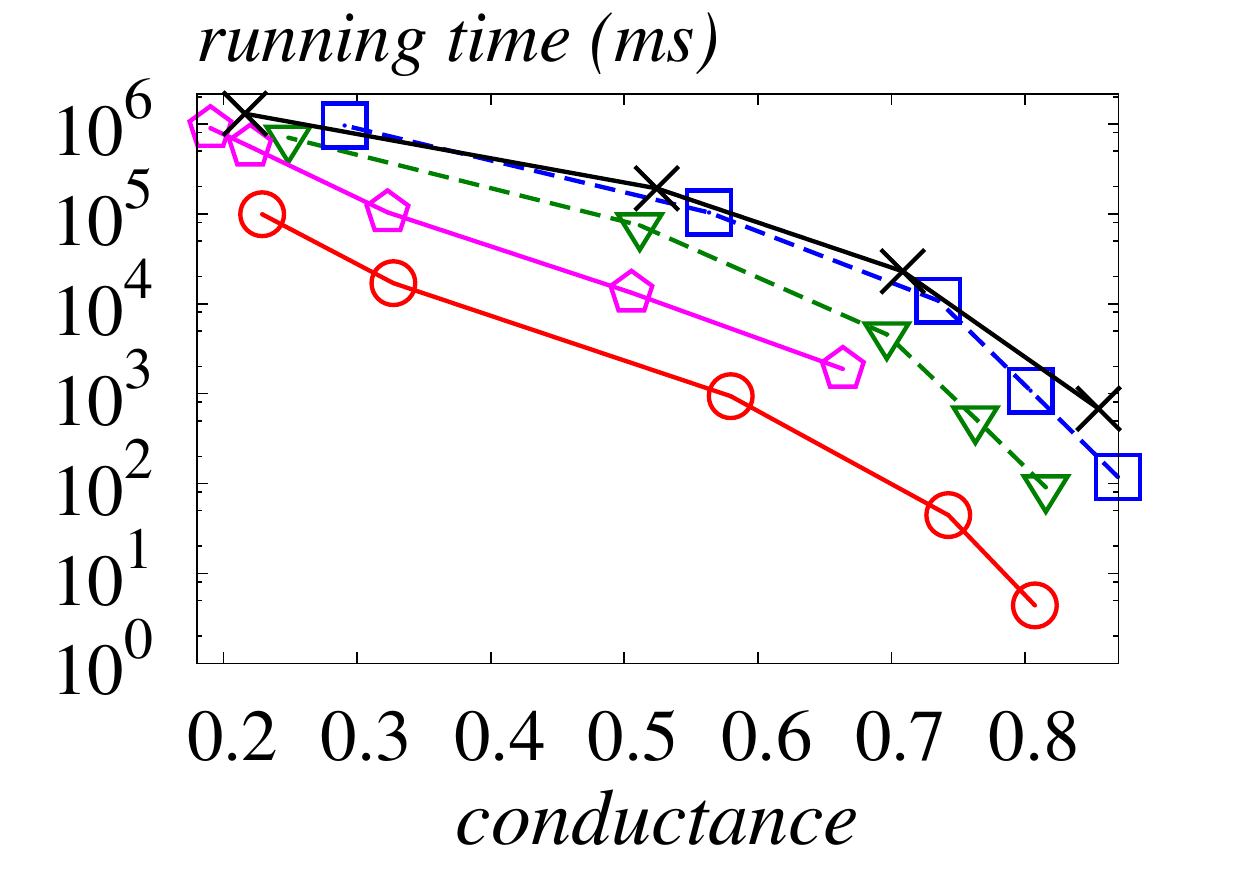}
			\\[-1mm]
			\hspace{-4mm} (e) {\em LiveJournal} &
			\hspace{-4mm} (f) {\em 3D-grid} &
			\hspace{-4mm} (g) {\em Twitter} &
			\hspace{-4mm} (h) {\em Friendster}
			\\[-1mm]
		\end{tabular}
		\vspace{-2mm}
		\caption{Running time vs conductance for local clustering queries (best viewed in color).} \label{fig:time-conductance}
		\vspace{-1mm}
	\end{small}
\end{figure*}

Figure~\ref{fig:time-opt} reports the computational time of \pukra and \pukraplus on all datasets when varying $\epsilon_r$ from $0.1$ to $0.9$ and fixing $\delta=10^{-6}$.  Observe that \pukraplus always outperforms \pukra markedly for all datasets. In particular, when the relative error threshold $\epsilon_r$ is large (e.g., $0.5-0.9$), \pukraplus is one to two orders of magnitude faster than \pukra on most of datasets, but only around $5$ times faster on {\em 3D-grid}. This is caused by the fact that each node in {\em 3D-grid} has six neighbors, thus residues will drop below the threshold quickly and both \pukra and \pukraplus require very few random walks. As we keep decreasing $\epsilon_r$, the gap between \pukra and \pukraplus is narrowed. Especially, when $\epsilon_r=0.1$, \pukraplus achieves around $5\times$ to $10\times$ speedup. The reasons are as follows. When the relative error thresholds are large, \pukraplus only needs a small number of push operations and random walks due to the new termination conditions of \hkpushplus and residue reduction method compared to \pukra. However, when the relative error thresholds are very small, the termination conditions of \hkpushplus are harder to satisfy and as a result incurs much higher costs to terminate. Furthermore, the residue reduction method is not able to reduce the number of random walks significantly since the residue sum is already small. Thus, both \pukra and \pukraplus perform many push operations and random walks, and as a result, the speedup is modest. The results demonstrate the power of new termination conditions of \hkpushplus and residue reducation method in \pukraplus, especially when the error thresholds are not very small. \textit{In summary, \pukraplus outperforms \pukra without sacrificing theoretical accuracies of HKPR values}.

\vspace{-1ex}\subsection{Comparisons with Competitors}\label{sec:exp-all}

We compare \pukra and \pukraplus against \chkpr, \simlocal \cite{veldt2016simple}, \crd \cite{wang2017capacity}, \mc and \hkrelax in terms of clustering quality and efficiency (or memory overheads). Recall that \mc is a random-walk-based approach as described in Section~\ref{sec:so}. \mc accepts two thresholds $\epsilon_r$ and $\delta$, and a failure probability $p_f$ as inputs. It performs $\frac{2(1+\epsilon_r/3)\log{(n/p_f)}}{\epsilon^2_r\cdot\delta}$ random walks from the seed node $s$ and returns a $(d, \epsilon_r,\delta)$-approximate HKPR vector for $s$ with probability at least $1-p_f$.
\hkrelax ensures $\epsilon_a$ absolute error in each $\frac{\ar[v]}{d(v)}$, which is incomparable with the accuracy guarantees of other three methods, that is $(d, \epsilon_r,\delta)$-approximation guarantees. Hence, we do not compare all algorithms under the same theoretical accuracy guarantees. Instead, we evaluate each method by its empirical clustering quality (i.e., conductance) and empirical running time (or memory overheads) with various parameter settings and find the method that achieves the best trade-off in terms of clustering quality and running time (or memory overheads). \hkrelax only has one internal parameter $\epsilon_a$, which is the absolute error threshold. We vary it in $\{10^{-8},10^{-7},10^{-6},10^{-5},10^{-4}\}$ in our experiments. Since \mc, \pukra, and \pukraplus have almost the same parameters, we set relative error threshold $\epsilon_r=0.5$, and $\delta$ is varied in $\{2\times 10^{-8}, 2\times 10^{-7}, 2\times 10^{-6}, 2\times 10^{-5}, 2\times 10^{-4}\}$ for all of them. \chkpr has one internal parameter $\epsilon$, which is varied in $\{0.005,0.01,0.02,0.05,0.1\}$. \simlocal has one locality parameter $\delta$. We vary it in $\{0.005,0.01,0.02,0.05,0.1\}$. In addition, we vary the number of iterations of \crd in $\{7,10,15,20,30\}$ and keep other parameters default. For fair comparison, we let the $x$-axis be the average conductance of the output clusters and the $y$-axis be the average running time (or memory overheads), depicting the empirical clustering quality and empirical efficiency (or memory overheads), respectively.

\begin{figure*}[!t]
	\centering
	\begin{small}
		\begin{tabular}{cccc}
			\multicolumn{4}{c}{\hspace{-4mm}
				\includegraphics[height=2.8mm]{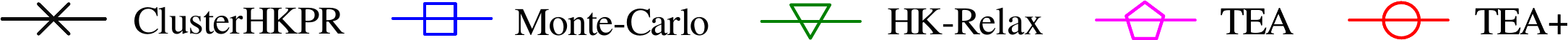}}  \\
			\hspace{-4mm} \includegraphics[height=32mm]{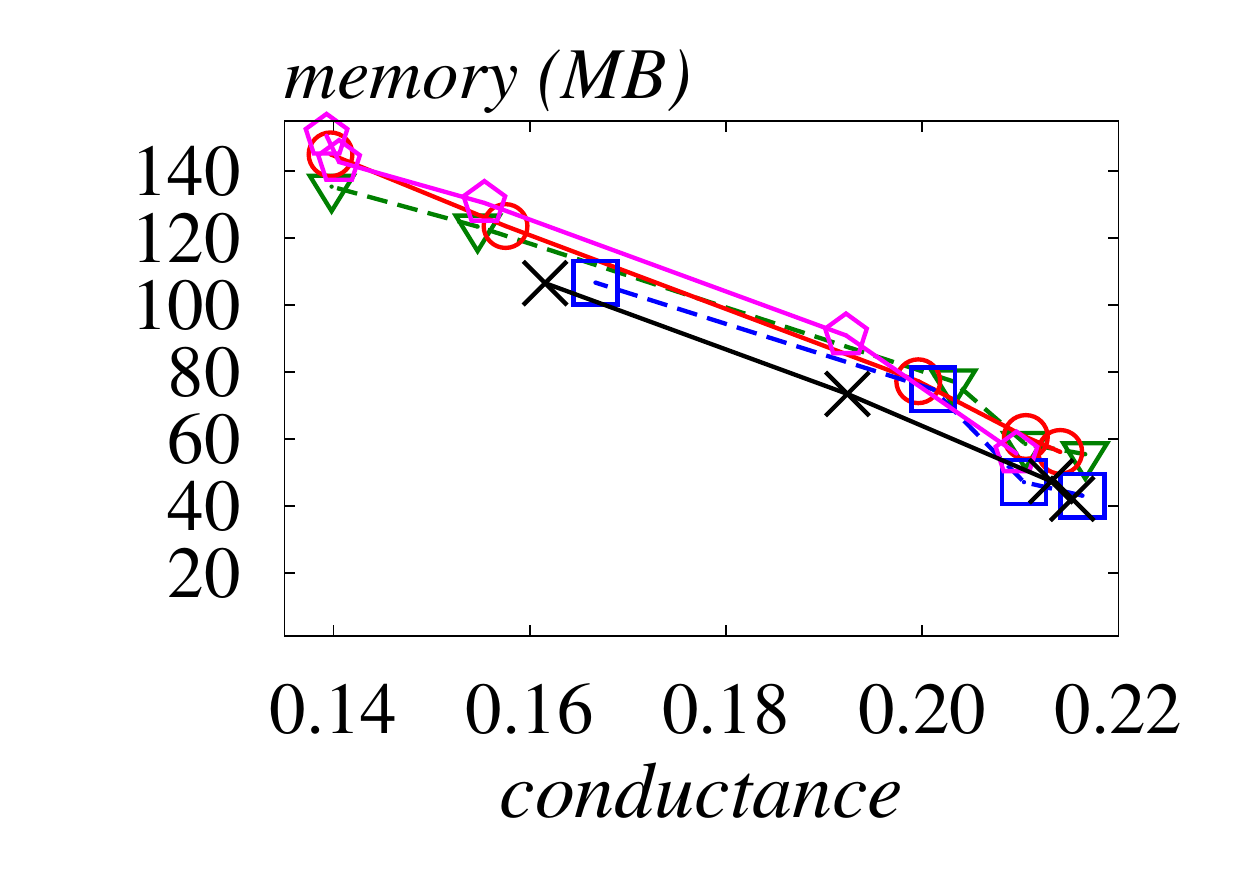} &
			\hspace{-4mm} \includegraphics[height=32mm]{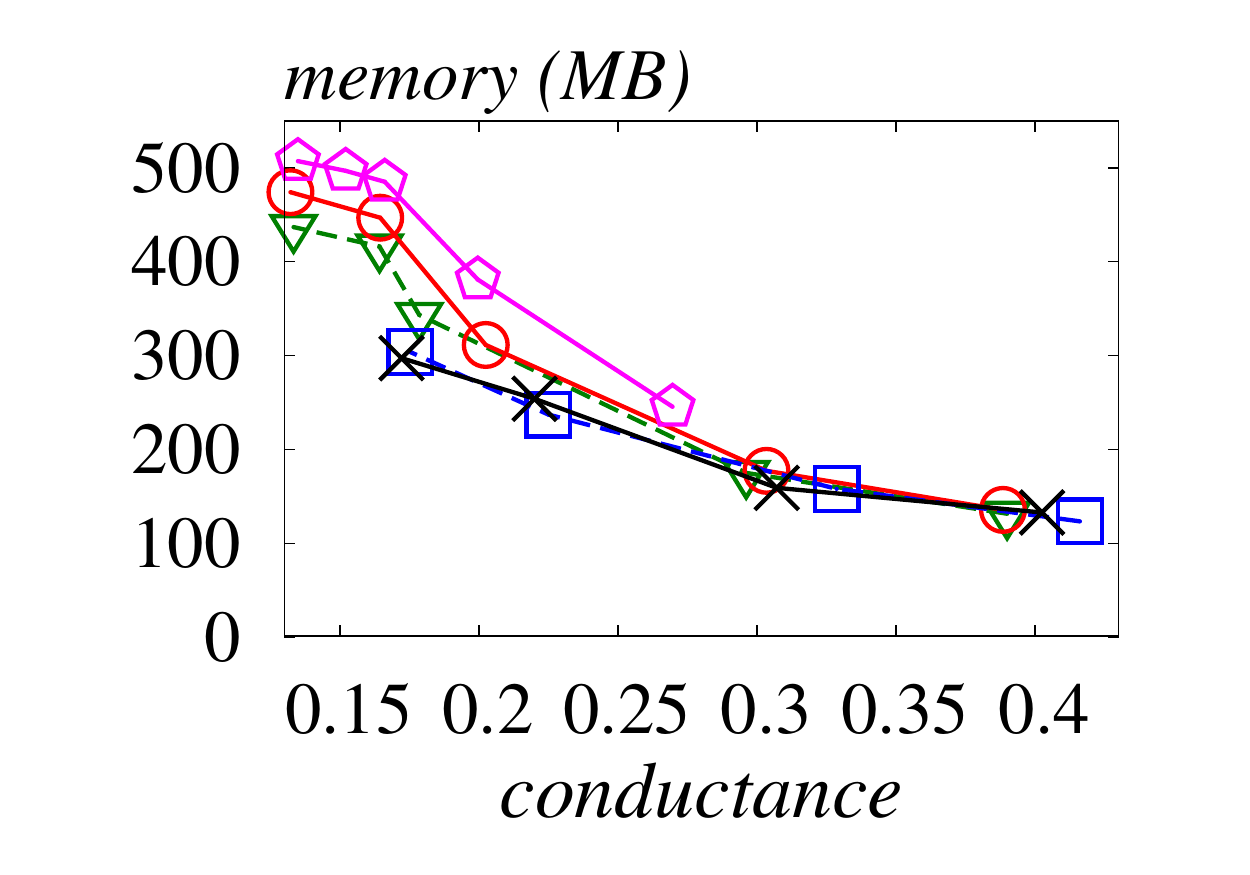} &
			\hspace{-4mm} \includegraphics[height=32mm]{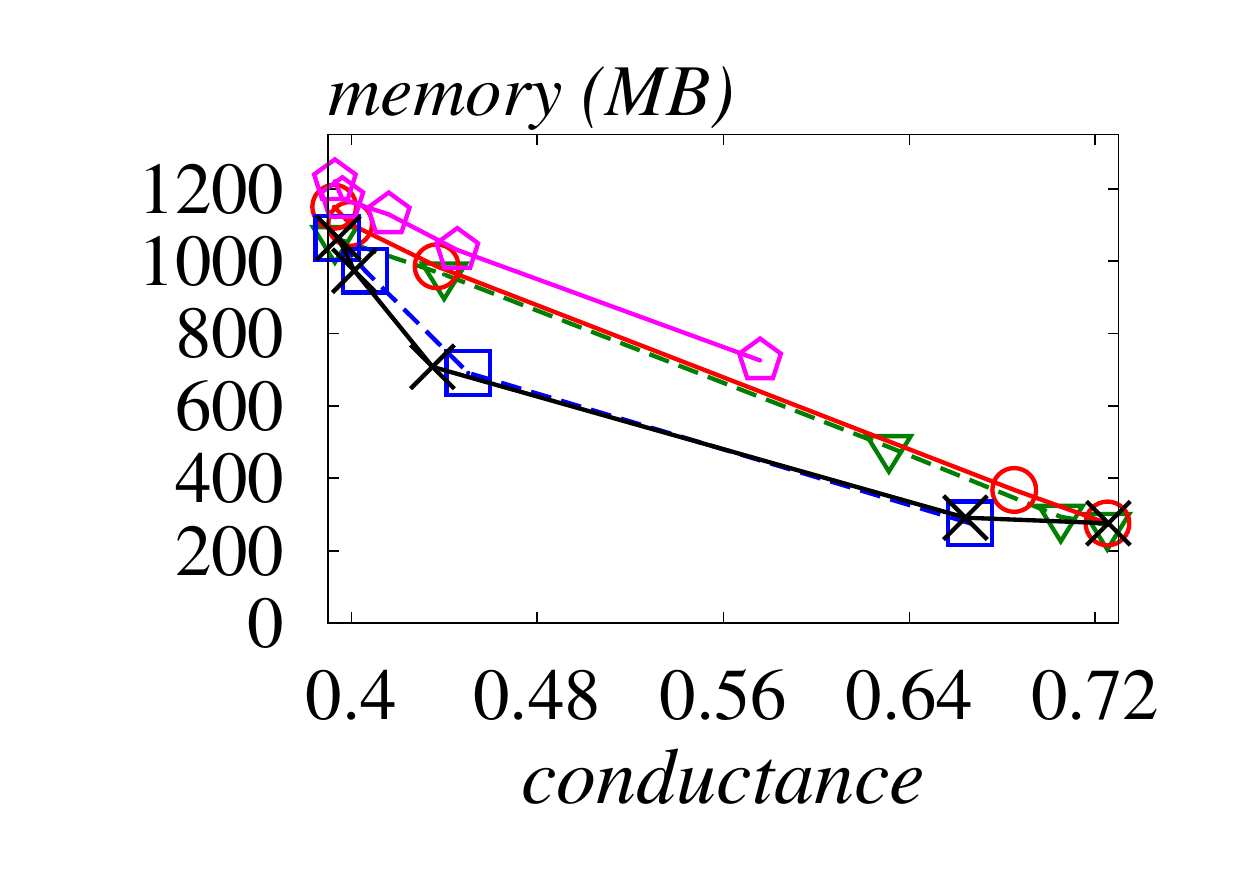} &
			\hspace{-4mm} \includegraphics[height=32mm]{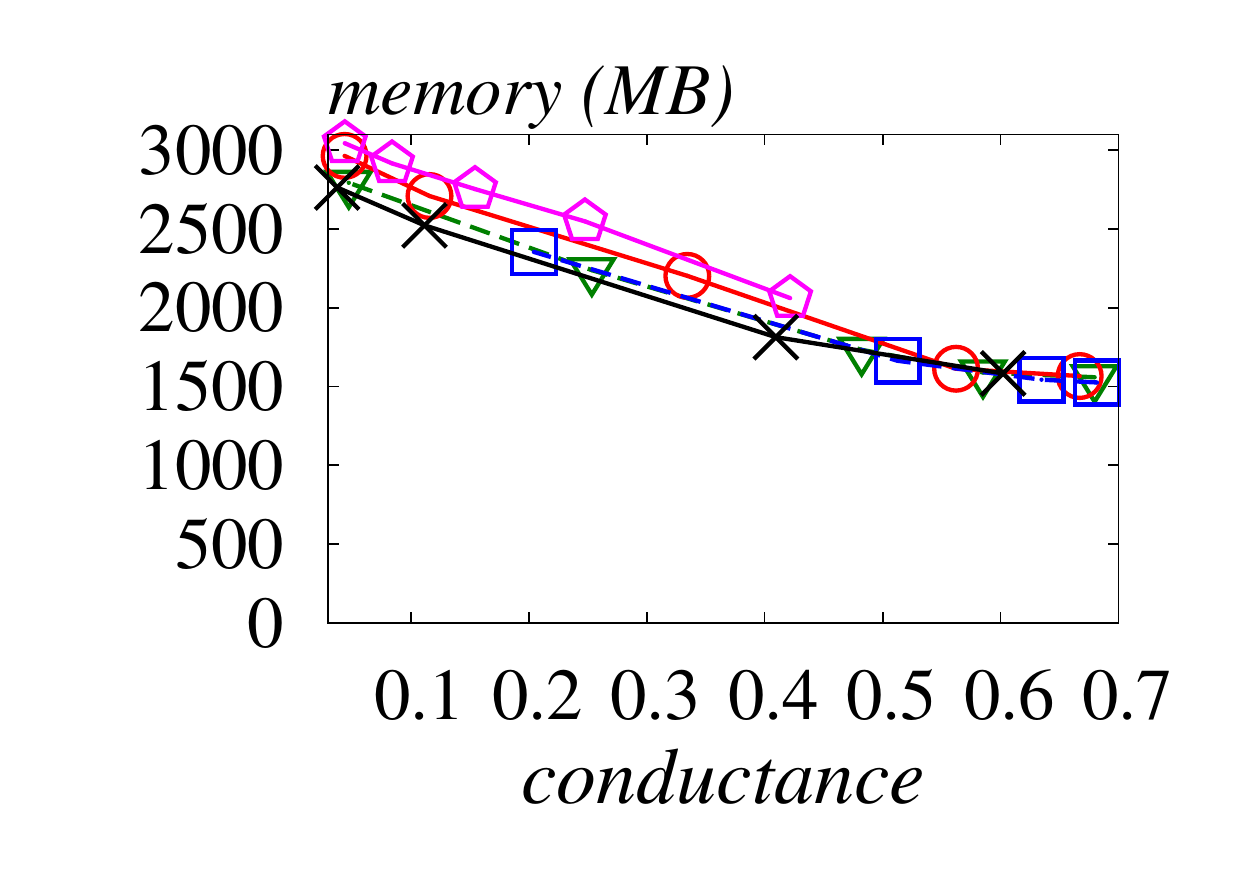}
			\\[-1mm]
			\hspace{-4mm} (a) {\em DBLP} &
			\hspace{-4mm} (b) {\em Youtube} &
			\hspace{-4mm} (c) {\em PLC} &
			\hspace{-4mm} (d) {\em Orkut}
			\\[0mm]
			
			\hspace{-4mm} \includegraphics[height=32mm]{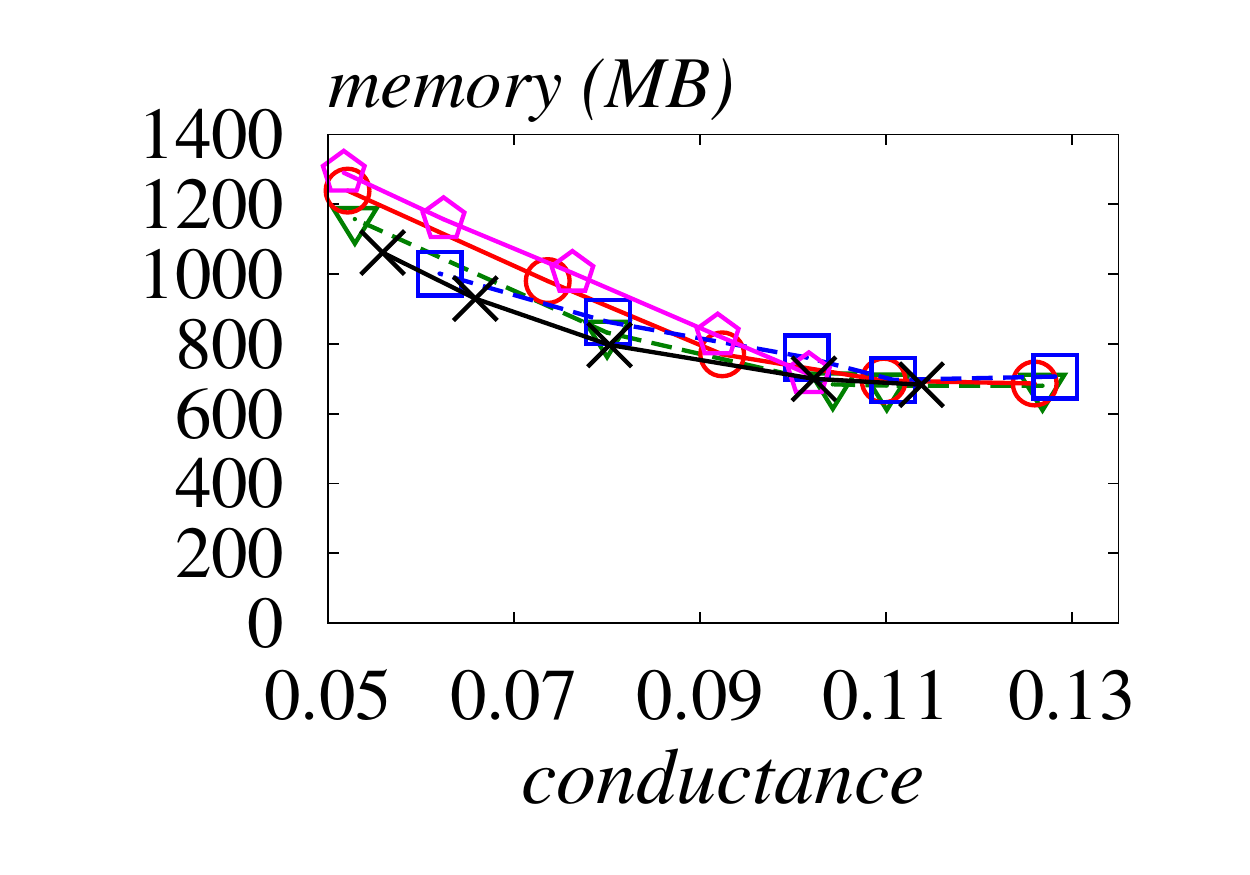} &
			\hspace{-4mm} \includegraphics[height=32mm]{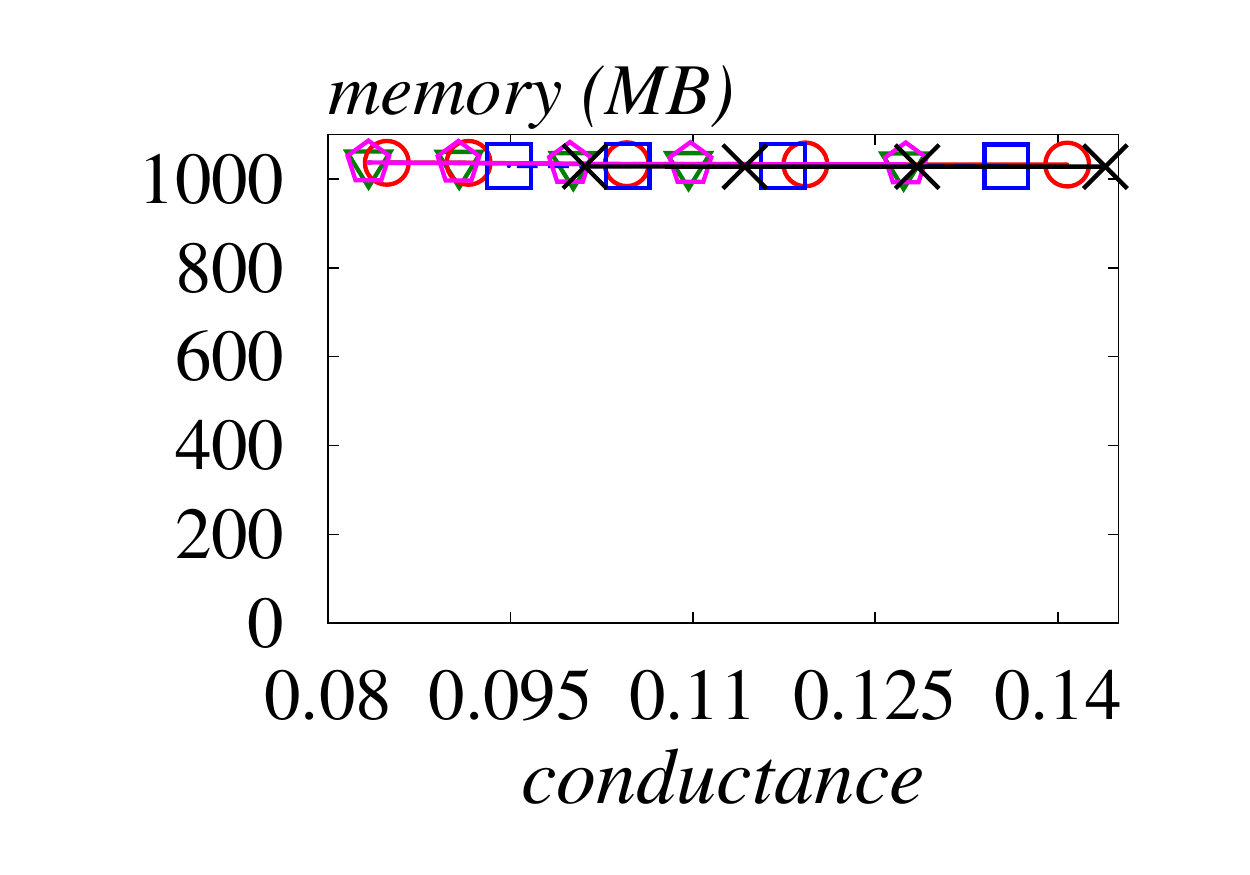} &
			\hspace{-4mm} \includegraphics[height=32mm]{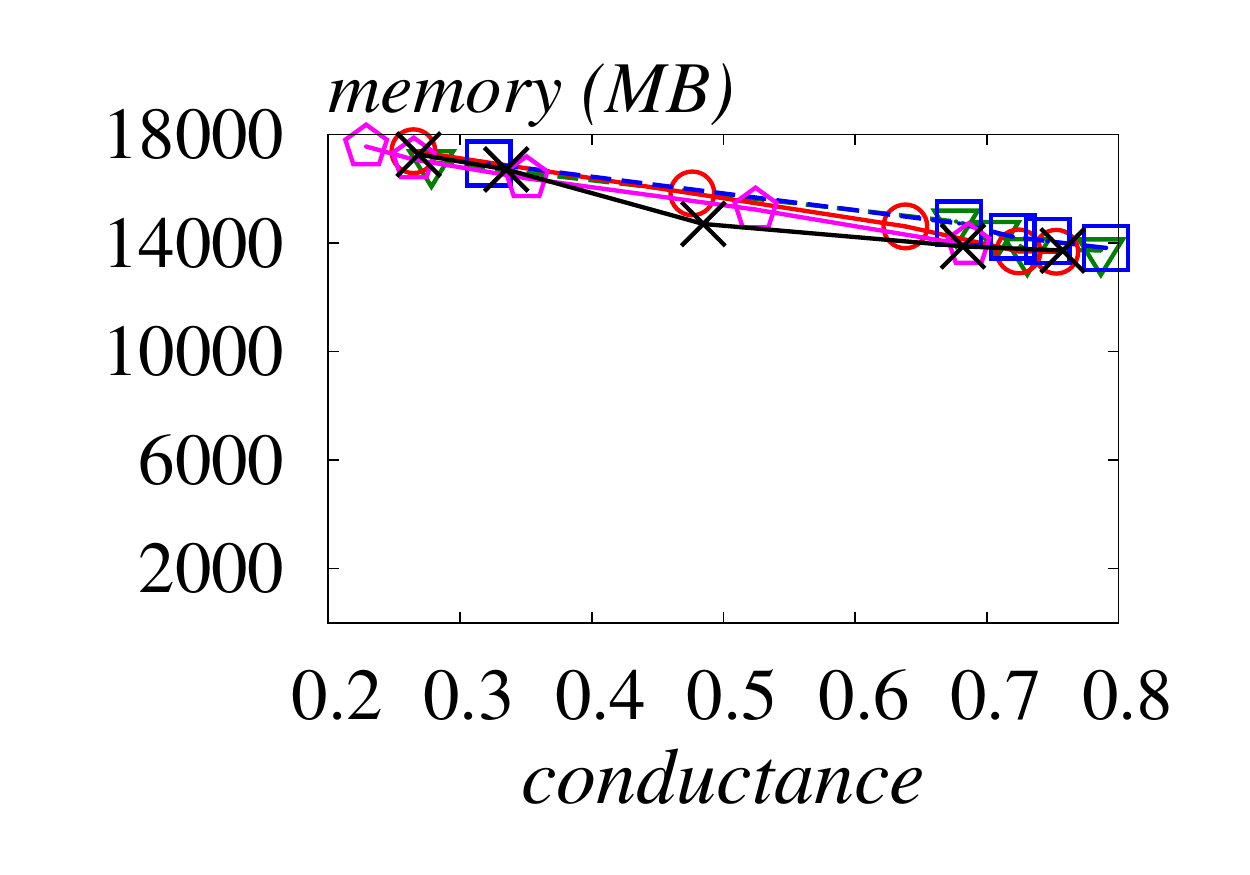} &
			\hspace{-4mm} \includegraphics[height=32mm]{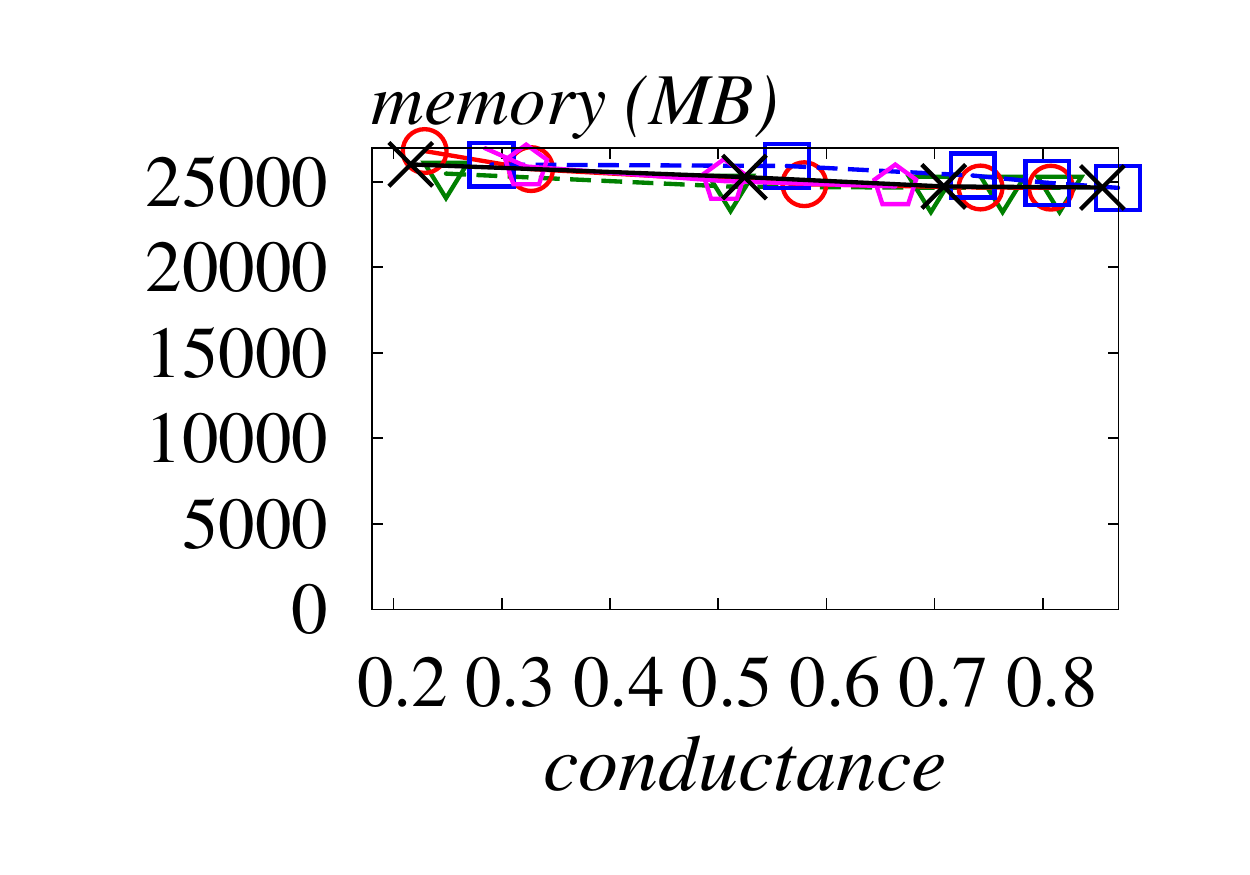}
			\\[-1mm]
			\hspace{-4mm} (e) {\em LiveJournal} &
			\hspace{-4mm} (f) {\em 3D-grid} &
			\hspace{-4mm} (g) {\em Twitter} &
			\hspace{-4mm} (h) {\em Friendster}
			\\[-1mm]
		\end{tabular}
		\vspace{-2mm}
		\caption{Memory cost vs.\ conductance.} \label{fig:mem-algo}
		\vspace{-1mm}
	\end{small}
\end{figure*}

\begin{figure*}[!t]
	\centering
	\begin{small}
		\begin{tabular}{cccc}
			\multicolumn{4}{c}{
				\hspace{-4mm}\includegraphics[height=2.8mm]{algolegendfig4.pdf}}  \\
			\hspace{-4mm} \includegraphics[height=32mm]{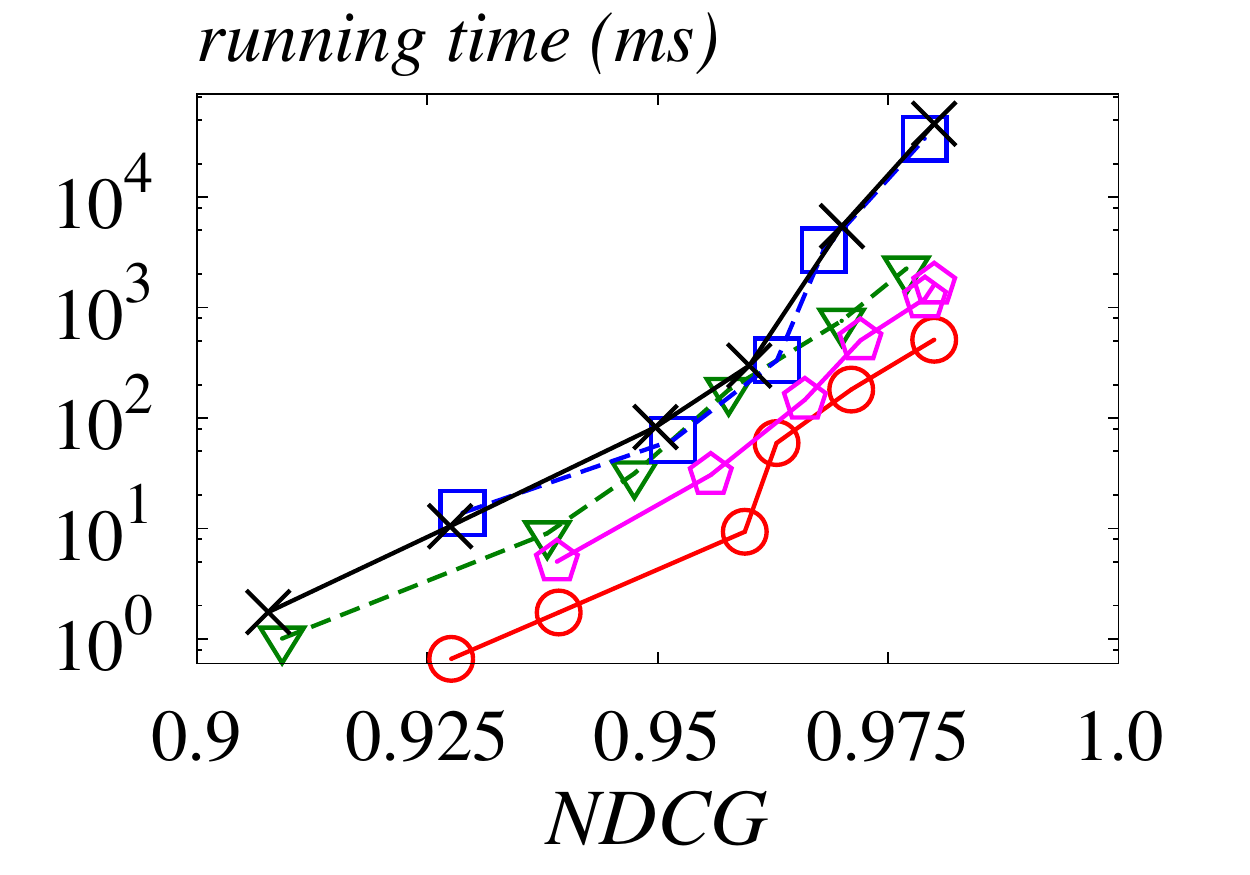} &
			\hspace{-4mm} \includegraphics[height=32mm]{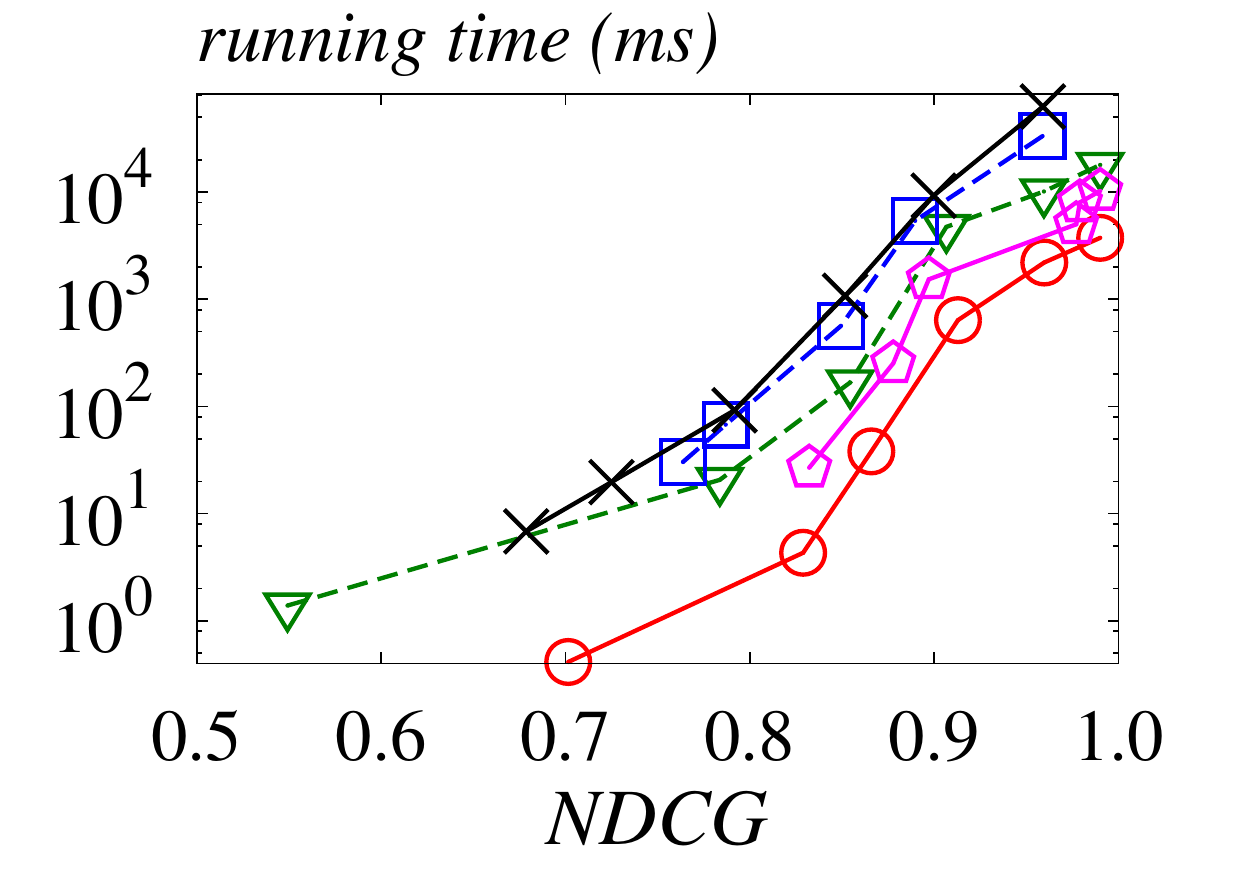} &
			\hspace{-4mm} \includegraphics[height=32mm]{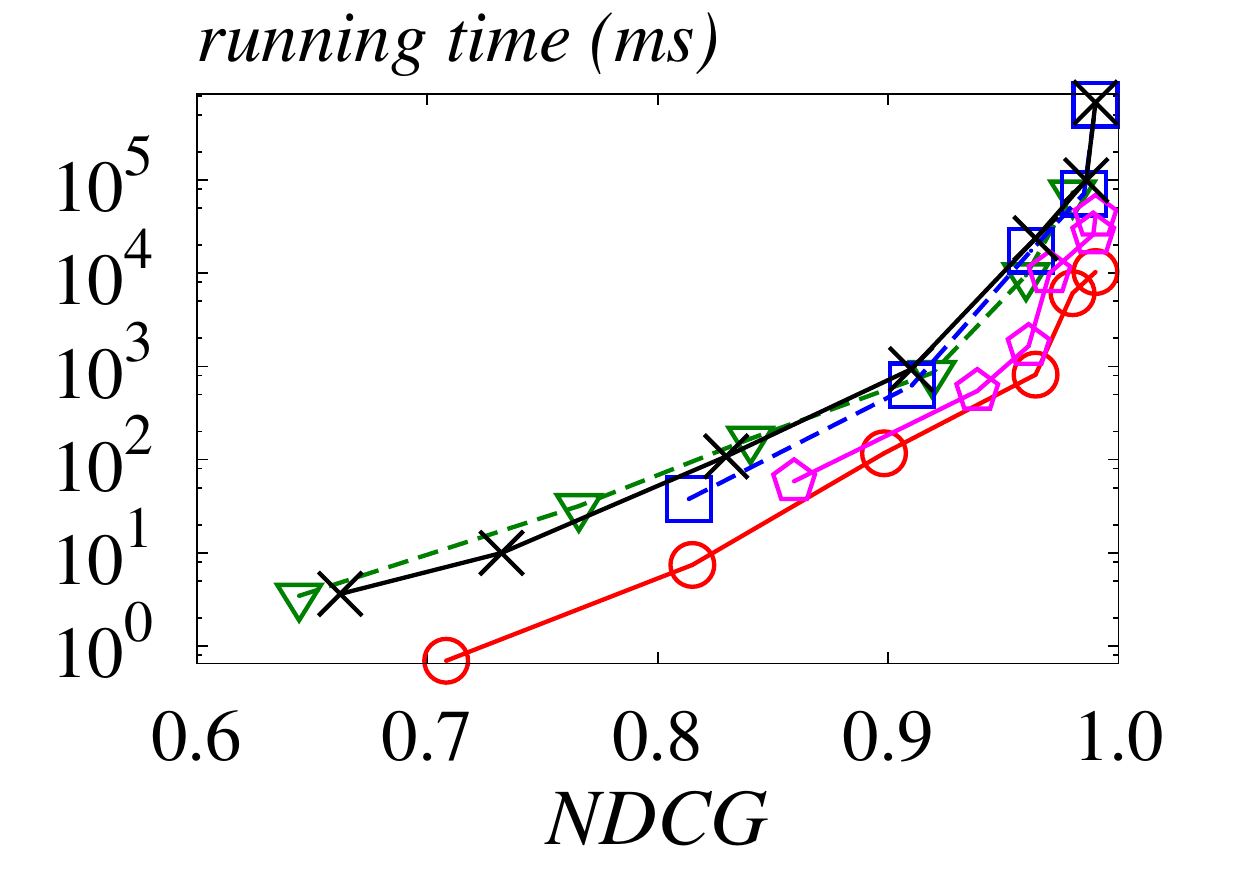} &
			\hspace{-4mm} \includegraphics[height=32mm]{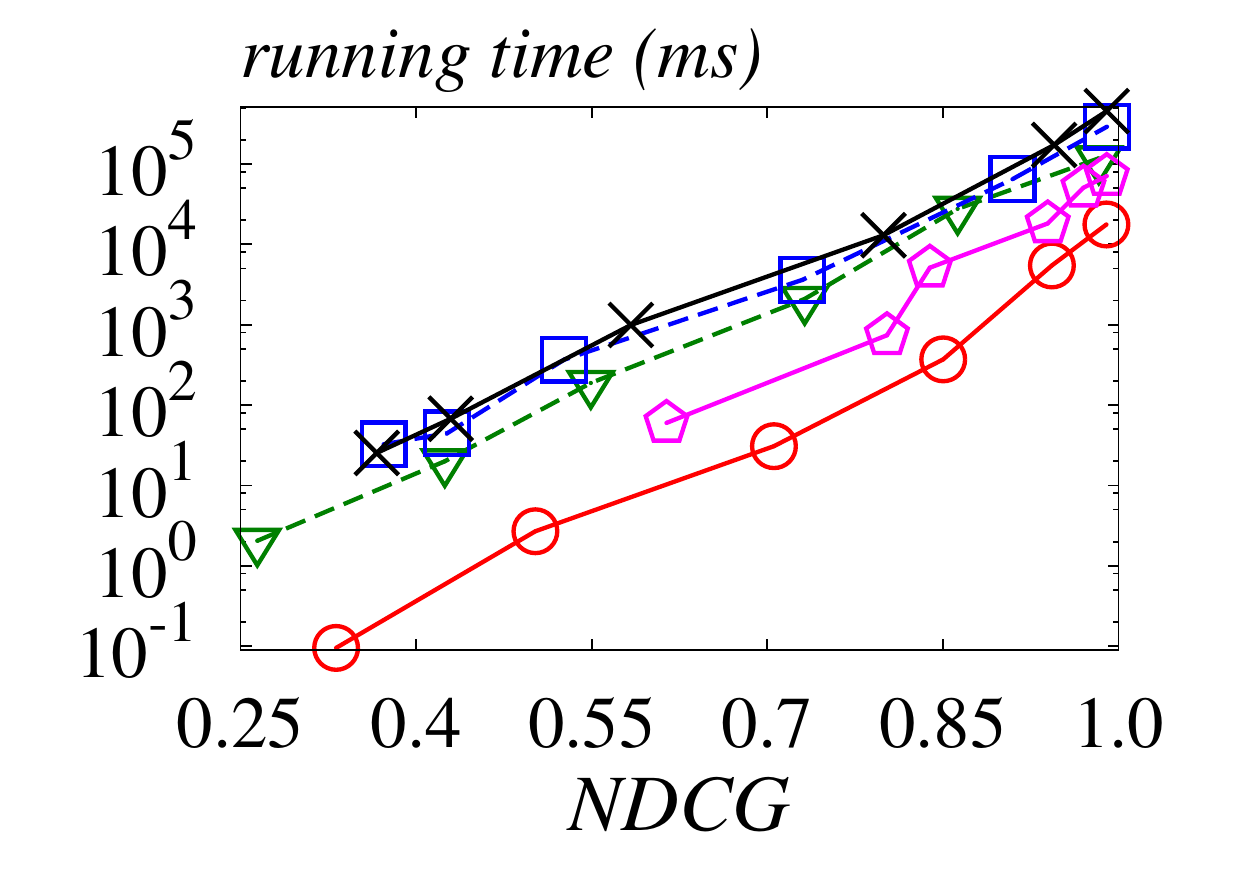}
			\\[-1mm]
			\hspace{-4mm} (a) {\em DBLP} &
			\hspace{-4mm} (b) {\em Youtube} &
			\hspace{-4mm} (c) {\em PLC} &
			\hspace{-4mm} (d) {\em Orkut}
		\end{tabular}
		\vspace{-2ex}
		\caption{Running time vs. NDCG for computing normalized HKPR (best viewed in color).} \label{fig:time-ndcg}
	\end{small}
	\vspace{-1ex}
\end{figure*}

\begin{table*}[t]
	\renewcommand{\arraystretch}{1}
	\centering
	\caption{The result of evaluating all algorithms on finding real-world communities.}
	\vspace{-2mm}
	\begin{footnotesize}
		\begin{tabular}{|c|c|c|c|c|c|c|c|c|c|c|}
			\hline
			\multirow{2}{*}{\bf Data} &
			\multicolumn{5}{c|}{\bf $F_{1}$-measure} &
			\multicolumn{5}{c|}{\bf Running Time (ms)}\\ \cline{2-11}
			& \chkpr & \mc & \hkrelax & \pukra & \pukraplus & \chkpr & \mc & \hkrelax & \pukra & \pukraplus  \\
			\hline
			{\em DBLP} & 0.13655 & 0.13631 & 0.13592 & {\bf 0.13679} & 0.136699 & 3053.95 & 2891.64 & 297.78 & 176 & 109.66 \\
			\hline
			{\em Youtube} & 0.10113 & 0.10097 & 0.09858 & 0.10133 & {\bf 0.10334} & 7.76 & 7.2 & 8.11 & 2.49 & 2\\
			\hline
			{\em LiveJournal} & 0.64644 & 0.65105 & 0.64516 & 0.64959 & {\bf 0.67} & 1.34665 & 1.2 & 3.57 & 0.55 & 0.29 \\
			\hline
			{\em Orkut} & 0.18497 & 0.18464 & 0.19375 & 0.19333 & {\bf 0.19636} & 29.95 & 29.35 & 62.17 & 24.78 & 14.86 \\
			\hline
		\end{tabular}
	\end{footnotesize}
	\label{tab:f1}
\end{table*}

Figure~\ref{fig:time-conductance} shows the average conductance of the output clusters and the the average running time when varying the aforementioned parameters for each algorithm. We can make the following observations. First, for each algorithm, when the error thresholds (i.e., $\epsilon_a, \epsilon$ and $\delta$) become smaller or the number of iterations increase, the conductance of the output clusters reduces (i.e., the quality of the output clusters improves), as well as the computational time goes up markedly, which accords with their theoretical time complexities. In particular, \simlocal incurs very high running time as well as poor cluster quality due to its high time complexity and the fact that it is mainly devised for recovering the cluster for a subset of nodes of the cluster rather than detecting a cluster for a single seed node. We also note that \crd shows a better performance than \simlocal. However, it is still much slower than HKPR-based methods. Hence, we omit the results of \simlocal on the remaining datasets and that of \crd on {\em Orkut, LiveJournal, 3D-grid, Twitter} and {\em Friendster}.

Second, we can see that \mc and \chkpr take several minutes to finish one local clustering query on all graph datasets in order to find clusters with small conductance. Hence, it is outperformed by \hkrelax, \pukra and \pukraplus by $1$ to $3$ orders of magnitude when they output clusters with almost the same conductance. This is due to the fact that \mc and \chkpr require performing a large number of random walks. In fact, this is consistent with the experimental results in prior work~\cite{chung2015computing,shun2016parallel}. Moreover, it can be observed that \hkrelax always runs faster than \mc and \chkpr and achieves more than $10\times$ speedup on all datasets except {\em Orkut}, {\em Twitter}, and {\em Friendster}. To explain this phenomenon, recall that \hkrelax requires iteratively pushing residuals to $1-K$-hop nodes from the seed node and $K$ is very large (see Section~\ref{sec:cpt}). However, on graphs with large average degrees ({\em Orkut}, {\em Twitter}, and {\em Friendster}) the number of push operations increases dramatically after several hops from the seed node.

Third, \pukraplus outperforms \hkrelax by more than $10\times$ speedup on {\em PLC}, {\em Orkut}, {\em Twitter}, and {\em Friendster}, and more than $4\times$ speedup on the rest of graphs. The speedup is achieved by new termination conditions for \hkpushplus and the residue reduction method for reducing the number random walks. However, we note that the speedup on {\em DBLP}, {\em Youtube}, {\em LiveJournal} and {\em 3D-grid} is not as significant as that on {\em PLC}, {\em Orkut}, {\em Twitter}, and {\em Friendster}. The reason is that these graphs either have large clustering coefficients~\cite{yang2012def} or small average degrees. The first one can also be observed in our experiments. With the same parameters as inputs to all three algorithms, the conductance values of output clusters from {\em PLC}, {\em Orkut}, {\em Twitter}, and {\em Friendster} are clearly greater than those from the remaining four graphs (i.e., {\em DBLP}, {\em Youtube}, {\em LiveJournal} and {\em 3D-grid}). This implies that nodes in these four graphs are more likely to cluster together, and then residues on these graphs tend to be propagated within a small cluster of nodes when \hkpushplus is performed. Now consider the second reason. Recall that \hkpushplus iteratively distributes residues to neighbors evenly before any termination condition is satisfied. Since the average degrees are small, the residues that each node receives will be large. Additionally, it needs more iterations to distribute the residues to more nodes, which may not be done before termination. As a result of these two factors, a few nodes will hold large residues rather than many nodes holding small residues. Consequently, the residue reduction method in \pukraplus fails to significantly reduce the number of random walks. Note that \hkrelax, \pukra and \pukraplus all terminate very quickly on {\em 3D-grid} (less than 10 milliseconds), which is consistent with the observation in \cite{shun2016parallel}. This is due to the fact that each node in {\em 3D-grid} has six neighbors and the residues will drop below the residue threshold quickly after performing several rounds of push operations.

In addition, we also note that \pukra fails to achieve considerable speedup compared with \hkrelax. Especially on {\em DBLP}, {\em Youtube}, {\em LiveJournal}, and {\em 3D-grid}, \pukra's performance degrades to the same level as that of \hkrelax. This is also caused by aforementioned high clustering coefficients and small average degrees of these graphs. \pukraplus is around $4\times$ faster than \pukra on {\em Orkut}, {\em Twitter}, and {\em Friendster}. \textit{In summary, our experiments demonstrate the power of new termination condition of \hkpushplus and residue reduction method, which reduce many push operations and random walks without sacrificing the cluster qualities}.

Figure \ref{fig:mem-algo} shows the memory overheads (including the space required to store the input graph) for all datasets by varying the error thresholds. First, we observe that the memory overheads increase with the reduction in error thresholds as more space is required to store residues and HKPR values. In particular, memory overheads on {\em 3D-grid} and {\em Friendster} remain almost stable for all algorithms. As each node in {\em 3D-grid} is connected to six neighbors, \chkpr, \mc, \hkrelax, \pukra and \pukraplus easily detect the large set of nodes around the seed nodes and the size of memory allocated for storing HKPR values for this set of nodes remains stable. In {\em Friendster}, the input graph size is too large and memory usage is not impacted remarkably. Second, we observe that the memory overheads of all algorithms are roughly comparable because space usage is dominated by the storage of the input graph, and hence, in terms of memory cost, there is no significant advantage to choose one algorithm over another.

\begin{figure*}[!t]
	\centering
	\begin{small}
		\begin{tabular}{cccc}
			\multicolumn{4}{c}{
				\hspace{-4mm}\includegraphics[height=2.8mm]{algolegendfig4.pdf}}  \\
			\hspace{-4mm}\includegraphics[height=32mm]{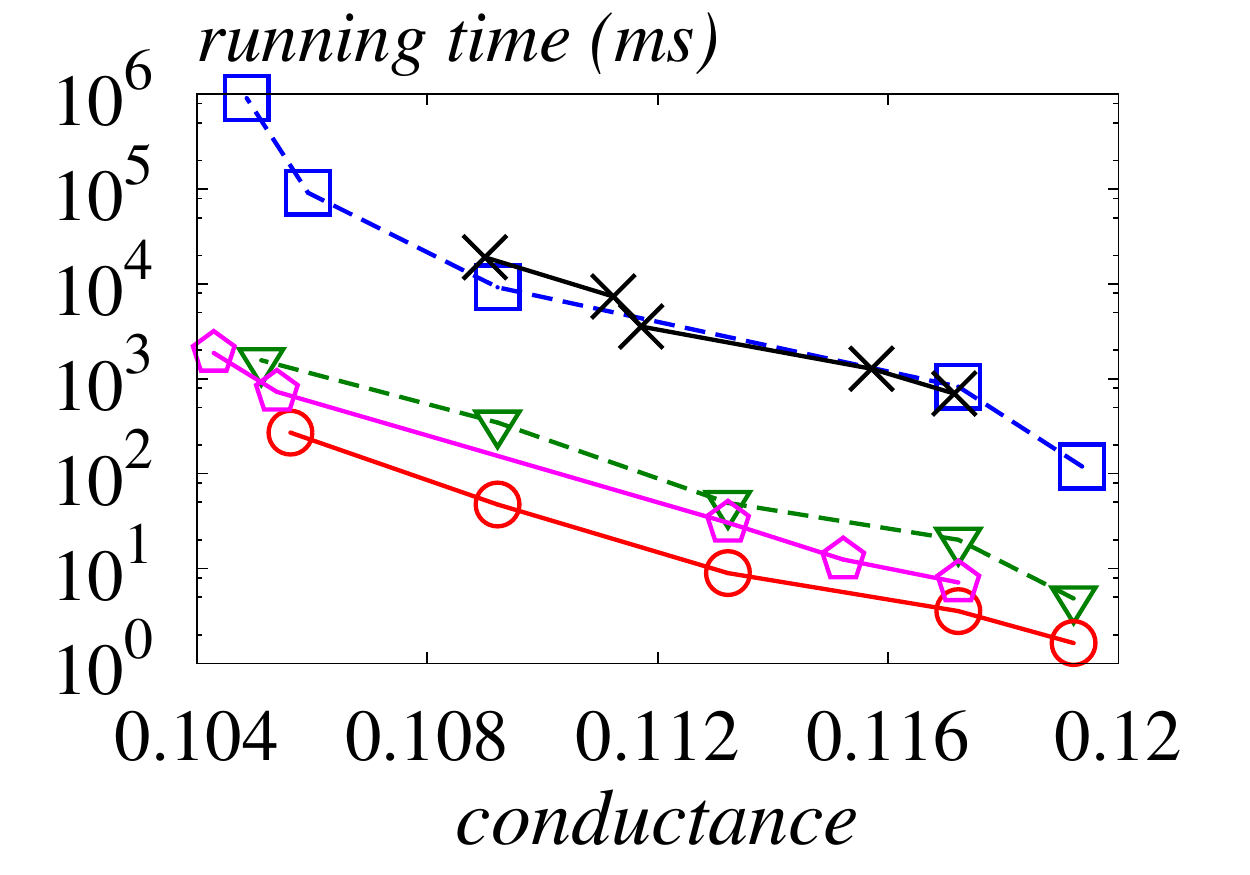} &
			\hspace{-4mm}\includegraphics[height=32mm]{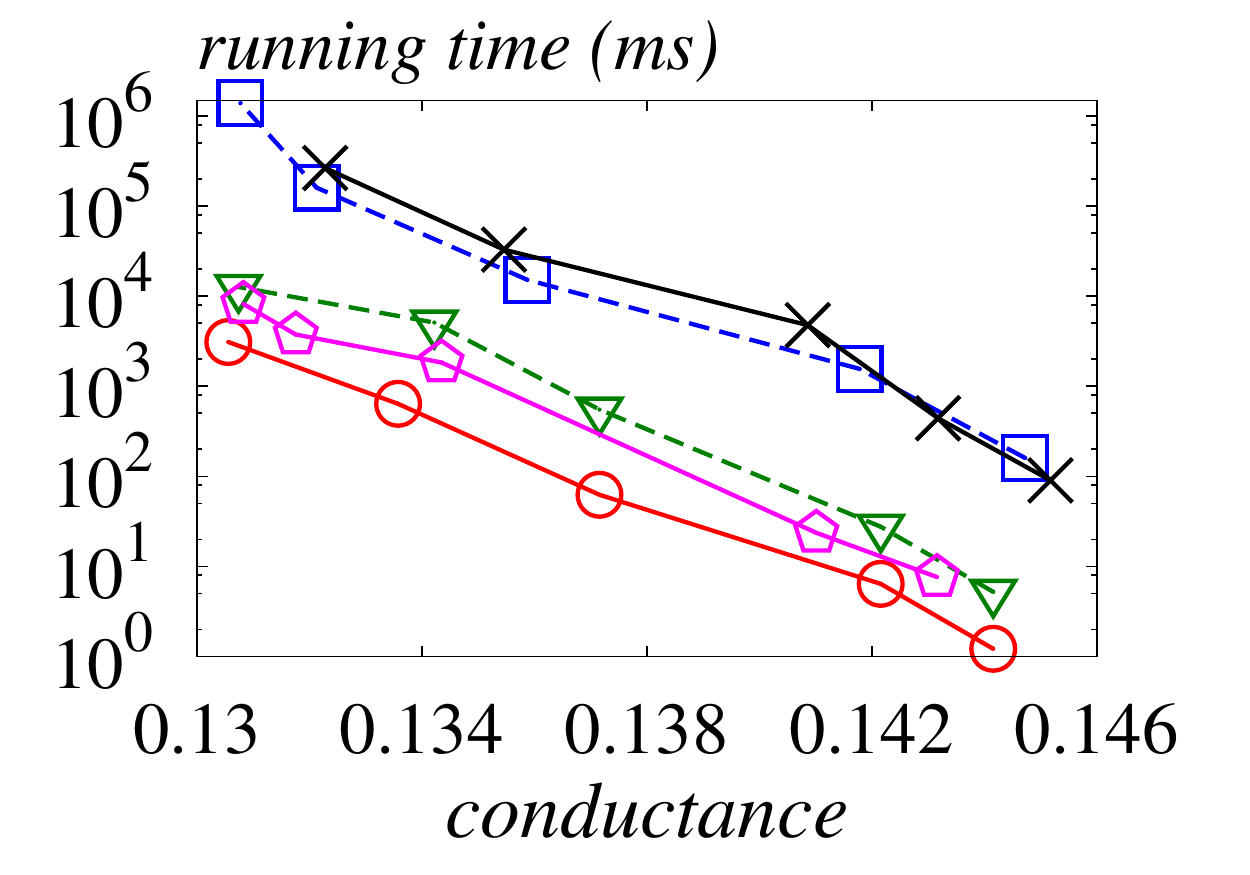} &
			\hspace{-4mm}\includegraphics[height=32mm]{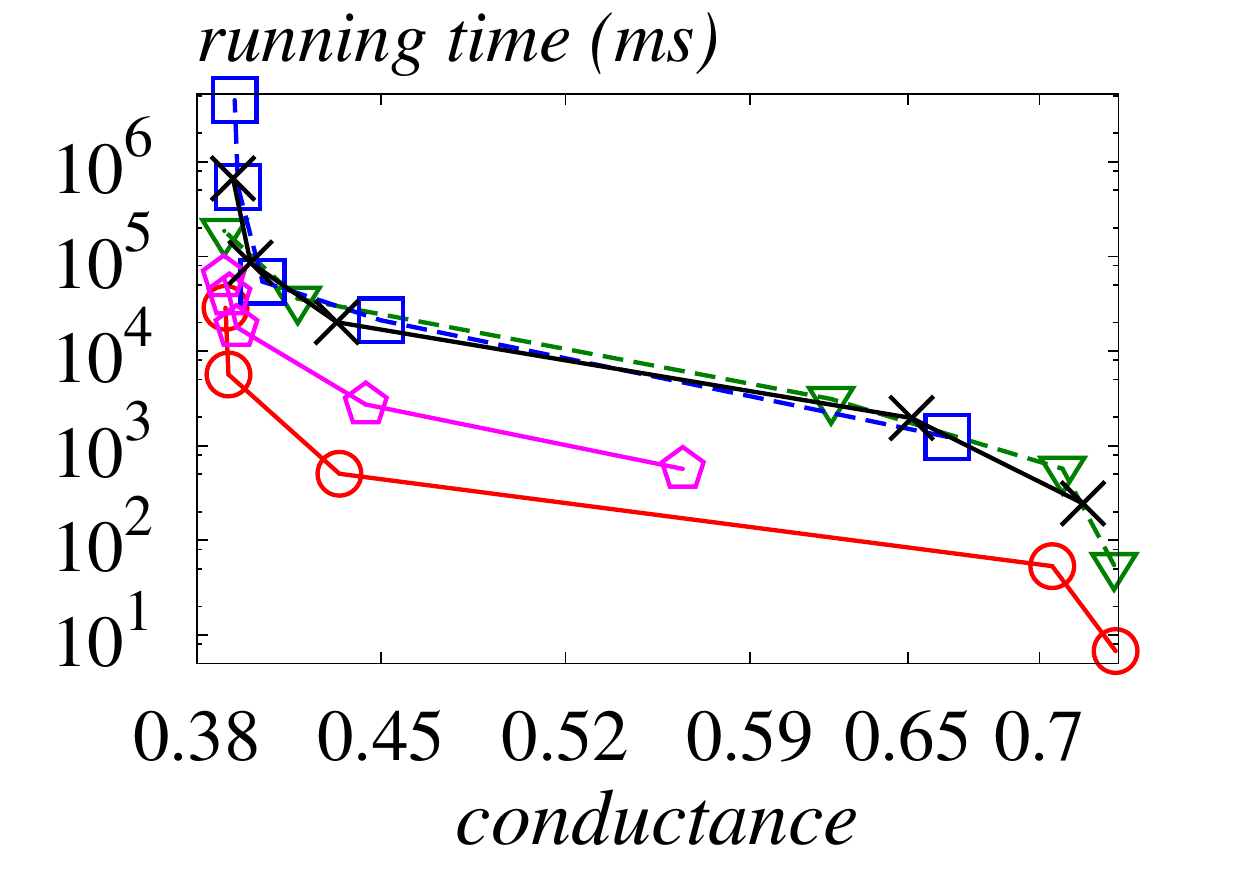} &
			\hspace{-4mm}\includegraphics[height=32mm]{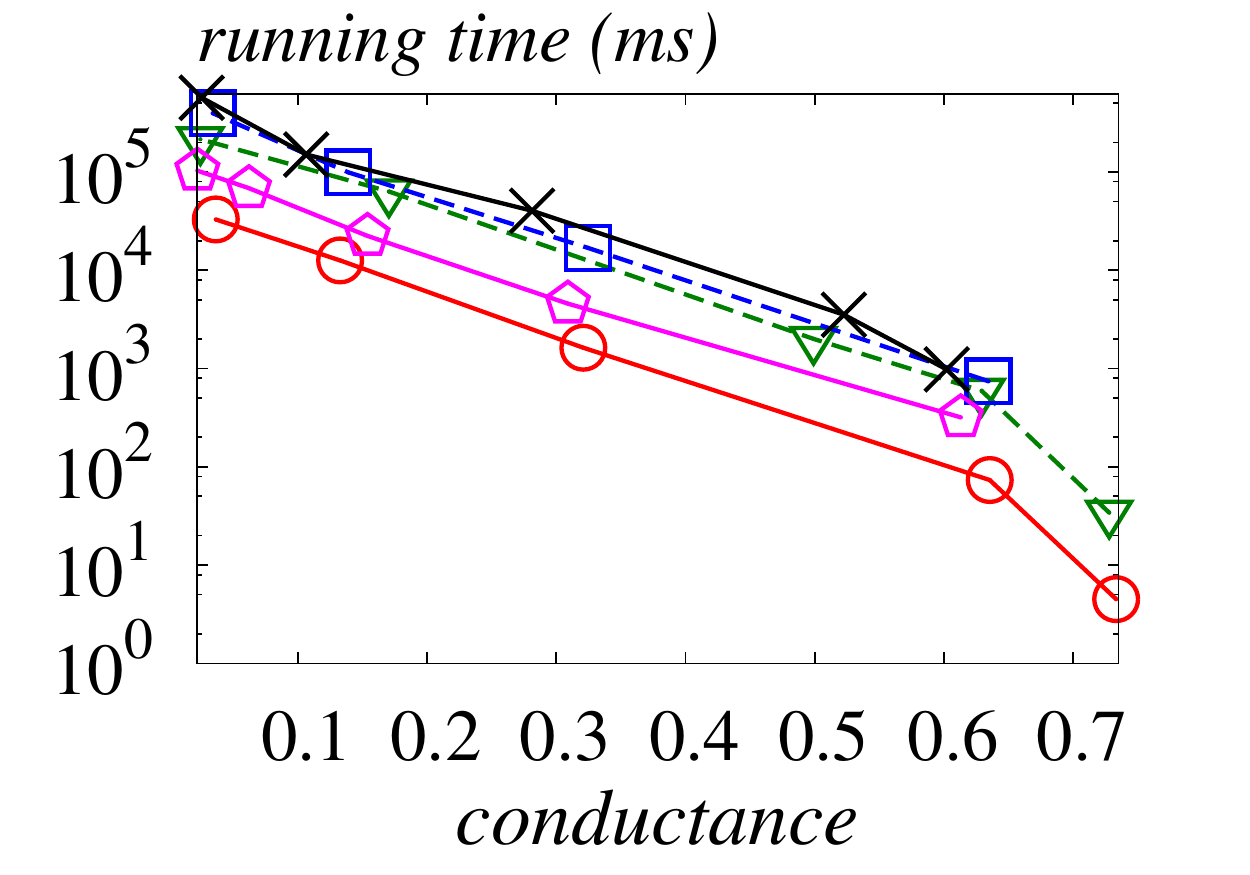}
			\\[1mm]
			\hspace{-4mm} (a) {\em DBLP} (high-density) &
			\hspace{-4mm} (b) {\em Youtube} (high-density)&
			\hspace{-4mm} (c) {\em PLC} (high-density)&
			\hspace{-4mm} (d) {\em Orkut} (high-density)
			\\[1mm]
			
			\hspace{-4mm}\includegraphics[height=32mm]{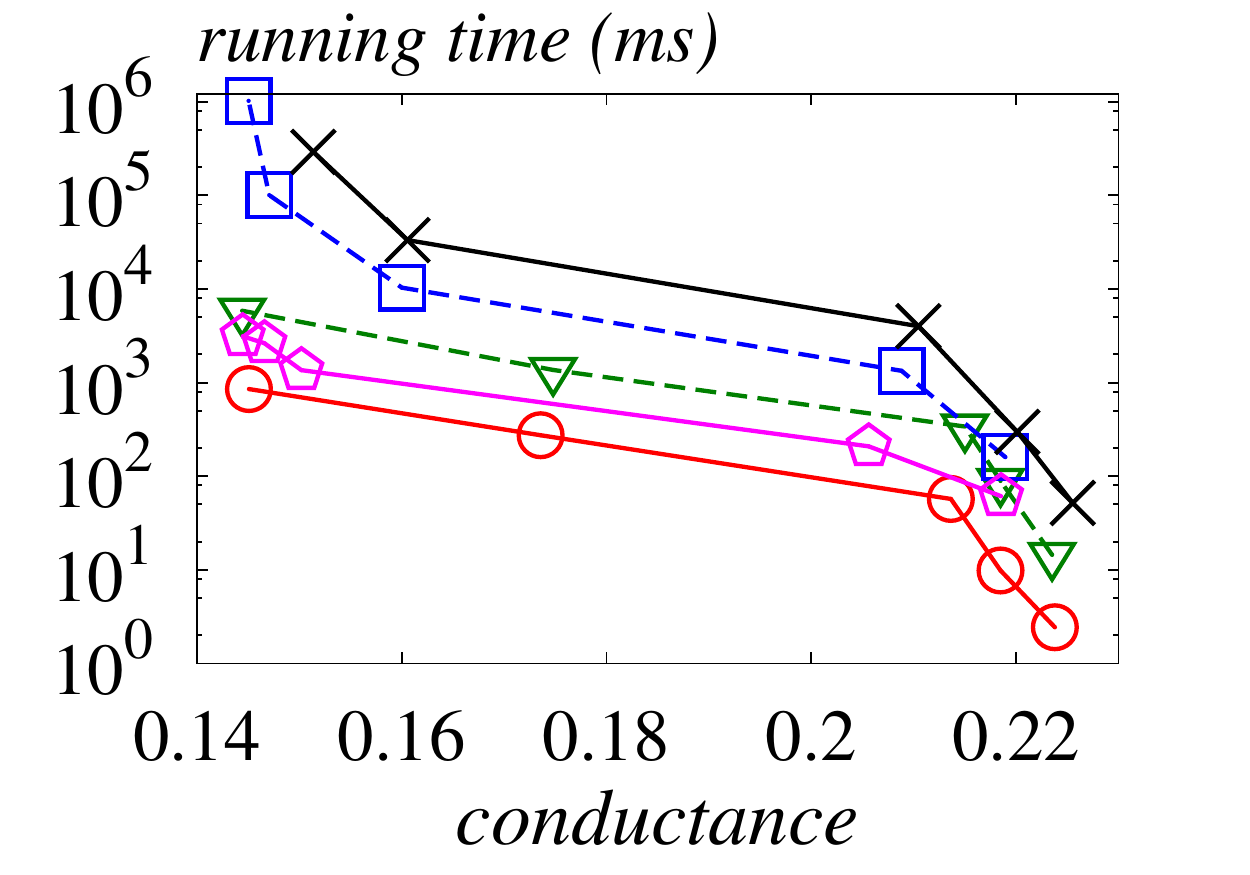} &
			\hspace{-4mm}\includegraphics[height=32mm]{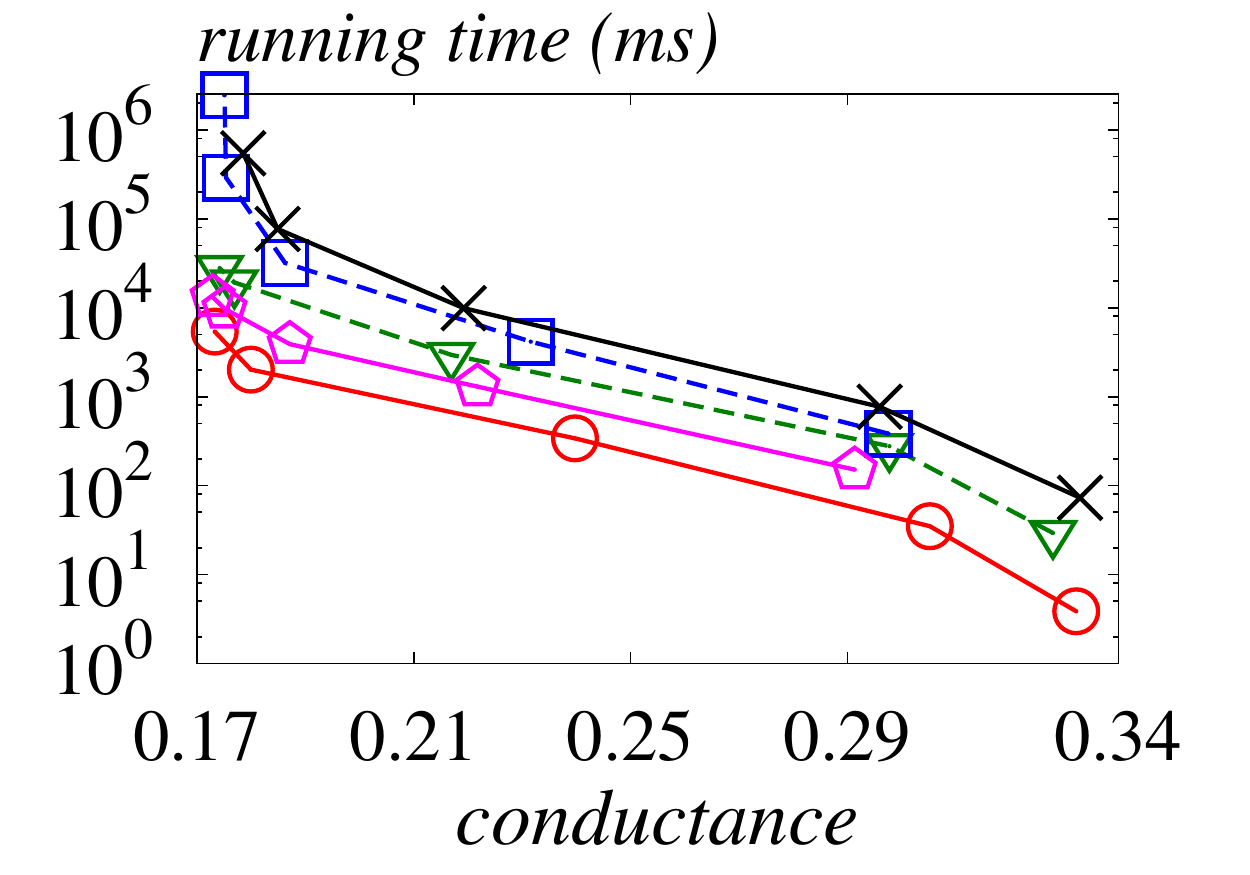} &
			\hspace{-4mm}\includegraphics[height=32mm]{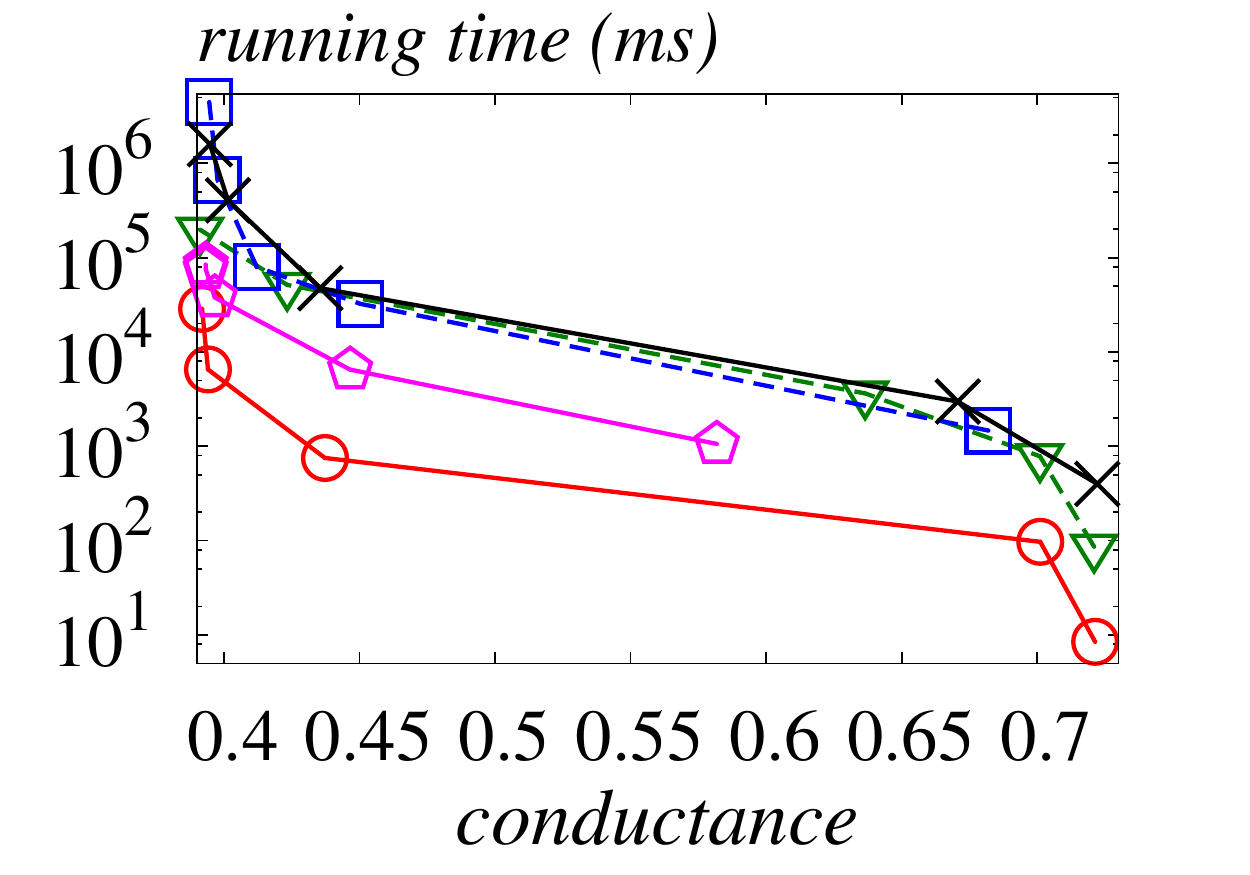} &
			\hspace{-4mm}\includegraphics[height=32mm]{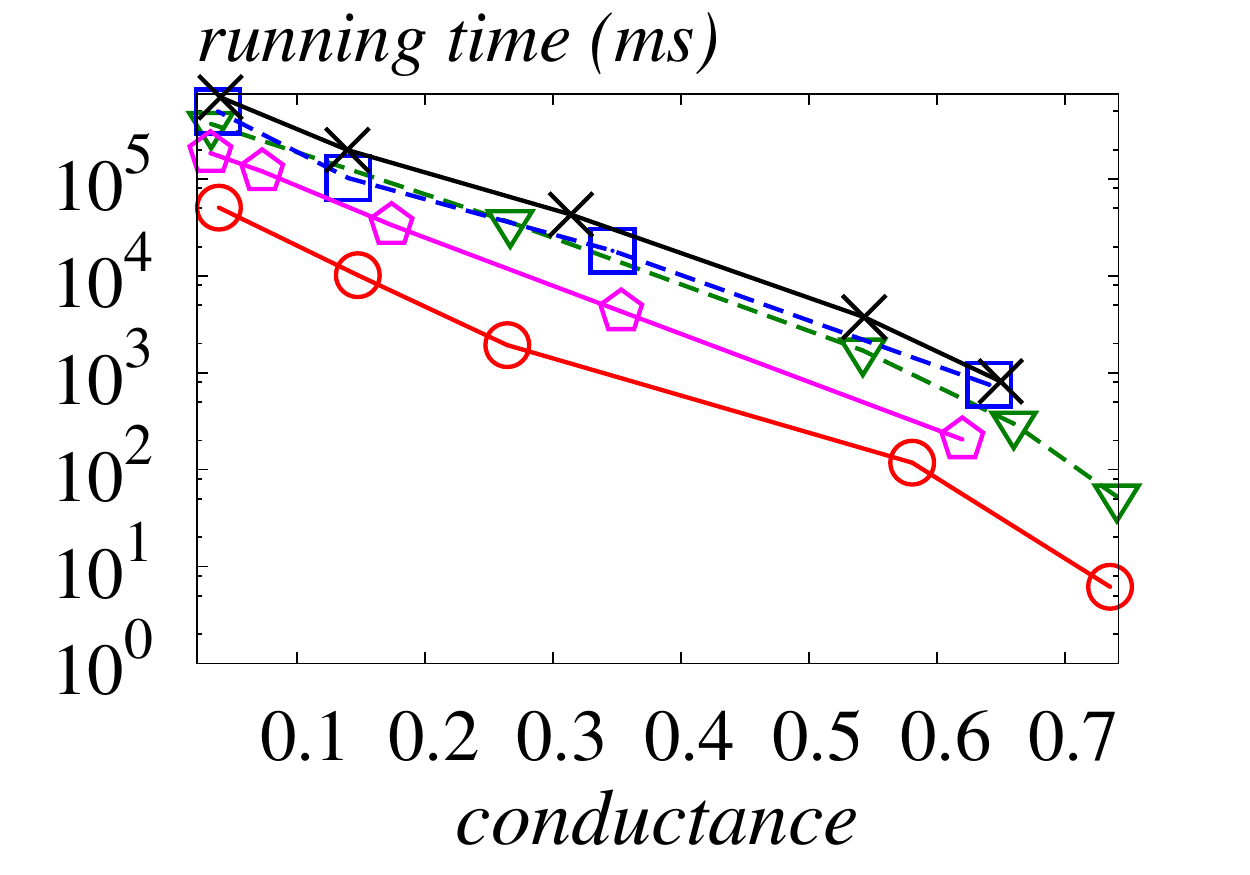} 
			\\[1mm]
			\hspace{-4mm} (e) {\em DBLP} (medium-density)&
			\hspace{-4mm} (f) {\em Youtube} (medium-density)&
			\hspace{-4mm} (g) {\em PLC} (medium-density)&
			\hspace{-4mm} (h) {\em Orkut} (medium-density)
			\\[1mm]
			
			\hspace{-4mm}\includegraphics[height=32mm]{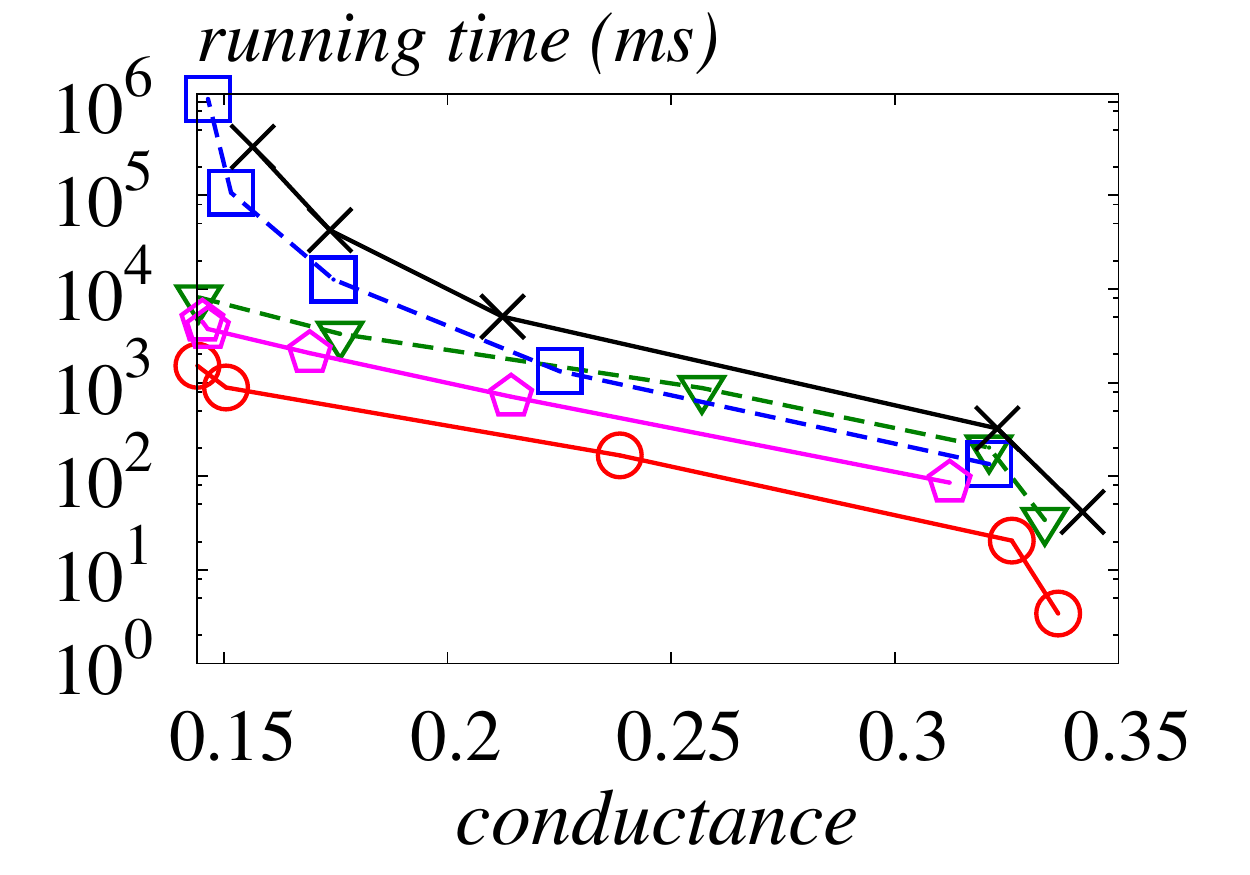} &
			\hspace{-4mm}\includegraphics[height=32mm]{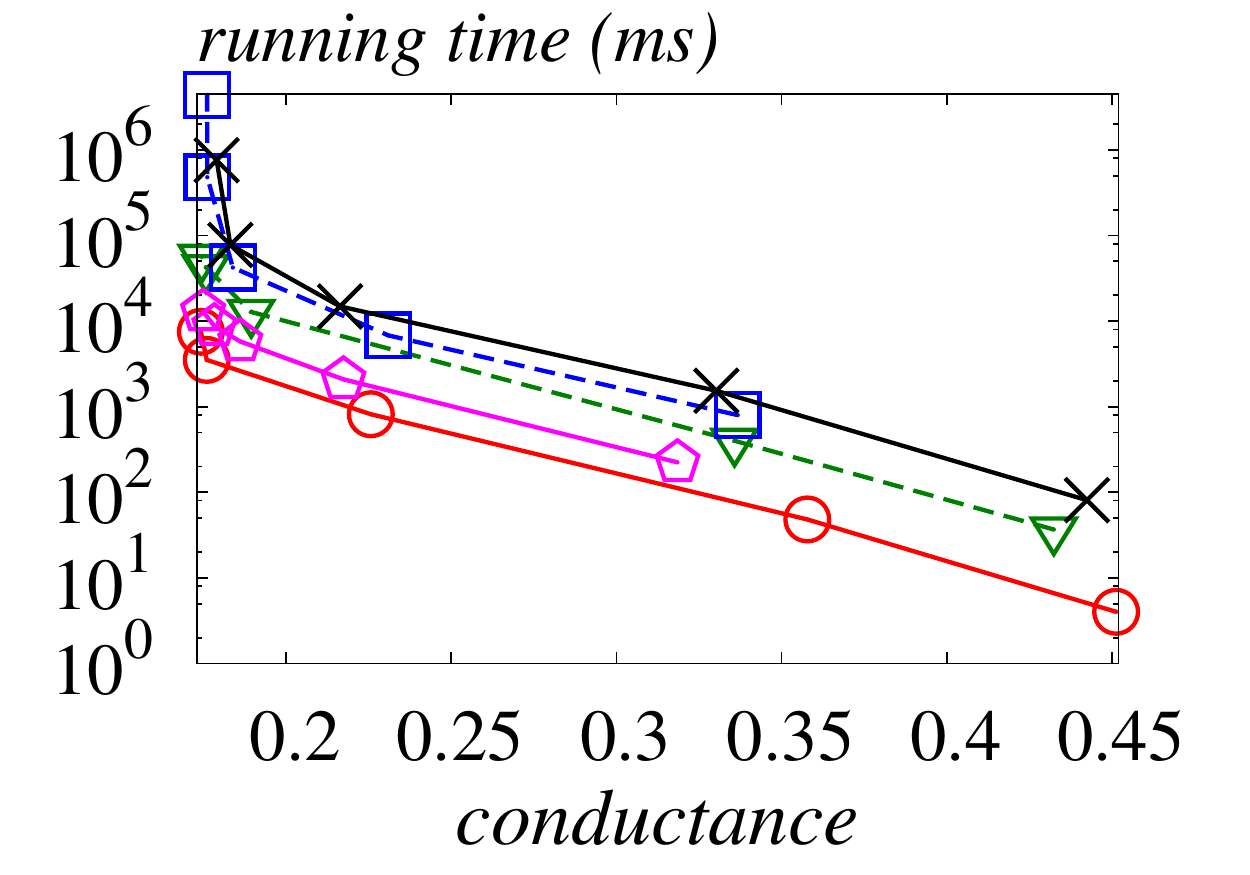} &
			\hspace{-4mm}\includegraphics[height=32mm]{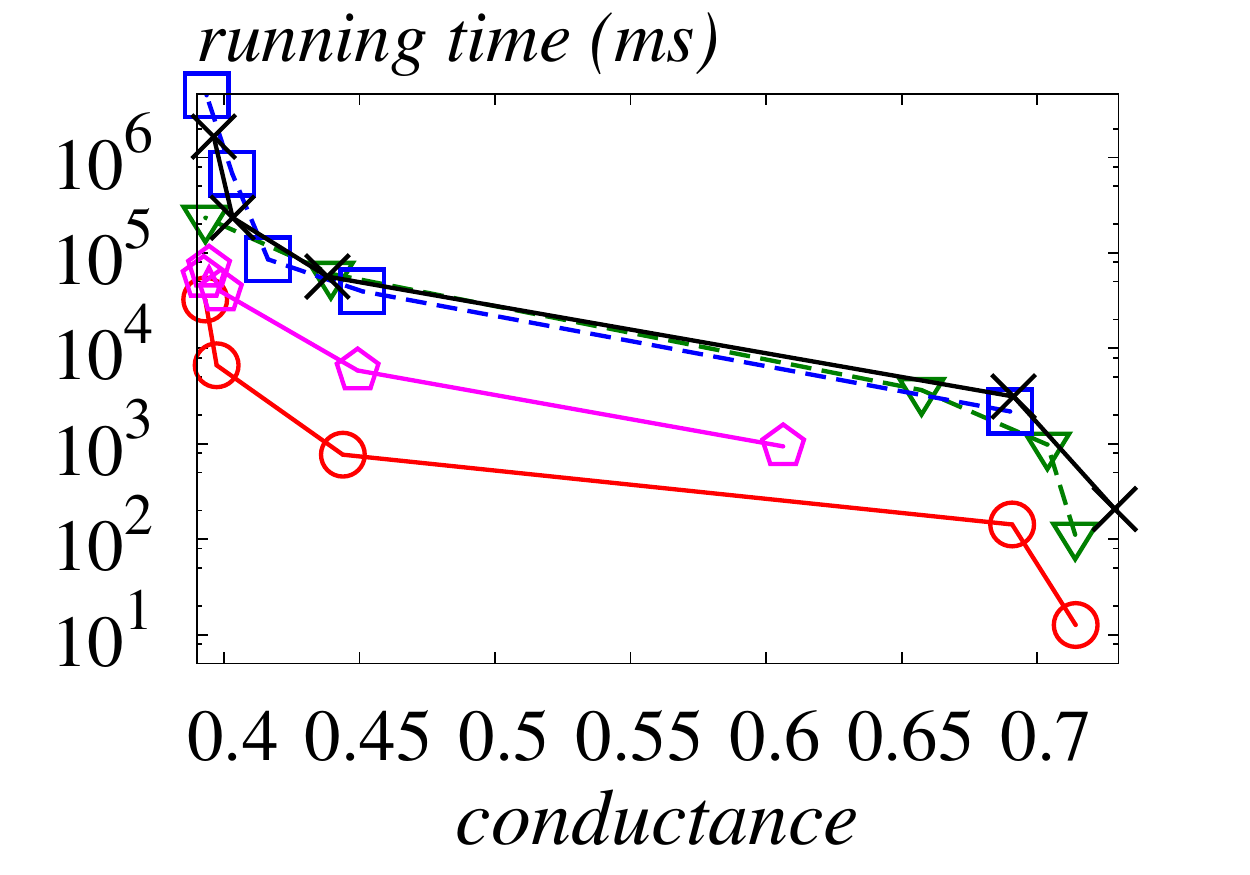} &
			\hspace{-4mm}\includegraphics[height=32mm]{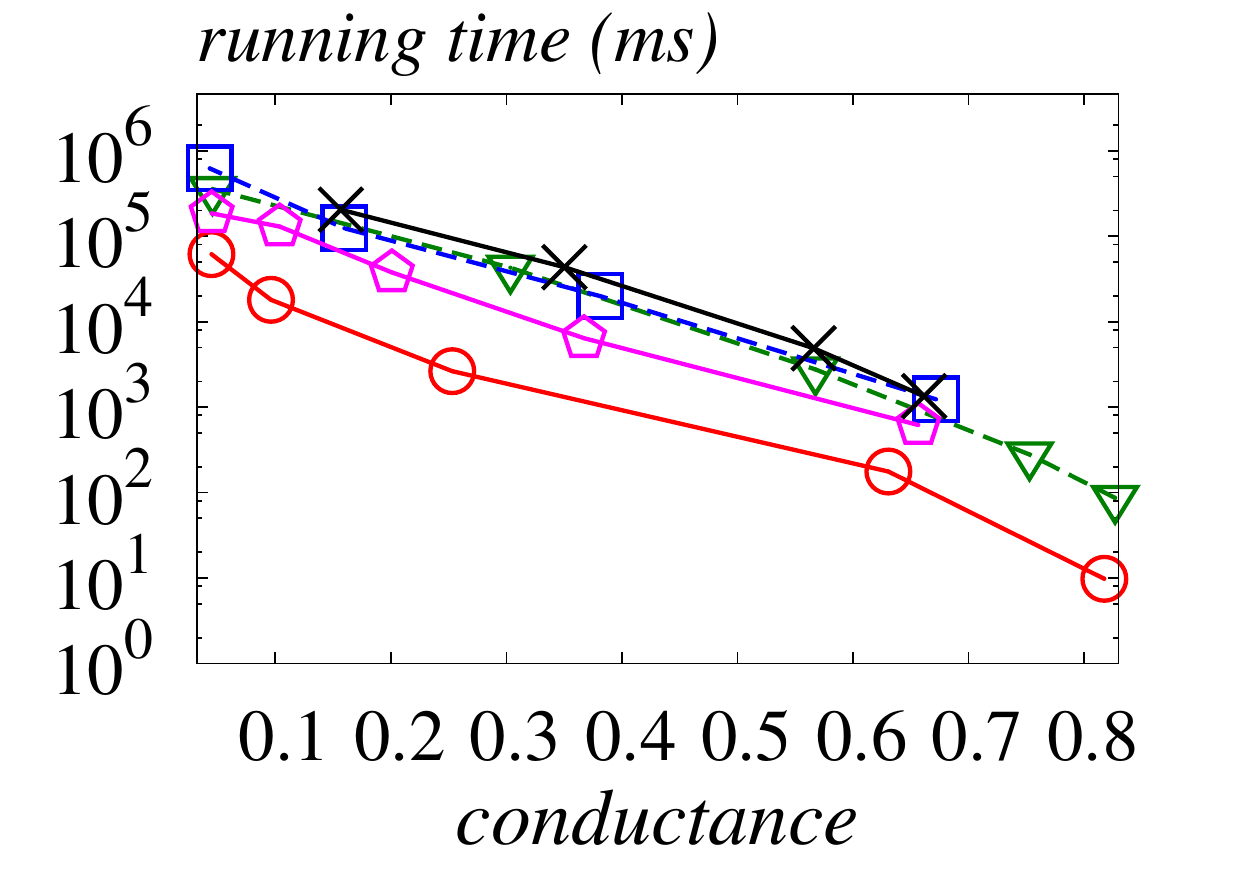} 
			\\[1mm]
			\hspace{-4mm} (i) {\em DBLP} (low-density)&
			\hspace{-4mm} (j) {\em Youtube} (low-density)&
			\hspace{-4mm} (k) {\em PLC} (low-density)&
			\hspace{-4mm} (l) {\em Orkut} (low-density)
		\end{tabular}
		\vspace{-1ex}
		\caption{Effect of subgraph densities (best viewed in color).} \label{fig:time-seed}
	\end{small}
\end{figure*}

\begin{figure*}[!t]
	\centering
	\begin{small}
		\begin{tabular}{cccc}
			\multicolumn{4}{c}{
				\hspace{-4mm}\includegraphics[height=2.8mm]{algolegendfig4.pdf}}  \\
			\hspace{-4mm} \includegraphics[height=32mm]{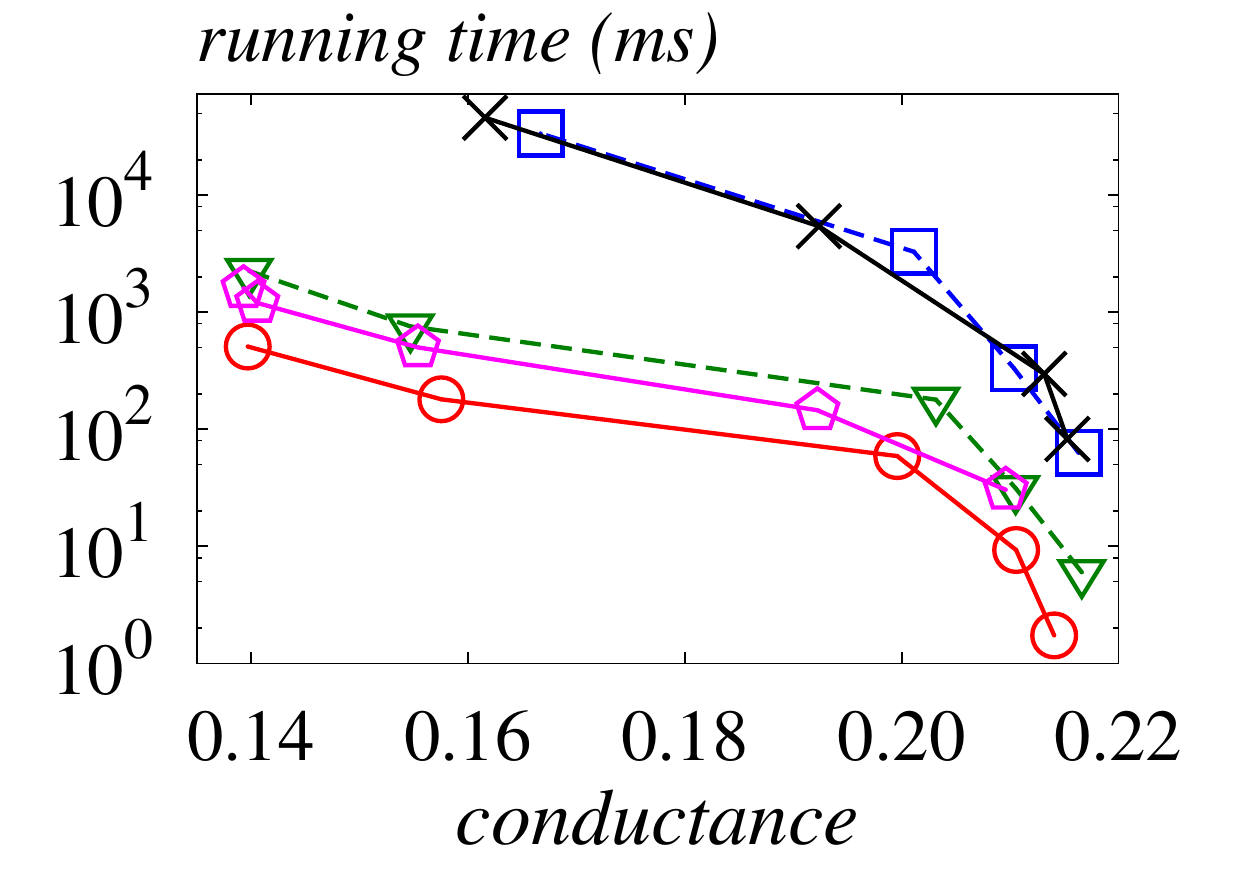} &
			\hspace{-4mm} \includegraphics[height=32mm]{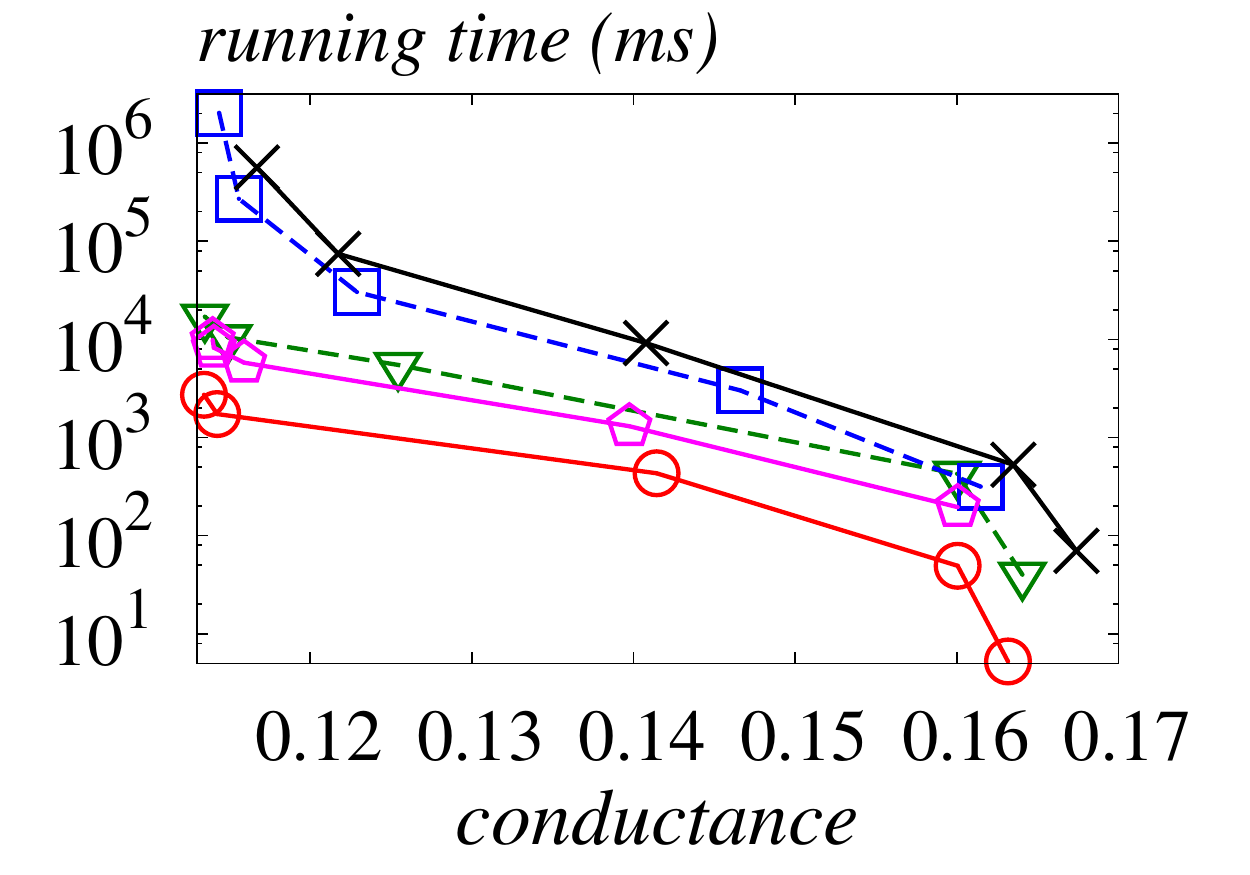} &
			\hspace{-4mm} \includegraphics[height=32mm]{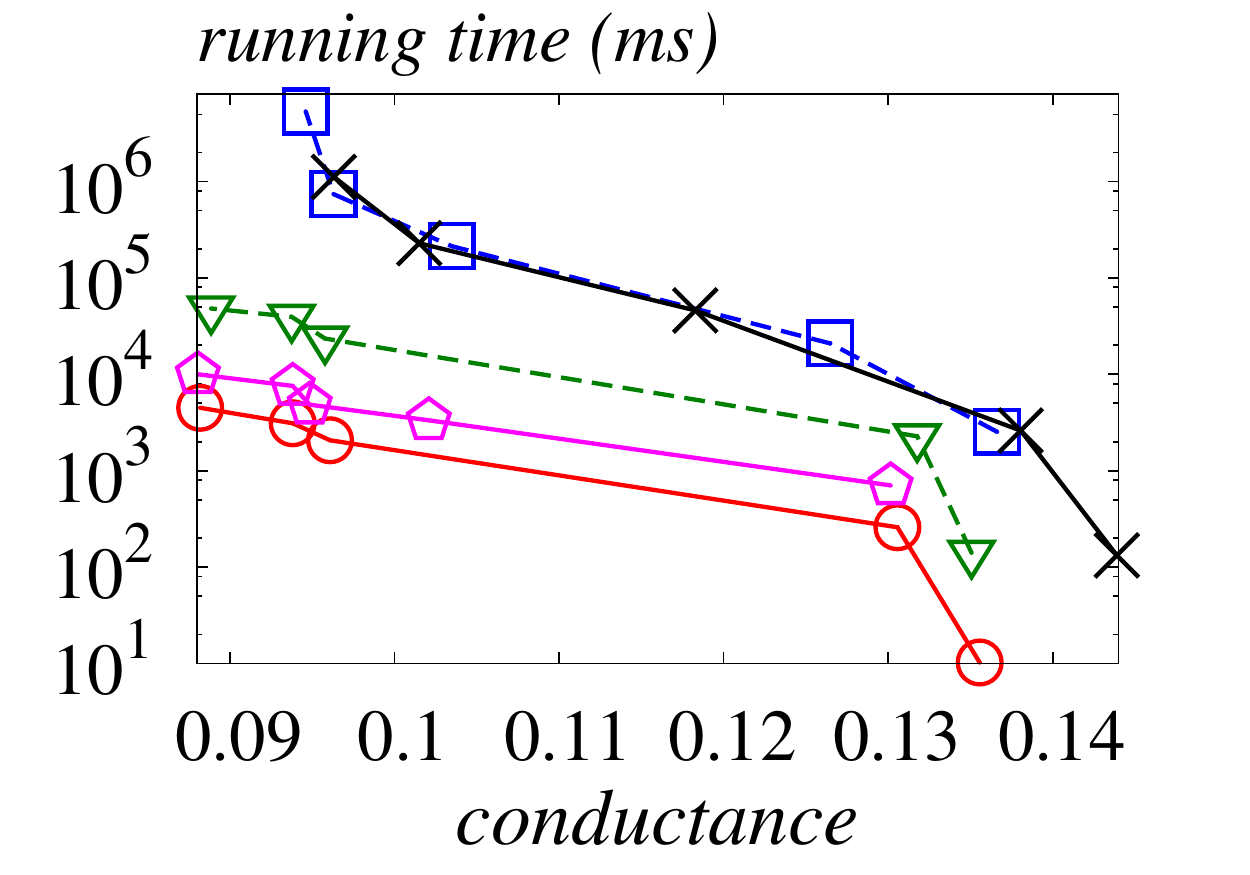} &
			\hspace{-4mm} \includegraphics[height=32mm]{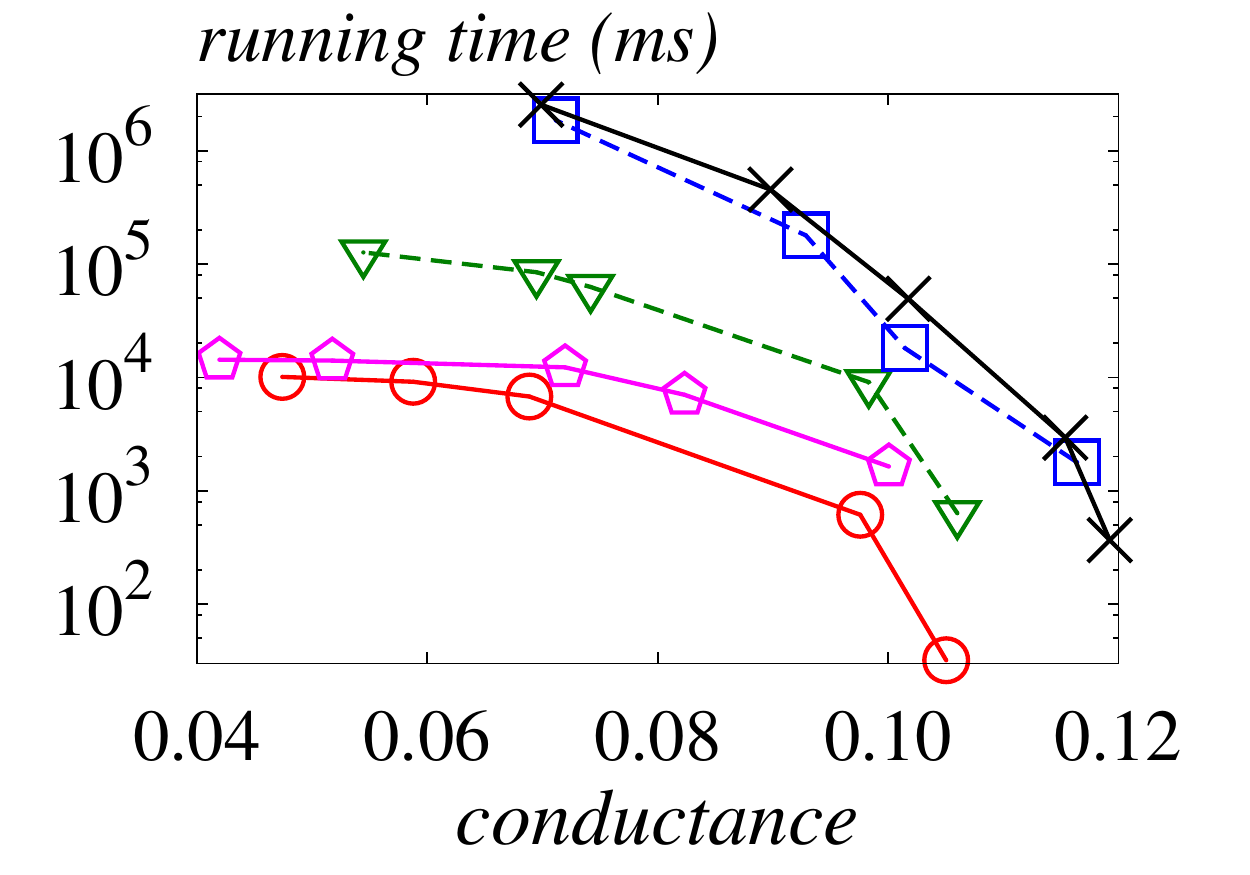}
			\\[-1mm]
			\hspace{-4mm} (a) $t=5$ &
			\hspace{-4mm} (b) $t=10$ &
			\hspace{-4mm} (c) $t=20$ &
			\hspace{-4mm} (d) $t=40$
		\end{tabular}
		\vspace{-2mm}
		\caption{Effect of heat constant $t$ on {\em DBLP} (best viewed in color).} \label{fig:time-dblp-t}
		\vspace{-2mm}
	\end{small}
\end{figure*}

\begin{figure*}[t]
	\centering
	\begin{small}
		\begin{tabular}{cccc}
			\multicolumn{4}{c}{
				\hspace{-4mm}\includegraphics[height=2.8mm]{./algolegendfig4.pdf}}  \\
			\hspace{-4mm} \includegraphics[height=32mm]{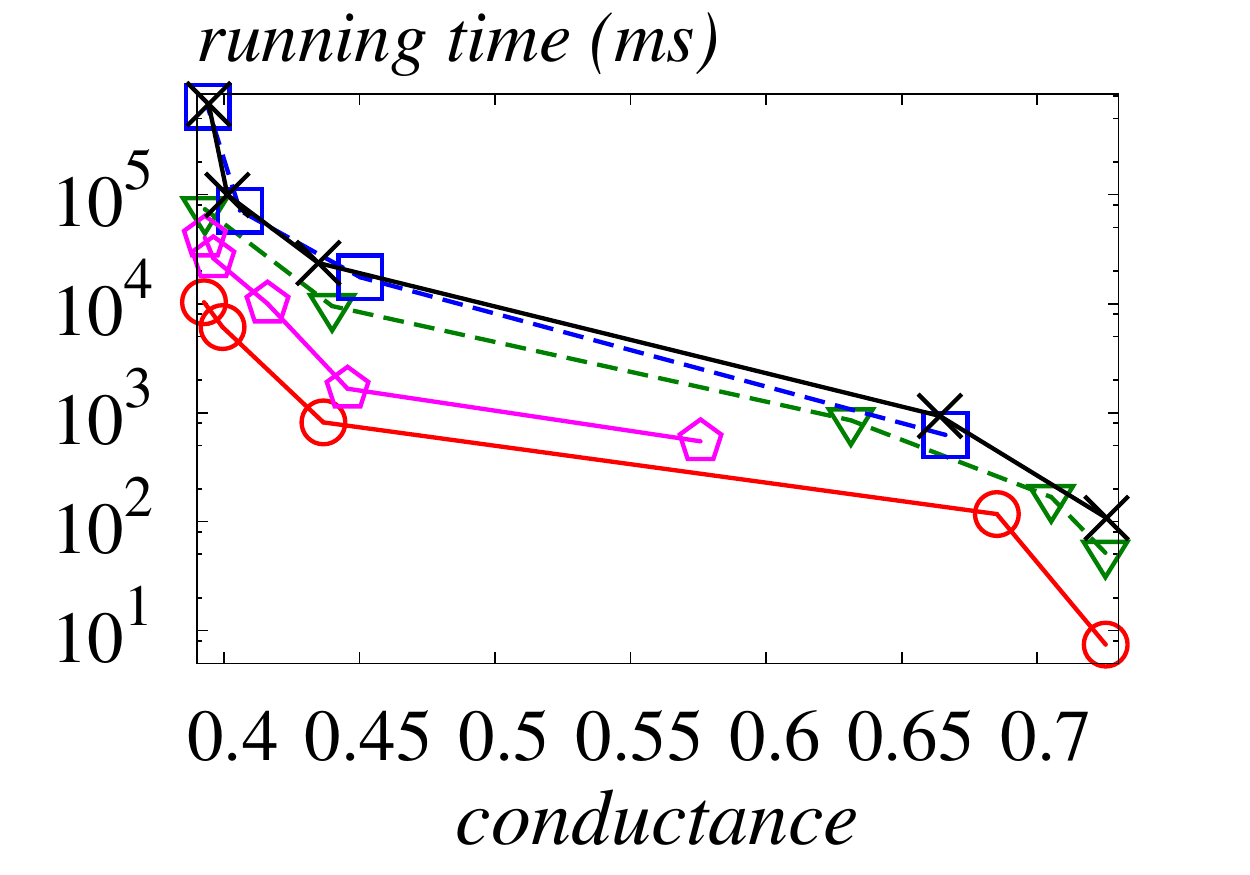} &
			\hspace{-4mm} \includegraphics[height=32mm]{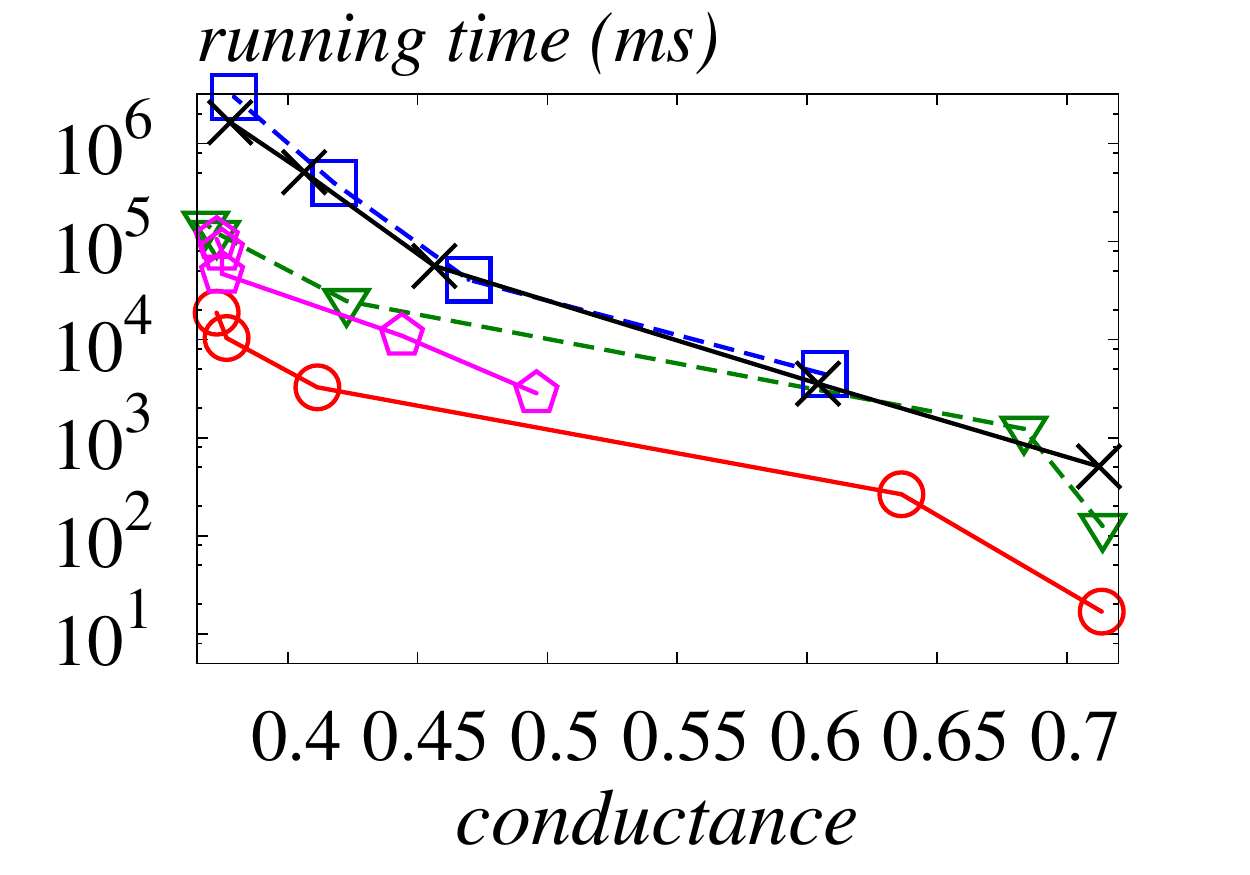} &
			\hspace{-4mm} \includegraphics[height=32mm]{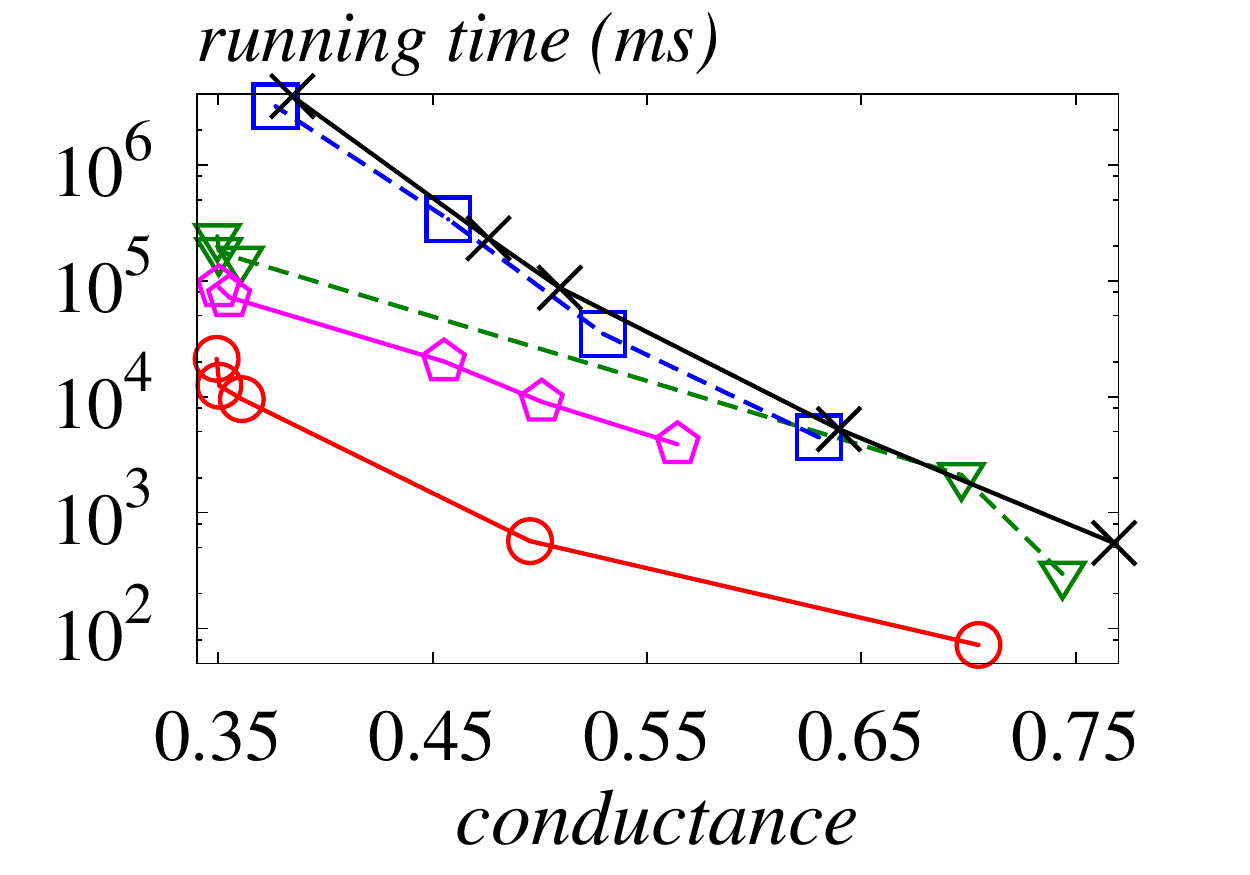} &
			\hspace{-4mm} \includegraphics[height=32mm]{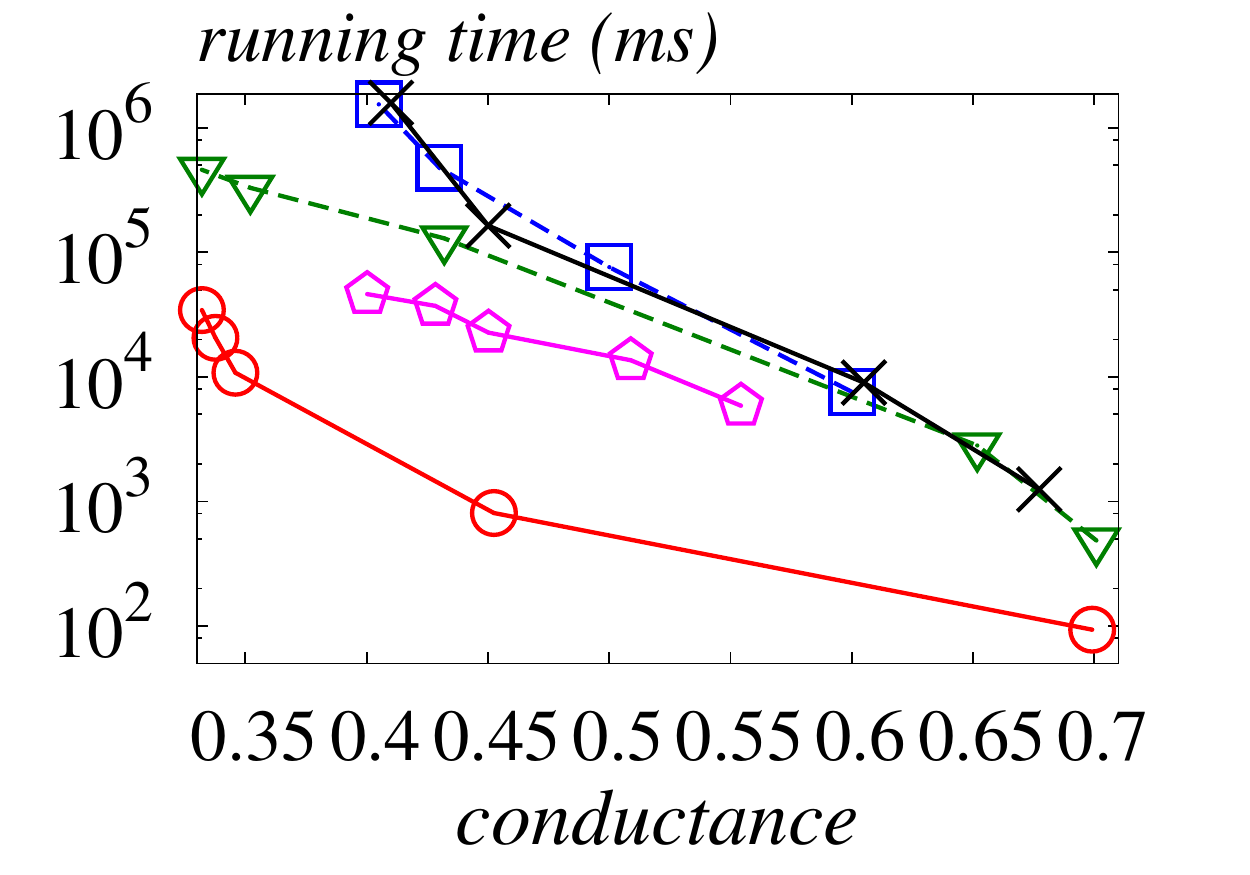}
			\\[-1mm]
			\hspace{-4mm} (a) $t=5$ &
			\hspace{-4mm} (b) $t=10$ &
			\hspace{-4mm} (c) $t=20$ &
			\hspace{-4mm} (d) $t=40$
		\end{tabular}
		\vspace{-2ex}
		\caption{Effect of heat constant $t$ on {\em PLC} (best viewed in color).} \label{fig:time-plc-t}
		\vspace{-2mm}
	\end{small}
	\vspace{-1ex}
\end{figure*}

\subsection{Ranking Accuracy of Normalized HKPR}
In this set of experiments, we evaluate the accuracy and efficiency of each method for computing normalized HKPR values (e.g., $\frac{\rs[v]}{d(v)}$). First, we randomly select $50$ seed nodes and apply the power method \cite{page1999pagerank} with $40$ iterations to compute the ground-truth normalized HKPR values (we omit the large datasets due to time and memory limitations). Following the experimental settings in Section \ref{sec:exp-set}, we run \hkrelax, \chkpr, \mc, \pukra and \pukraplus to generate normalized HKPR values for the selected seed nodes with varied error thresholds. Specifically, we vary $\epsilon$ in $\{10^{-8},10^{-7},10^{-6},10^{-5},10^{-4},10^{-3}\}$ for \hkrelax, $\epsilon$ in $\{0.01,0.02,0.05,0.1,0.2,0.3\}$ for \chkpr, and we set $\epsilon_r=0.5$ and $\delta$ is varied in $\{2\times 10^{-8}, 2\times 10^{-7}, 2\times 10^{-6}, 2\times 10^{-5}, 2\times 10^{-4}, 2\times 10^{-3}\}$ for \mc, \pukra and \pukraplus, respectively. Then, we evaluate the accuracy of each method by using {\em Normalized Discounted Cumulative Gain (NDCG)} \cite{jarvelin2000ir}, which is a classic metric for evaluating ranking results.

Figure \ref{fig:time-ndcg} reports the performance of each method on four datasets. We can make the following observations. First, as we reduce the error thresholds, both the running time and NDCG of each method increase markedly, which is consistent with their theoretical guarantees. Second, \pukraplus consistently incurs least running time while achieving the same NDCG compared to the competing methods. In addition, \pukra is $2\times-8\times$ slower than \pukraplus while \hkrelax runs even more slowly. Especially on {\em PLC} and {\em Orkut} datasets, \hkrelax's performance degrades to the same level of \chkpr and \mc. Although \chkpr and \mc also provide relative-error guarantees, they still incur the highest overheads because they require a large number of random walks. Third, we note that the efficiency and ranking accuracy results accord with the efficiency and clustering quality results reported in Section \ref{sec:exp-all}. This demonstrates the relationship between ranking accuracy of normalized HKPR and the quality of HKPR-based clustering algorithms, emphasizing why our methods produce clusters with smaller conductance than the competing ones.

\vspace{-1ex}
\subsection{Clusters Produced vs. Ground-truth}\label{sec:ground-truth}
We collect the top 5,000 ground-truth communities in {\em DBLP}, {\em Youtube}, {\em LiveJournal} and {\em Orkut} datasets from \cite{linksnap}. We select 100 seed nodes from 100 known communities of size greater than 100 randomly as the query set. For all algorithms, we vary $t$ from $3$ to $10$ ($t>10$ would give us clusters with substantially lower quality) and their error thresholds respectively to produce clusters with highest average $F_1$-measure (i.e.,  harmonic mean of precision and recall). More specifically, we vary $\epsilon$ from $0.005$ to $0.35$ for \chkpr, $\epsilon$ from $10^{-8}$ to $10^{-1}$ for \hkrelax. In addition, we fix $\epsilon_r=0.5$ and vary $\delta$ from $10^{-8}$ to $10^{-1}$ for \mc, \pukra and \pukraplus respectively.

Table~\ref{tab:f1} reports the highest $F_1$ measure of each algorithm and their corresponding running times. \pukraplus consistently produces clusters with the best average $F_1$-measures and least running times for all datasets except {\em DBLP}. On {\em DBLP}, \pukra produces clusters with the best average $F_1$-measure while \pukraplus produces  clusters with slightly smaller $F_1$ measure but significantly faster. We also observe that \chkpr and \mc generate very similar results for all datasets. They run significantly slower than \pukra and \pukraplus and also produce clusters with slightly smaller average $F_1$-measures than our methods. In addition, \hkrelax has the worst performance on most datasets. The only exception occurs on {\em Orkut}, where it produces clusters with the second best $F_1$-measure but $4\times$ slower than \pukraplus.

\subsection{Sensitivity Analysis to the Subgraph Characteristics}
Next, we study the impact of query sets generated from subgraphs of different characteristics on clustering quality and efficiency. First, from each dataset of {\em Youtube}, {\em PLC} and {\em Orkut}, we select 250 subgraphs with different densities \cite{lawler2001combinatorial} randomly. Then we sort the subgraphs  by their densities in descending order (denoted as $\{SG_1,SG_2,\cdots, SG_{250}\}$). We pick 50 nodes from $SG_1,\cdots,SG_{50}$ respectively to form a query set referred to as \textit{high-density} seed nodes, 50 nodes from $SG_{100},\cdots,SG_{150}$ respectively as \textit{medium-density} seed nodes, and 50 nodes from $SG_{200},\cdots,SG_{250}$ respectively as \textit{low-density} seed nodes. We run \chkpr, \mc, \hkrelax, \pukra and \pukraplus with the same parameter settings as in Section \ref{sec:exp-all} on these three query sets.

Figure \ref{fig:time-seed} plots the average conductance of the output clusters and the average running times of all algorithms under different error thresholds for the three query sets. We report the results on {\em DBLP, Youtube, PLC} and {\em Orkut} here.  We can make the following observations. First, \pukra and \pukraplus are consistently faster than the existing approaches for all query sets. Second, the conductance values of each graph in Figures \ref{fig:time-seed}(i)-(l) are higher than the rest. This is because subgraphs with high densities have low conductance. Also, both \chkpr and \mc show similar results on all query sets for all datasets whereas \hkrelax, \pukra and \pukraplus are sensitive to the subgraph densities. Since seed nodes picked from subgraphs with high densities would have many neighbors, the residues in \hkrelax, \pukra and \pukraplus will drop quickly as push operations are performed, making them terminate quickly.

\vspace{-1ex}
\subsection{Effects of Heat Constant $t$}
Lastly, we investigate the impact of the heat constant $t$. Using the same parameter settings and query set in Section \ref{sec:exp-all}, we run all algorithms on {\em DBLP} and {\em PLC} datasets by varying $t$ in $\{5, 10, 20, 40\}$. Figures \ref{fig:time-dblp-t} and \ref{fig:time-plc-t} plot the average running time and average conductance of the output clusters of each algorithm on {\em DBLP} and {\em PLC} respectively.  Observe that the running time of each algorithm increases as we increase $t$, which is consistent with their time complexities. \chkpr and \mc are the slowest as $t$ changes. We further observe that the advantage of \pukraplus over competing methods is more prominent as $t$ becomes larger. More specifically, \pukraplus is around $4$ times faster than \hkrelax when $t=5$ and the speedup goes up to more than one order of magnitude when $t=40$ on {\em DBLP}. Similar observations can be made on {\em PLC}. In addition, we find that the conductance values of clusters produced by each algorithm with larger $t$ are smaller than those with smaller $t$. This shows that we can obtain clusters with small conductance by choose a large $t$. However, our \textit{``Clusters Produced vs. Ground Truth''} experiment (see Section \ref{sec:ground-truth}) reveals that clusters produced by all algorithms with a large $t$ are very different from the ground-truth. This is because algorithms with a large $t$ tend to give us a cluster of nodes that are far from the seed node. As a result, choosing a good $t$ is paramount for finding high quality clusters.

\vspace{-1ex}
\section{Conclusions} \label{sec:ccl}
In this paper, we propose two novel heat-kernel-based local clustering algorithms, \pukra and \pukraplus, for computing approximate HKPR values and local graph clustering efficiently. Our algorithms bridge deterministic graph traversal with Monte-Carlo random walks in a non-trivial way, thereby overcoming the drawbacks of both and achieving significant gain in performance in comparison to the state-of-the-art local clustering techniques. 
Our experiments demonstrate that \pukraplus significantly outperforms the state-of-the-art heat-kernel-based algorithm by at least $4$ times on small graphs and up to one order of magnitude on large graphs in terms of computational time when producing clusters with the same qualities. 

\vspace{-1ex}
\section{ACKNOWLEDGEMENTS}
This research is supported in part by MOE, Singapore under grant MOE2015-T2-2-069 and grant RG128/18 (M4012086.020), by NUS, Singapore under an SUG, by NTU, Singapore under grant M4082311.020, and by National Natural Science Foundation of China (No. 61832017, No. 61772346 and U1809206).


\bibliographystyle{ACM-Reference-Format}

\begin{thebibliography}{99}

\bibitem{linksnap} {\url http://snap.stanford.edu}.

\bibitem{andersen2006local} Reid Andersen, Fan Chung, Kevin Lang. 2006. Local Graph Partitioning Using Pagerank Vectors. In \textit{FOCS}, pages 475-486.

\bibitem{andersen2009finding} Reid Andersen, Yuval Peres. 2009. Finding Sparse Cuts Locally Using Evolving Sets. In \textit{STOC}, pages 235-244.

\bibitem{avron2015community} Haim Avron, Lior Horesh. 2015. Community Detection Using Time-dependent Personalized Pagerank. In \textit{ICML}, pages 1795-1803.

\bibitem{banerjee2015fast} Siddhartha Banerjee, Peter Lofgren. 2017. Fast Bidirectional Probability Estimation in Markov Models. In \textit{NIPS}, pages 1423-1431.

\bibitem{bollobas1998modern} Bela Bollobas. Modern Graph Theory. 1998.

\bibitem{chung2007heat} Fan Chung. 2007. The Heat Kernel as the Pagerank of a Graph. In \textit{PNAS}, pages 19735-19740.

\bibitem{chung2009local} Fan Chung. 2009. A Local Graph Partitioning Algorithm Using Heat Kernel Pagerank. \textit{Internet Mathematics}, pages 315-330.

\bibitem{chung2006concentration} Fan Chung, Linyuan Lu. 2006. Concentration Inequalities and Martingale Inequalities: A Survey. \textit{Internet Mathematics}, pages 79-127.

\bibitem{chung2014hklocal} Fan Chung, Olivia Simpson. 2014. Computing Heat Kernel Pagerank and a Local Clustering Algorithm. \textit{IWOCA}, pages 110-121.

\bibitem{chung2015computing} Fan Chung, Olivia Simpson. 2015. Computing Heat Kernel Pagerank and a Local Clustering Algorithm. \textit{arXiv preprint arXiv:1503.03155}.

\bibitem{chung2015distributed} Fan Chung, Olivia Simpson. 2015. Distributed Algorithms for Finding Local Clusters Using Heat Kernel Pagerank. In \textit{WAW}, pages 177-189.

\bibitem{diaconis1990strong} Persi Diaconis, James Allen Fill. 1990. Strong Stationary Times via a New Form of Duality. \textit{The Annals of Probability}.

\bibitem{jeh2003scaling} Glen Jeh, Jennifer Widom. 2003. Scaling Personalized Web Search. In \textit{WWW}, pages 271-279.

\bibitem{kwak2010twitter} Haewoon Kwak, Changhyun Lee, Hosung Park, Sue Moon. 2010. What is Twitter, a Social Network or a News Media? In \textit{WWW}, pages 591-600.

\bibitem{kloster2014heat} Kyle Kloster, David Gleich. 2014. Heat Kernel Based Community Detection. In \textit{KDD}, pages 1386-1395.

\bibitem{liao2009isorankn} Chung-Shou Liao, Kanghao Lu, Michael Baym, Rohit Singh, Bonnie Berger. 2009. IsoRankN: Spectral Methods for Global Alignment of Multiple Protein Networks. In \textit{Bioinformatics}, pages 253-258.

\bibitem{lofgren2016personalized} Peter Lofgren, Siddhartha Banerjee, Ashish Goel. 2016. Personalized Pagerank Estimation and Search: A Bidirectional Approach. In \textit{WSDM}, pages 163-172.

\bibitem{page1999pagerank} Lawrence Page, Sergey Brin, Rajeev Motwani, Terry Winograd. 1999. The PageRank Citation Ranking: Bringing Order to the Web.

\bibitem{spielman2004nearly} Daniel A Spielman, Shang-Hua Teng. 2004. Nearly-linear Time Algorithms for Graph Partitioning, Graph sparsification, and Solving Linear Systems. In \textit{STOC}, pages 81-90.

\bibitem{shun2016parallel} Julian Shun, Farbod Roosta-Khorasani, Kimon Fountoulakis, Michael W. Mahoney. 2016. Parallel Local Graph Clustering. In \textit{VLDB}, pages 1041-1052.

\bibitem{tolliver2006graph} David A Tolliver, Gary L Miller. 2006. Graph Partitioning by Spectral Rounding: Applications in Image Segmentation and Clustering. In \textit{CVPR}, pages 1053-1060.

\bibitem{wang2015community} Meng Wang, Chaokun Wang, Jeffrey Xu Yu, Jun Zhang.  2015. Community Detection In Social Networks: An In-depth Benchmarking Study with a Procedure-oriented Framework. In \textit{VLDB}, pages 998-1009.

\bibitem{wang2016hubppr} Sibo Wang, Youze Tang, Xiaokui Xiao, Yin Yang, Zengxiang Li. 2016. HubPPR: Effective Indexing for Approximate Personalized Pagerank. In \textit{VLDB}, pages 205-216.

\bibitem{wang2017capacity} Di Wang, Kimon Fountoulakis, Monika Henzinger, Michael W. Mahoney, Satish Rao. 2017. Capacity Releasing Diffusion for Speed and Locality. In \textit{ICML}, pages 3598-3607.

\bibitem{wang2017fora} Sibo Wang, Renchi Yang, Xiaokui Xiao, Zhewei Wei, Yin Yang. 2017. FORA: Simple and Effective Approximate Single-Source Personalized PageRank. In \textit{KDD}, pages 505-514.

\bibitem{wei2018topppr} Zhewei Wei, Xiaodong He, Xiaokui Xiao, Sibo Wang, Shuo Shang, Ji-Rong Wen. 2018. Topppr: Top-k Personalized Pagerank Queries with Precision Guarantees on Large Graphs. In \textit{SIGMOD}, pages 441-456.

\bibitem{felzenszwalb2004efficient} Pedro F Felzenszwalb, Daniel P Huttenlocher. 2004. Efficient Graph-based Image Segmentation. In \textit{IJCV}, pages 167-181.

\bibitem{fortunato2010community} Santo Fortunato. 2010. Community Detection in Graphs. \textit{Physics Reports}, pages 75-174.

\bibitem{gharan2012approximating} Shayan Oveis Gharan, Luca Trevisan. 2012. Approximating the Expansion Profile and Almost Optimal Local Graph Clustering. In \textit{FOCS}, pages 187-196.

\bibitem{haveliwala2002topic} Taher H. Haveliwala. 2002. Topic-sensitive Pagerank. In \textit{WWW}, pages 517-526.

\bibitem{jarvelin2000ir} Kalervo J{\"a}rvelin, Jaana Kek{\"a}l{\"a}inen. 2000. IR Evaluation Methods for Retrieving Highly Relevant Documents. In \textit{SIGIR}, pages 41-48.

\bibitem{lawler2001combinatorial} Eugene L Lawler. 2001. Combinatorial Optimization: Networks and Matroids.

\bibitem{leskovec2010empirical} Jure Leskovec, Kevin J Lang, Michael Mahoney. 2010. Empirical Comparison of Algorithms for Network Community Detection. In \textit{WWW}, pages 631-640.

\bibitem{orecchia2014flow} Lorenzo Orecchia, Zeyuan Allen Zhu. 2014. Flow-based Algorithms for Local Graph Clustering. In \textit{SODA}, pages 1267-1286.

\bibitem{pons2005computing} Pascal Pons, Matthieu Latapy. 2005. Computing Communities in Large Networks Using Random Walks. In \textit{ISCIS}, pages 284-293.

\bibitem{spielman2013local} Daniel A Spielman, Shang-Hua Teng. 2013. A Local Clustering Algorithm for Massive Graphs and Its Application to Nearly Linear Time graph partitioning. In \textit{SICOMP}, pages 1-26.

\bibitem{veldt2016simple} Nate Veldt, David Gleich, Michael Mahoney. 2016. A Simple and Strongly-local Flow-based Method for Cut Improvement. In \textit{ICML}, pages 1938-1947.

\bibitem{voevodski2009finding} Konstantin Voevodski, Shang-Hua Teng, Yu Xia. 2009. Finding Local Communities in Protein Networks. In \textit{BMC Bioinformatics}, page 297.

\bibitem{walker1974new} Alastair J. Walker. 1974. New Fast Method for Generating Discrete Random Numbers with Arbitrary Frequency Distributions. \textit{Electronics Letters}, page 127-128.

\bibitem{yang2012def} Jaewon Yang, Jure Leskovec. 2012. Defining and Evaluating Network Communities Based on Ground-truth. In \textit{KDD Workshop}.

\bibitem{zhu2013local} Zeyuan Allen Zhu, Silvio Lattanzi, Vahab S Mirrokni. 2013. A Local Algorithm for Finding Well-Connected Clusters. In \textit{ICML}, pages 396-404.
\end{thebibliography}

\balance
\appendix
\section{Choosing $K$ for TEA+}\label{sec:chooseK}
Recall that $K$ limits the push operations within $K$ hops from the seed node, which is an important parameter controlling the termination of \hkpushplus. If $K$ is too small, it will lead to a small overhead for \hkpushplus but a significant cost for \rswk, and vice versa. Consequently, it is crucial to find an appropriate value for $K$ so as to strike a good balance between \hkpushplus and \mc. In particular, there are two factors we need to take into account when choosing $K$:
\begin{enumerate}
	\item $K$ ought to be changed along with $\epsilon_r \cdot \delta$;
	\item We need large $K$ for graphs with small degrees, and conversely, small $K$ for large-degree graphs.
\end{enumerate}
This reason is as follows. If $K$ is a fixed constant and small, the number of push operations performed by \hkpushplus will be bounded by a maximum number. Recall that in each iteration of \hkpushplus, we will perform $d(v)$ push operations, where $d(v)$ is the number of neighbors of current node $v$. This implies that the total number of push operations performed within $K$ hops from $s$ on the graph with small average degree will be rather limited compared with that of the graph with high average degree. As a result, if the number of push operations is not large enough, \pukraplus will degrade to \mc and the overhead for random walks will be significant especially when the value of $\epsilon_r \cdot \delta$ is small. Considering these two factors, we propose to choose $K$ by the following equation:
\begin{equation}\label{eq:chooseK}
\left({1}/{\overline{d}}\right)^{\frac{K}{c}}= \epsilon_r \cdot \delta,
\end{equation}
where $\bar{d}$ is the average degree of the input graph. The intuition behind Equation~(\ref{eq:chooseK}) is that, after $K$-hops of push operations, the residues tend to approach $\epsilon_r \cdot \delta$, and hence, the value for $\alpha$ would not be large.

\section{Proofs}\label{sec:proof}
\subsection{Proof of Lemma \ref{lem:fwdeq}}
\begin{proof}
	This proof is based on induction. Given reserve vector $\qs$ and residue vectors $\mathbf{r}^{(0)}_s, \ldots \mathbf{r}^{(K)}_s$ constructed by the end of any interation in Algorithm~\ref{alg:hkpush}, we define $\mathbf{f}_s[v]$ as
	\begin{equation*}
		\textstyle \mathbf{f}_s[v]=\textstyle \qs[v]+\sum_{u\in V}{\sum_{k=0}^{K}{\rks[u]\cdot\mathbf{h}^{(k)}_u[v]}}.
	\end{equation*}
	First, let us consider the initial condition, in which all entries of $\mathbf{r}^{(0)}_s$ are zero, except $\mathbf{r}^{(0)}_s[s]=1$, and other vectors are zero vectors. This implies that
	\begin{align*}
		\textstyle \mathbf{f}_s[v]&=\textstyle \qs[v]+\sum_{u\in V}\sum_{k=0}^{K}{\left[\rks[u]\cdot \sum_{\ell=0}^{\infty}{\frac{\eta(k+\ell)}{\psi(k)}\mathbf{P}^{\ell}[u,v]}\right]}\\
		&=\textstyle \sum_{\ell=0}^{\infty}{\eta(\ell)\cdot\mathbf{P}^{\ell}[s,v]}=\rs[v].
	\end{align*}
	Namely, Equation~(\ref{eq:hkpush-invar}) holds in the initial case.
	
	Furthermore, we assume that after $j$ iterations, reserve vector $\qs$ and residue vectors $\mathbf{r}^{(0)}_s, \ldots \mathbf{r}^{(K)}_s$ satisfy Equation~(\ref{eq:hkpush-invar}), namely $\mathbf{f}_s[v]=\rs[v]$. In the beginning of the $(j+1)$-th iteration, we will conduct push operations with an entry $(w,i)$, where $w\in V $ and $i<K$. Then, we define the change of $\mathbf{f}_s[v]$ after this iteration as
	\begin{align*}
		\textstyle \Delta(w,i)=\textstyle \left(\mathbf{\widehat{q}}_s[v]+\sum_{u\in V}\sum_{k=0}^{K}{\mathbf{\widehat{r}}^{(k)}_s[u]\cdot\mathbf{h}^{(k)}_u[v]}\right)\\ \textstyle\ \ -\left(\qs[v]+\sum_{u\in V}\sum_{k=0}^{K}{\rks[u]\cdot\mathbf{h}^{(k)}_u[v]}\right),
	\end{align*}
	where $\mathbf{\widehat{q}}_s$ is the updated reserve vector, $\mathbf{\widehat{r}}^{(i)}_s$ and $\mathbf{\widehat{r}}^{(i+1)}_s$ are the updated residue vectors after performing push operations in this interation. Then, by Lines 4-7 of Algorithm~\ref{alg:hkpush}, we have $$\textstyle \mathbf{\widehat{q}}_s[w]-\qs[w]=\mathbf{r}^{(i)}_s[w]\cdot\frac{\eta(i)}{\psi(i)},$$
	$$\textstyle \mathbf{\widehat{r}}^{(i)}_s[w]-\mathbf{r}^{(i)}_s[w]=-\mathbf{r}^{(i)}_s[w],$$
	and $\forall{u\in N(w)}$, $\textstyle \mathbf{\widehat{r}}^{(i+1)}_s[u]-\mathbf{r}^{(i+1)}_s[u]=\left(1-\frac{\eta(i)}{\psi(i)}\right)\cdot\mathbf{r}^{(i)}_s[w]\cdot\mathbf{P}[w,u].$ Hence, the change of $\mathbf{f}_s[v]$ after this iteration is
	\begin{align*}
		\Delta(w,i)&=\textstyle (\mathbf{\widehat{q}}_s[v]-\qs[v])+(\mathbf{\widehat{r}}^{(i)}_s[w]-\mathbf{r}^{(i)}_s[w])\cdot\mathbf{h}^{(i)}_w[v]\\
		&\ \ \ \ \textstyle +\sum_{u\in\mathcal{N}(w)}(\mathbf{\widehat{r}}^{(i+1)}_s[u]-\mathbf{r}^{(i+1)}_s[u])\cdot\mathbf{h}^{i+1}_u[v].
	\end{align*}
	Notice that
	\begin{align*} &\textstyle \sum_{u\in\mathcal{N}(w)}(\mathbf{\widehat{r}}^{(i+1)}_s[u]-\mathbf{r}^{(i+1)}_s[u])\cdot\mathbf{h}^{i+1}_u[v]\\
		&=\textstyle \left(1-\frac{\eta(i)}{\psi(i)}\right)\mathbf{r}^{(i)}_s[w]\cdot\sum_{u\in\mathcal{N}(w)}{\mathbf{P}[w,u]\cdot\mathbf{h}^{i+1}_u[v]}\\
		&=\textstyle \mathbf{r}^{(i)}_s[w]\sum_{u\in\mathcal{N}(w)}{\left[\mathbf{P}[w,u]\sum_{l=0}^{\infty}{\frac{\eta(i+1+l)}{\psi(i)}\mathbf{P}^{l}{[u,v]}}\right]}\\
		&=\textstyle \mathbf{r}^{(i)}_s[w]\cdot \left(\mathbf{h}^{(i)}_w[v]-\frac{\eta(i)}{\psi(i)} \right).
	\end{align*}
	Therefore, $\textstyle \Delta(w,i)= \textstyle \mathbf{\widehat{q}}_s[v]-\qs[v]-\mathbf{r}^{(i)}_s[w]\cdot\mathbf{h}^{(i)}_w[v] +\mathbf{r}^{(i)}_s[w]\cdot \left( \mathbf{h}^{(i)}_w[v]-\frac{\eta(i)}{\psi(i)} \right)=0,$
	which implies $\mathbf{f}_s[v]$ is invariant and still equals $\rs[v]$ after the $(j+1)$-th iteration. This completes the proof of this lemma.
\end{proof}

%

\subsection{Proof of Lemma~\ref{lem:srrw-approx}}
\begin{proof}
	Recall that Algorithm~\ref{alg:skipped-walk} consists of several iterations (see Lines 3-8). In the $\ell$-th ($\ell=0,1,\cdots$) iteration, the algorithm terminates with $\frac{\eta(k+\ell)}{\psi(k+\ell)}$ probability (Line 4); with the other $1-\frac{\eta(k+\ell)}{\psi(k+\ell)}$ probability, it samples a neighbor node of current node and set the sampled node as current node (Lines 6-8). Let $v_i$ be the node at the beginning of the $i$-th iteration, and ${p}(v,\ell)$ be the probability that the algorithm terminates at the $\ell$-th iteration with $v$ as the returned node. We will prove that, given $G,s=u,k$ as the input to Algorithm~\ref{alg:skipped-walk}, we have
	\begin{equation}\label{eq:prob-ell}
	\textstyle  {p}(v,\ell)=\frac{\eta(k+\ell)}{\psi(k)}\cdot\mathbf{P}^{\ell}[u,v].
	\end{equation}
	Note that if Equation~(\ref{eq:prob-ell}) holds, then the overall probability that $v$ is sampled as returned node is
	\begin{equation*}
		\textstyle  \sum_{\ell=0}^{\infty}{{p}(v,\ell)}=\sum_{\ell=0}^{\infty}{\frac{\eta(k+\ell)}{\psi(k)}\cdot\mathbf{P}^{\ell}[u,v]}=\mathbf{h}^{(k)}_u[v],
	\end{equation*}
	which establishes the lemma.
	
	We prove by induction. For the base case where $\ell=0$, we have $v=s$, and the probability that the $0$-th iteration of Algorithm~\ref{alg:skipped-walk} terminates is $\frac{\eta(k)}{\psi(k)}$. In that case, $p(v,\ell)=\frac{\eta(k)}{\psi(k)}$ if $v=u$; otherwise $p(v,\ell)=0$. Meanwhile, it can be verified that when $\ell=0$, the r.h.s of Equation~(\ref{eq:prob-ell}) equals $\frac{\eta(k)}{\psi(k)}$ if $v=u$, and $0$ otherwise. Therefore, Equation~(\ref{eq:prob-ell}) holds when $\ell=0$.
	
	Assume that Equation~(\ref{eq:prob-ell}) holds when $\ell=i$. Then, for any node $w\in V$, the probability that $w$ is at the beginning of $i$-th iteration is
	\begin{equation*}
		\textstyle  p(w,\ell)\cdot\frac{\psi(k+i)}{\eta(k+i)}=\frac{\psi(k+i)}{\psi(k)}\cdot\mathbf{P}^{i}[u,w].
	\end{equation*}
	Now consider the $(i+1)$-th iteration. Since Algorithm~\ref{alg:skipped-walk} does not terminate at $i$-th iteration, for any node $v$, the probability that $v$ is picked as current node is
	\begin{align*}
		& \textstyle  \left(1-\frac{\eta(k+i)}{\psi(k+i)}\right)\cdot \frac{\psi(k+i)}{\psi(k)}\cdot \sum_{w\in V}{\mathbf{P}^{i}[u,w]\cdot \mathbf{P}[w,v]}\\
		& \textstyle =\frac{\psi(k+i+1)}{\psi(k)}\cdot \mathbf{P}^{i+1}[u,v].
	\end{align*}
	In the $(i+1)$-th iteration, Algorithm~\ref{alg:skipped-walk} terminates with probability $\frac{\eta(k+i+1)}{\psi(k+i+1)}$, which implies that $v$ is returned with probability
	\begin{align*}
		p(v,i+1) &\textstyle  = \frac{\eta(k+i+1)}{\psi(k+i+1)}\cdot \frac{\psi(k+i+1)}{\psi(k)}\cdot \mathbf{P}^{i+1}[u,v]\\
		&\textstyle  = \frac{\eta(k+i+1)}{\psi(k)}\cdot\mathbf{P}^{i+1}[u,v].
	\end{align*}
	Therefore, the lemma is proved.
\end{proof}

\subsection{Proof of Lemma~\ref{lem:time-push}}
\begin{proof} Since the initialization of vectors in Algorithm~\ref{alg:hkpush} can be implemented in $O(1)$ cost by using {\em hashmap}, the total cost of Algorithm~\ref{alg:hkpush} is mainly determined by the number of push operations performed before termination.
	Recall that in each iteration of Algorithm~\ref{alg:hkpush} (Lines 4-7), if there exists an entry $(u,k)$ such that $u$'s current $k$-hop residue is greater than $r_{max}\cdot d(u)$, \hkpush will convert $\frac{\eta(k)}{\psi(k)}$ fraction of the current $k$-hop residue into its reserve. This implies that in each interation occurred on entry $(u,k)$, its reserve will be increased by at least $\frac{\eta(k)}{\psi(k)}\cdot r_{max} \cdot d(u)$. Let $\rks[u]$ denote the total $k$-hop residue that $u$ receives and $x_{u,k}$ be the number of interations occurred on entry $(u,k)$. Then, we have
	$$\textstyle \frac{\eta(k)}{\psi(k)}\cdot r_{max} \cdot d(u)\cdot x_{u,k} \le \frac{\eta(k)}{\psi(k)}\cdot \rks[u],$$
	and thus $d(u)\cdot x_{u,k} \le \cdot \frac{\rks[u]}{r_{max}}$.
	Since in each iteration occurred on entry $(u,k)$, \hkpush will perform $d(u)$ push operations, the total number of push operations occurred on all $(u,k)$ entries is then bounded by
	\begin{align*}
		\textstyle  n_p & = \sum_{u\in V}{\sum_{k=0}^{\infty}{d(u)\cdot x_{u,k}}}
		\textstyle  \le \sum_{u\in V}{\sum_{k=0}^{\infty}{\frac{\rks[u]}{r_{max}}}}\\
		\textstyle  & \textstyle  = \frac{1}{r_{max}}\cdot \sum_{k=0}^{\infty}{\sum_{u\in V}{\rks[u]}}=\frac{1}{r_{max}}.
	\end{align*}
	Since each push operation on entry $(u,k)$ will increase the value of $\mathbf{r}^{(k)}_s[u]$, the number of non-zero elements in residue vectors $\mathbf{r}^{(0)}_s, \ldots, \mathbf{r}^{(K)}_s$ returned by Algorithm~\ref{alg:hkpush} is bounded by the total number of push operations, i.e., $O(\frac{1}{r_{max}})$. 
	
	Excluding the space required by the input graph, the space cost of Algorithm~\ref{alg:hkpush} is incurred by the reserve vector and the non-zero elements in residue vectors, which is $O(\frac{1}{r_{max}})$. Therefore, the lemma is proved.
\end{proof}

\subsection{Proof of Lemma~\ref{lem:time-krandwalk}}
\begin{proof}
	With regard to each invocation of \rswk, the time complexity relies on the total number of nodes it visits in the random walk. According to Lemma~\ref{lem:srrw-approx}, the probability that any node $v\in V$ is returned as the end node at $\ell$-th iteration is $p(v,\ell)$. If an invocation of \rswk terminates at $v$ in $\ell$-th iteration, then $\ell$ nodes are visited before it returns $v$. Considering all nodes in the graph, we have the expected time of each invocation of \rswk as follows
	\begin{align}
		\textstyle  \sum_{\ell=0}^{\infty}{\sum_{v\in V}{p(v,\ell)\cdot \ell}}&= \textstyle \sum_{\ell=0}^{\infty}{\frac{\eta(k+\ell)}{\psi(k)}\cdot \ell \cdot \sum_{v\in V}{\mathbf{P}^{\ell}[u,v]}}\nonumber\\
		&\textstyle =\sum_{\ell=0}^{\infty}{\frac{\eta(k+\ell)}{\psi(k)}\cdot \ell}\nonumber\\
		& \textstyle= \sum_{\ell=0}^{\infty}{\left[\frac{\eta(k+\ell)}{\psi(k)}\cdot (k+\ell)-\frac{\eta(k+\ell)}{\psi(k)}\cdot k\right]}\nonumber\\
		&\textstyle=\frac{1}{\psi(k)}\cdot\left(\sum_{\ell=0}^{\infty}{\eta(k+\ell)\cdot(k+\ell)}\right)-k\nonumber\\
		&\textstyle=\frac{t}{\psi(k)}\cdot \sum_{\ell=k-1}^{\infty}{\eta(\ell)}-k\nonumber\\
		&\textstyle=\frac{t}{\psi(k)}\cdot \psi(k-1)-k\nonumber\\
		&\textstyle= t+t\cdot\frac{\eta(k-1)}{\psi(k)}-k \label{eq:k-exp-len}.
	\end{align}
	Note that $\frac{t}{k}\cdot\eta(k-1)=\eta(k)\le \sum_{\ell=k}^{\infty}{\eta(\ell)} = \psi(k)$. Then $\textstyle t\cdot\frac{\eta(k-1)}{\psi(k)}\le k.$ This implies that Equation~(\ref{eq:k-exp-len}) is no greater than $t$, which finishes our proof.

	
	%
	
\end{proof} 

\vspace{-4ex}
\subsection{Proof of Theorem~\ref{thrm:pukra}}
\begin{proof}
	Let $Y_i$ be as defined in the context of Equation~(\ref{eq:pukra_asum}), and let $Y=\sum_{i=1}^{n_r}{Y_i}$. By Line 12 of \pukra, $\textstyle \ar[v]=\qs[v]+\frac{Y\cdot \alpha}{n_r}.$
	By Equation (\ref{eq:pukra-exp-xi}), the expectation of $Y$ is
	\begin{equation*}
		\textstyle \E[Y]=\E[\sum_{i=1}^{n_r}{Y_i}]=\frac{n_r}{\alpha}\left({\rs[v]-\qs[v]}\right) \le\frac{n_r}{\alpha}\cdot \rs[v].
	\end{equation*}
	Let $\lambda=\frac{n_r\epsilon_r}{\alpha}\cdot \rs[v]$  and $n_r$ be as defined in Line 8 of \pukra.
	By the Chernoff bound (see Lemma~\ref{lem:chernoff}), for any node $v$ in $V$ with $\rs[v]> d(v)\cdot\delta$, we have
	\begin{align*}
		\textstyle \mathbb{P}[ |Y-\mathbb{E}[Y]| \ge\lambda] &\textstyle =\mathbb{P}\left[\left| \ar[v]-\rs[v]\right| \ge \epsilon_r \cdot{\rs[v]}\right]\\
		&\textstyle \le\exp\left(-\frac{n_r\cdot\epsilon^2_r\cdot\rs[v]}{2\alpha\cdot (1+\epsilon_r/3)}\right)\le (p^{\prime}_f)^{d(v)}.
	\end{align*}
	On the other hand, let $\lambda=\frac{n_r\epsilon_r\delta d(v)}{\alpha}$. Then, for any node $v$ in $V$ with $\rs[v] \le d(v)\cdot \delta$,
	\begin{align*}
		\textstyle \mathbb{P}[ |Y-\mathbb{E}[Y]| \ge \lambda] &\textstyle  =\mathbb{P}\left[\left|\ar[v]-\rs[v]\right| \ge d(v)\cdot\epsilon_r\delta\right]\\
		&\textstyle  \le\exp\left(-\frac{n_r\cdot\epsilon^2_r\delta^2 d^2(v)}{2\alpha(1+\epsilon_r/3)\cdot\rs[v]}\right)\le (p^{\prime}_f)^{d(v)}.
	\end{align*}
	By the union bound, for $V_1=\{v| v\in V \ s.t. \ \rs[v]> d(v)\cdot\delta\}$, we have
	\begin{equation*}
		\textstyle \mathbb{P}\left[\bigcup_{v\in V_1}\left\{\left| \ar[v]-\rs[v]\right| \ge \epsilon_r \cdot{\rs[v]}\right\}\right] \le \sum_{v\in V_1}{(p^{\prime}_f)^{d(v)}},
	\end{equation*}
	and for $V_2=\{v| v \in V \ s.t. \ \rs[v] \le d(v)\cdot\delta\}$, we have
	\begin{equation*}
		\textstyle \mathbb{P}\left[\bigcup_{v\in V_2}\left\{\left| \frac{\ar[v]}{d(v)}-\frac{\rs[v]}{d(v)}\right| \ge \epsilon_r\delta \right\}\right] \le \sum_{v\in V_2}{(p^{\prime}_f)^{d(v)}}.
	\end{equation*}
	%
	Therefore, we have the following results, respectively.
	With probability at least $1-\sum_{v\in V}{(p^{\prime}_f)^{d(v)}}$, for every node $v$ in $V$ with $\frac{\rs[v]}{d(v)}> \delta$, $\textstyle \left| \frac{\ar[v]}{d(v)}-\frac{\rs[v]}{d(v)}\right|\le \epsilon_r \cdot\frac{\rs[v]}{d(v)},$ and for every node $v$ in $V$ with $\frac{\rs[v]}{d(v)}\le \delta$, $\textstyle \left| \frac{\ar[v]}{d(v)}-\frac{\rs[v]}{d(v)}\right| \le \epsilon_r\delta.$
	
	By the definition of $p^{\prime}_f$ in Equation~(\ref{eq:pfail}), if $\sum_{v\in V}{p_f^{d(v)-1}} \le 1$, then $p^{\prime}_f=p_f$, which leads to $\textstyle \sum_{v\in V}{(p^{\prime}_f)^{d(v)}} = \sum_{v\in V}{p_f^{d(v)}}\le p_f;$ otherwise, $p^{\prime}_f=\frac{p_f}{\sum_{v\in V}{p_f^{d(v)-1}}}$, hence $\textstyle \sum_{v\in V}{(p^{\prime}_f)^{d(v)}} < \sum_{v\in V}{\frac{p_f^{d(v)}}{\sum_{v\in V}{p_f^{d(v)-1}}}}=p_f.$
	Namely, $\ar$ is a $(d,\epsilon_r,\delta)$-approximate HKPR vector with probability at least $1-p_f$.
\end{proof}

\vspace{-1ex}\subsection{Proof of Theorem~\ref{thrm:hkpush}}
\begin{proof}
	By Lemma~\ref{lem:mkrev} and the definition of $\mathbf{h}^{(k)}_u[v]$ in Equation~\ref{eq:hsum-def}, we have $\textstyle  \frac{\mathbf{h}^{(k)}_u[v]}{d(v)} = \frac{\mathbf{h}^{(k)}_v[u]}{d(u)}.$
	Then we can rewrite Equation~(\ref{eq:hkpush-invar}) as follows
	\begin{align*}
		\textstyle \rs[v]-\qs[v]&=\textstyle d(v)\cdot \sum_{u\in V}\sum_{k=0}^{K}{\left[\frac{\rks[u]}{d(u)}\cdot \mathbf{h}^{(k)}_v[u]\right]}\\
		& \textstyle \le d(v)\cdot \sum_{k=0}^{K}{\left[\max_{u\in V}{\left\{\frac{\rks[u]}{d(u)}\right\}}\cdot\sum_{u\in V}{\mathbf{h}^{(k)}_v[u]}\right]}.
	\end{align*}
	Now we prove that $\sum_{u\in V}{\mathbf{h}^{(k)}_v[u]}=1$ for any node $v\in V$ and $k\in [0,K]$.
	\begin{align*}
		\textstyle \sum_{u\in V}{\mathbf{h}^{(k)}_v[u]} &\textstyle=\sum_{\ell=0}^{\infty}{\left[\frac{\eta(k+\ell)}{\psi(k)}\sum_{u\in V}{\mathbf{P}^{\ell}[v,u]}\right]}\\
		&\textstyle=\sum_{\ell=0}^{\infty}{\frac{\eta(k+\ell)}{\psi(k)}}=1.
	\end{align*}
	Hence, $\textstyle \rs[v]-\qs[v]\le d(v)\cdot \sum_{k=0}^{K}{\max_{u\in V}{\left\{\frac{\rks[u]}{d(u)}\right\}}}.$
	Once Inequality (\ref{eq:hkpush-con3}) held, we have $\textstyle  \frac{\rs[v]}{d(v)}-\frac{\qs[v]}{d(v)} \le \epsilon_a,$ which completes the proof.
\end{proof}

\vspace{-1ex}\subsection{Proof of Theorem~\ref{thrm:pukraplus}}
\begin{proof}
	First, consider the case where \pukraplus terminates at Line 7, i.e., $\sum_{k=0}^{K}{\max_{u\in V}{\left\{\frac{\rks[u]}{d(u)}\right\}}} \le \epsilon_a$. In that case, by Theorem \ref{thrm:hkpush}, for any node $v\in V$, $\textstyle \left|\frac{\ar[v]}{d(v)}-\frac{\rs[v]}{d(v)}\right| \le \epsilon_a = \epsilon_r\cdot \delta.$
	This indicates that, for any node $v$ with $\rs[v]> d(v)\cdot\delta$, $\textstyle \left|\frac{\ar[v]}{d(v)}-\frac{\rs[v]}{d(v)}\right| \le \epsilon_r \cdot\frac{\rs[v]}{d(v)}.$ In addition, for any node $v$ with $\rs[v]\le d(v)\cdot\delta$, $\textstyle \left|\frac{\ar[v]}{d(v)}-\frac{\rs[v]}{d(v)}\right| \le \epsilon_r\cdot \delta.$ Thus, $\ar$ is a $(d,\epsilon_r,\delta)$-approximate HKPR vector.
	
	Now consider the case where \pukraplus does not terminate at Line 7. Let $Y_i$ be as defined in the context of Equation~(\ref{eq:pukra_asum}), and let $Y=\sum_{i=1}^{n_r}{Y_i}$. By Lines 17 and 19 of \pukraplus,
	\begin{equation*}
		\textstyle \ar[v]=\qs[v]+\frac{Y\cdot \alpha}{n_r} +\frac{\epsilon_r\delta}{2}\cdot d(v).
	\end{equation*}
	By Equation (\ref{eq:pukra-exp-xi}), the expectation of $Y$ is
	\begin{equation*}
		\textstyle \E[Y]=\E[\sum_{i=1}^{n_r}{Y_i}]=\frac{n_r}{\alpha}\left({\rs[v]-\qs[v]-\bs[v]}\right) \le\frac{n_r}{\alpha}\cdot {\rs[v]}.
	\end{equation*}
	Let $\lambda=\frac{n_r\epsilon_r}{2\cdot\alpha}\cdot \rs[v]$  and $n_r$ be as defined in Line 13 of \pukraplus.
	By the Chernoff bound (see Lemma~\ref{lem:chernoff}), for any node $v$ in $V$ with $\rs[v]> d(v)\cdot \delta$, we have
	\begin{align*}
		&\textstyle \mathbb{P}[ |Y-\mathbb{E}[Y]| \ge\lambda]\\
		&\textstyle =\mathbb{P}\left[\left| {{\ar[v]-\rs[v]+\bs[v]}-\frac{\epsilon_r\delta}{2}\cdot d(v)}\right| \ge \frac{\epsilon_r}{2} \cdot{\rs[v]}\right]\\
		&\textstyle \le\exp\left(-\frac{n_r\cdot\epsilon^2_r\cdot \rs[v]}{8\alpha\cdot (1+\epsilon_r/6)}\right)\le (p^{\prime}_f)^{d(v)}.
	\end{align*}
	On the other hand, let $\lambda=\frac{n_r\epsilon_r\delta d(v)}{2 \alpha}$. Then, for any node $v$ in $V$ with $\rs[v]\le d(v)\cdot\delta$, we have
	\begin{align*}
		&\textstyle \mathbb{P}[ |Y-\mathbb{E}[Y]| \ge \lambda] \\ &\textstyle  =\mathbb{P}\left[\left|{{\ar[v]-\rs[v]+\bs[v]}-\frac{\epsilon_r\delta}{2}\cdot d(v)}\right| \ge \frac{\epsilon_r\delta}{2}\cdot d(v)\right]\\
		&\textstyle  \le\exp\left(-\frac{n_r\cdot \epsilon^2_r\delta^2 d^2(v)}{8\alpha(1+\epsilon_r/6)\cdot\rs[v]}\right)\le (p^{\prime}_f)^{d(v)}.
	\end{align*}
	
	By union bound, for $V_1=\{v| v\in V \ s.t. \ \rs[v]> d(v)\cdot\delta\}$, we have
	\begin{align*}
		&\textstyle \mathbb{P}\left[\bigcup_{v\in V_1}\left\{\left| {\frac{\ar[v]-\rs[v]+\bs[v]}{d(v)}-\frac{\epsilon_r\delta}{2}}\right| \ge \frac{\epsilon_r}{2} \cdot\frac{\rs[v]}{d(v)} \right\}\right] \\
		&\textstyle \le \sum_{v\in V_1}{(p^{\prime}_f)^{d(v)}},
	\end{align*}
	and for $V_2=\{v| v \in V \ s.t. \ \rs[v] \le d(v)\cdot\delta\}$, we have
	\begin{align*}
		\textstyle \mathbb{P}\left[\bigcup_{v\in V_2}\left\{\left|{\frac{\ar[v]-\rs[v]+\bs[v]}{d(v)}-\frac{\epsilon_r\delta}{2}}\right| \ge \frac{\epsilon_r\delta}{2}\right\}\right] \le \sum_{v\in V_2}{(p^{\prime}_f)^{d(v)}}.
	\end{align*}
	
	%

	By Inequality~(\ref{eq:bsbound}), $\left|\frac{\bs[v]}{d(v)}-\frac{\epsilon_r\delta}{2}\right| \le \frac{\epsilon_r\delta}{2}$. Then, we have the following results, respectively.
	With probability at least $1-\sum_{v\in V}{(p^{\prime}_f)^{d(v)}}$, for every node $v$ in $V$ with $\rs[v]> d(v)\cdot \delta$,
	\begin{equation*}
		\textstyle \left| \frac{\ar[v]}{d(v)}-\frac{\rs[v]}{d(v)}\right|\le \frac{\epsilon_r}{2} \cdot\frac{\rs[v]}{d(v)}+\frac{\epsilon_r\delta}{2}\le \epsilon_r \cdot\frac{\rs[v]}{d(v)},
	\end{equation*}
	and for every node $v$ in $V$ with $\rs[v]\le d(v)\cdot\delta$,
	$$\textstyle \left| \frac{\ar[v]}{d(v)}-\frac{\rs[v]}{d(v)}\right| \le \epsilon_r\delta.$$
	
	By the definition of $p^{\prime}_f$ in Equation~(\ref{eq:pfail}), the total failure probability will be at most $\sum_{v\in V}{(p^{\prime}_f)^{d(v)}} \le p_f$. Therefore, $\ar$ is a $(d,\epsilon_r,\delta)$-approximate HKPR vector with probability at least $1-p_f$. 
\end{proof}

\end{document}